\documentclass[pra,superscriptaddress,showpacs,aps,10pt]{revtex4-1}

\usepackage{xcolor}
\usepackage{colortbl}
\usepackage{graphicx}
\usepackage{amssymb,amsmath}
\usepackage{subfigure}
\usepackage{enumitem}
\usepackage{bm}
\usepackage{textcomp}
\usepackage{gensymb}
\usepackage{braket}
\usepackage{siunitx}

\usepackage[T1]{fontenc}

\usepackage[%
  bookmarks=true,
  colorlinks,
  linkcolor=blue,
  urlcolor=blue,
  citecolor=blue,
  plainpages=false,
  pdfpagelabels,
  final,
  breaklinks=true,
]{hyperref}
\hypersetup{
  pdftitle={Symphony on Strong Field Approximation}, 
  pdfauthor={Kasra Amini, Jens Biegert, Francesca Calegari, Alexis Chac\'on, Marcelo F. Ciappina, Alexandre Dauphin, Dmitry K. Efimov, Carla Figueira de Morisson Faria, Krzysztof Giergiel, Piotr Gniewek, Alexandra S. Landsman, Micha{\l} Lesiuk, Micha{\l} Mandrysz, Andrew S. Maxwell, Robert Moszy\'nski, Lisa Ortmann, Jose Antonio P\'erez-Hern\'andez, Antonio Pic\'on, Emilio Pisanty, Jakub Prauzner-Bechcicki, Krzysztof Sacha, Noslen Su\'arez, Amelle Za\"ir, Jakub Zakrzewski and Maciej Lewenstein}
}

\usepackage{xmpincl}
\includexmp{metadata} 
\makeatletter
\renewcommand\frontmatter@abstractwidth{\dimexpr\textwidth-1.15in\relax}
\makeatother

\newcommand{\orcid}[1]{%
  \href{https://orcid.org/#1}{\,\protect\includegraphics[width=7pt]{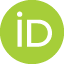}}%
}

\graphicspath{{Figures/}{}}

\usepackage[sort&compress]{natbib}

\begin{document}

\title{Symphony on Strong Field Approximation}

%Kasra Amini, Jens Biegert, Francesca Calegari, Alexis Chacón, Marcelo F. Ciappina, Alexandre Dauphin, Dmitry K. Efimov, Carla Figueira de Morisson Faria, Krzysztof Giergiel, Piotr Gniewek, Alexandra S. Landsman, Michał Lesiuk, Michał Mandrysz, Andrew S. Maxwell, Robert Moszyński, Lisa Ortmann, Jose Antonio Pérez-Hernández, Antonio Picón, Emilio Pisanty, Jakub Prauzner-Bechcicki, Krzysztof Sacha, Noslen Suárez, Amelle Zaïr, Jakub Zakrzewski and Maciej Lewenstein

\author{Kasra Amini\orcid{0000-0002-2491-1672}}
\affiliation{Faculty of Chemistry, University of Warsaw, Pasteura 1, 02-093 Warsaw, Poland}   
\affiliation{ICFO -- Institut de Ciencies Fotoniques, The Barcelona Institute of Science and Technology, 08860 Castelldefels (Barcelona), Spain}

\author{Jens Biegert\orcid{0000-0002-7556-501X}}
\affiliation{ICFO -- Institut de Ciencies Fotoniques, The Barcelona Institute of Science and Technology, 08860 Castelldefels (Barcelona), Spain}
\affiliation{ICREA, Pg. Llu\'{\i}s Companys 23, 08010 Barcelona, Spain}

\author{Francesca Calegari\orcid{0000-0003-3234-7298}}
\affiliation{Center for Free-Electron Laser Science, DESY, Notkestr. 85, 22607 Hamburg, Germany}
\affiliation{CNR, Instituto di Fotonica e Nanotecnologie Milano, Piazza L. da Vinci 32, 20133 Milano, Italy}

\author{Alexis Chac\'on\orcid{0000-0002-9279-4463}}
\affiliation{Center for Nonlinear Studies and Theoretical Division, Los Alamos National Laboratory, Los Alamos, New Mexico 87545, USA}

\author{Marcelo F. Ciappina\orcid{0000-0002-1123-6460}}
\affiliation{Institute of Physics of the ASCR, ELI-Beamlines project, Na Slovance 2, 182 21 Prague, Czech Republic}

\author{Alexandre Dauphin\orcid{0000-0003-4996-2561}}
\affiliation{ICFO -- Institut de Ciencies Fotoniques, The Barcelona Institute of Science and Technology, 08860 Castelldefels (Barcelona), Spain}

\author{Dmitry K. Efimov\orcid{0000-0002-1264-4044}}
\affiliation{Instytut Fizyki imienia Mariana Smoluchowskiego, Uniwersytet Jagiello\'nski, \L{}ojasiewicza 11, 30-348 Krak\'ow, Poland}

\author{Carla Figueira de Morisson Faria\orcid{0000-0001-8397-4529}}
\affiliation{Department of Physics \& Astronomy, University College London, Gower Street London WC1E 6BT, United Kingdom}

\author{Krzysztof Giergiel\orcid{0000-0003-3297-796X}}
\affiliation{Instytut Fizyki imienia Mariana Smoluchowskiego, Uniwersytet Jagiello\'nski, \L{}ojasiewicza 11, 30-348 Krak\'ow, Poland}

\author{Piotr Gniewek}
\affiliation{Faculty of Chemistry, University of Warsaw, Pasteura 1, 02-093 Warsaw, Poland}

\author{Alexandra S. Landsman\orcid{0000-0002-8194-8439}}
\affiliation{Max-Planck Institut f\"ur Physik Komplexer Systeme, N\"othnitzer-Strasse 38, D-01187 Dresden, Germany}
\affiliation{Department of Physics, Max Planck Postech, Pohang, Gyeongbuk 37673, Republic of Korea}

\author{Micha\l~Lesiuk\orcid{0000-0002-7928-4450}}
\affiliation{Faculty of Chemistry, University of Warsaw, Pasteura 1, 02-093 Warsaw, Poland}   

\author{Micha\l\ Mandrysz}
\affiliation{Instytut Fizyki imienia Mariana Smoluchowskiego, Uniwersytet Jagiello\'nski, \L{}ojasiewicza 11, 30-348 Krak\'ow, Poland}

\author{Andrew S. Maxwell\orcid{0000-0002-6503-4661}}
\affiliation{Department of Physics \& Astronomy, University College London, Gower Street London WC1E 6BT, United Kingdom}

\author{Robert Moszy\'nski}
\affiliation{Faculty of Chemistry, University of Warsaw, Pasteura 1, 02-093 Warsaw, Poland}

\author{Lisa Ortmann\orcid{0000-0002-2665-8471}}
\affiliation{Max-Planck Institut f\"ur Physik komplexer Systeme, N\"othnitzer-Strasse 38, D-01187 Dresden, Germany}

\author{Jose Antonio P\'erez-Hern\'andez\orcid{0000-0002-1117-0190}}
\affiliation{Centro de L\'aseres Pulsados (CLPU), Parque Cient\'ifico, E-37185 Villamayor, Salamanca, Spain}

\author{Antonio Pic\'on\orcid{0000-0002-6142-3440}}
\affiliation{ICFO -- Institut de Ciencies Fotoniques, The Barcelona Institute of Science and Technology, 08860 Castelldefels (Barcelona), Spain}
 \affiliation{Departamento de Qu\'imica, Universidad Aut\'onoma de Madrid, 28049, Madrid, Spain}

\author{Emilio Pisanty\orcid{0000-0003-0598-8524}}
\affiliation{ICFO -- Institut de Ciencies Fotoniques, The Barcelona Institute of Science and Technology, 08860 Castelldefels (Barcelona), Spain}

\author{Jakub Prauzner-Bechcicki\orcid{0000-0003-3986-1898}}
\affiliation{Instytut Fizyki imienia Mariana Smoluchowskiego, Uniwersytet Jagiello\'nski, \L{}ojasiewicza 11, 30-348 Krak\'ow, Poland}

\author{Krzysztof Sacha\orcid{0000-0001-6463-0659}}
\affiliation{Instytut Fizyki imienia Mariana Smoluchowskiego, Uniwersytet Jagiello\'nski, \L{}ojasiewicza 11, 30-348 Krak\'ow, Poland}
\affiliation{Mark Kac Complex Systems Research Center, Jagiellonian University, \L{}ojasiewicza 11, 30-348 Krak\'ow, Poland}

\author{Noslen Su\'arez\orcid{0000-0002-3705-741X}}
\affiliation{ICFO -- Institut de Ciencies Fotoniques, The Barcelona Institute of Science and Technology, 08860 Castelldefels (Barcelona), Spain}

\author{Amelle Za\"ir\orcid{0000-0003-1687-5453}}
\affiliation{King's College London, Department of Physics, London WC2R 2LS, United Kingdom}

\author{Jakub Zakrzewski\orcid{0000-0003-0998-9460}}
\affiliation{Instytut Fizyki imienia Mariana Smoluchowskiego, Uniwersytet Jagiello\'nski, \L{}ojasiewicza 11, 30-348 Krak\'ow, Poland}
\affiliation{Mark Kac Complex Systems Research Center, Jagiellonian University, \L{}ojasiewicza 11, 30-348 Krak\'ow, Poland}

\author{Maciej Lewenstein\orcid{0000-0002-0210-7800}}
\email[]{maciej.lewenstein@icfo.eu}
\affiliation{ICFO -- Institut de Ciencies Fotoniques, The Barcelona Institute of Science and Technology, 08860 Castelldefels (Barcelona), Spain}
\affiliation{ICREA, Pg. Llu\'{\i}s Companys 23, 08010 Barcelona, Spain}

\date{15 October 2019}
%\pacs{32.80.Rm,33.20.Xx,42.50.Hz}

\begin{abstract}

This paper has been prepared by the Symphony collaboration (University of Warsaw, Uniwersytet Jagiello\'nski, DESY/CNR and ICFO) on the occasion of the 25th anniversary of the ``simple man's models'' which underlie most of the phenomena that occur when intense ultrashort laser pulses interact with matter. 
The phenomena in question include High-Harmonic Generation (HHG), Above-Threshold Ionization (ATI), and Non-Sequential Multielectron Ionization (NSMI). ``Simple man's models'' provide, both an intuitive basis for understanding the numerical solutions of the time-dependent Schr\"odinger equation, and the motivation for the powerful analytic approximations generally known as the Strong Field Approximation (SFA). 
In this paper we first review the SFA in the form developed by us in the last 25 years. 
In this approach the SFA is a method to solve the TDSE, in which the non-perturbative interactions are described by including continuum-continuum interactions in a systematic perturbation-like theory.
In this review we focus on recent applications of the SFA to HHG, ATI and NSMI from multi-electron atoms and from multi-atom molecules. 
The main novel part of the presented theory concerns generalizations of the SFA to: 
(i) time-dependent treatment of two-electron atoms, allowing for studies of an interplay between Electron Impact Ionization (EII) and Resonant Excitation with Subsequent Ionization (RESI); 
(ii) time-dependent treatment in the single active electron (SAE) approximation of ``large'' molecules and targets which are themselves undergoing dynamics during the HHG or ATI processes. 
In particular, we formulate the general expressions for the case of arbitrary molecules, combining input from quantum chemistry and quantum dynamics. 
We formulate also theory of time-dependent separable molecular potentials to model analytically the dynamics of realistic electronic wave packets for molecules in strong laser fields.

$\ $

\noindent
We dedicate this work to the memory of Bertrand Carr\'e, who passed away in March 2018 at the age of 60.

$\ $
\vspace{7.75pt}

\noindent
Accepted Manuscript for 
\href{https://dx.doi.org/10.1088/1361-6633/ab2bb1}%
  {\textit{Rep. Prog. Phys.} \textbf{82} no.~11, 116001 (2019)}, 
\href{https://arxiv.org/abs/1812.11447}%
  {arXiv:1812.11447},
available under the %
\href{https://creativecommons.org/licenses/by-nc-nd/3.0/}%
  {CC BY-NC-ND} license.
\end{abstract}

\maketitle

\section{Introduction}
\label{sec:1-introduction}

\subsection{Motivation}
\label{subsec:1a-motivation}

In the last three decades, we have witnessed incredible advances in laser technology and in the understanding of nonlinear laser-matter interactions, crowned recently by the award of the Nobel prize to G\'erard Mourou and Donna Strickland~\cite{Strickland1985, CPA-Nobel-2018}. 
It is now routinely possible to  produce few-cycle femtosecond ($\SI{1}{fs} = \SI{e-15}{s}$) laser pulses in the visible and mid-infrared regimes~\cite{Nisoli1997, Hemmer2013}. By focusing such ultrashort laser pulses on gas or solid targets, possibly in the presence of nano-structures \cite{Ciappina2017}, the targets are subjected to an ultra-intense electric field, with peak field strengths approaching the binding field inside the atoms themselves.
Such fields permit the exploration of the interaction between  strong  electromagnetic coherent radiation and an atomic or molecular system  with unprecedented spatial and temporal resolution~\cite{Krausz2009}. 
On one hand, HHG nowadays can be used to generate attosecond  pulses in the extreme ultraviolet~\cite{Drescher2001, Chang2012}, or even in the soft X-ray regime~\cite{Silva2015}. Such pulses themselves may be used for dynamical spectroscopy of matter; despite carrying modest pulse energies, they exhibit excellent coherence properties \cite{Pascalcoh, AnneML}. 
Combined with femtosecond pulses they can also be used to extract  information about the laser pulse electric field itself~\cite{Paulus2001}. HHG sources therefore offer an important alternative to other sources of XUV and X-ray radiation: synchrotrons, free electron lasers, X-ray lasers, and laser plasma sources.
Moreover, HHG pulses can provide information about the structure of the target atom, molecule or solid~\cite{Itatani2004, Blaga2012, Pullen2015}. Of course, to decode such information from a highly nonlinear HHG signal is a challenge, and that is why a possibly perfect, and possibly ``as analytical as possible'' theoretical understanding of these processes is in high demand. Here is the first instance where SFA offers its basic services.

In HHG, an electron is liberated from an atom or molecule through ionization, which occurs close to the maximum of the electric field. Within the oscillating field, the electron can thus accelerate along oscillating trajectories, which may result in recollision with the parent ion, roughly when the laser field approaches a zero value. 
Since electronic motion is governed by the waveform of the laser electric field, an important quantity to describe the electric field shape is the so-called absolute phase or carrier-envelope phase (CEP). 
Control over the CEP is paramount for extracting information about  electron dynamics, and to retrieve structural information from atoms and molecules~\cite{Baltuska2003, Itatani2004, Wolter2014}. 
Control over the CEP is particularly important for HHG, when targets are driven by laser pulses comprising only one or two optical cycles. 
In that situation the CEP determines the relevant electron trajectories, i.e.~the CEP determines whether emission results in a single or in multiple attosecond bursts of radiation~\cite{Baltuska2003, Sola2006}.

The influence of the CEP on electron emission is also extremely important. It  was demonstrated for instance in an anti-correlation experiment, in which the number of ATI electrons emitted in opposite directions was measured~\cite{Paulus2001, Milosevic2006}. 
Since the first proof-of-principle experiment~\cite{Paulus2001}, the stereo ATI technique has established itself as a direct measure of the CEP, and demonstrated its ability for single-shot measurements even at multi-kHz laser repetition rates. 
Both bound-free and the rescattering continuum-continuum transitions are CEP sensitive;
hence, the photoelectron distribution of ATI can also be used to extract structural information about the target. 
Again, theoretical understanding of these processes and the SFA, which are ``as analytical as possible'', are more than welcome in this line of research.

Laser-induced electron diffraction (LIED) is a technique that uses the doubly-differential elastic scattering cross-sections to extract structural information~\cite{Zuo1996, Lein2002, Lin2010}. 
Meeting the requirements to extract structural information has, however, proven difficult due to the stringent prerequisites on the laser parameters. During
recent years, the development of new laser sources has dramatically advanced, leading to the first demonstrations of the technique~\cite{Xu2014, Meckel2008, Morishita, Xu2012, Pullen2015}, and the successful retrieval of the bond distances in simple diatomic molecules with fixed-angle broadband electron scattering~\cite{Xu2014}. 
Recently, Pullen et al.~\cite{Pullen2015} have exploited the full double differential cross section to image the entire structure of a polyatomic molecule for the first time. 
Again, to exploit the full potential of the recollision physics and the intrinsic time resolution of LIED (and eventually HHG), we need the comprehensive and complete understanding of the ATI and HHG processes and its theoretical, possibly analytical
description~\cite{Milosevic2006, Faisal1973, Reiss1980, Lewenstein1995, Starace2014, Becker2002, Morishita2010}. Here is the third instance where SFA is indispensable.

The key for gaining dynamical structural information and for realizing nonlinear dynamical spectroscopy with HHG, ATI and, last but not least, NSMI,  however, consists in generalizations of the existing theories to the case, when the target in question itself undergoes dynamics beyond its SAE electronic structure.
For NSMI this requires including two, or even more electrons in the SFA theory. For molecules that would mean, for instance, developing theory that takes into account vibration or dissociation processes, occurring on the time scale comparable with the laser pulse duration. {\it Ab initio} theory of this kind is generally too computationally intensive for numerical simulation. Generalizations of the SFA to include many electrons and/or nuclear motion are thus more than welcome.

\vfill

\subsection{History of the SFA}
\label{subsec:1b-history}

The history of the SFA has at least five beginnings, many intertwined and entangled paths, and many interfering endings. The point of view presented here is dominated by the opinion of one of us (M.L.), who started to work on the subject in 1980.
The five beginnings of SFA, according to this point of view, are the following:

\begin{itemize}

\item {\bf Keldysh ionization theory.} 
In 1964 L.V. Keldysh~\cite{Keldysh1965} attempted and solved analytically a very fundamental problem of tunnelling ionization of atoms in strong low-frequency oscillating electric fields. In the epochal paper entitled ``Ionization in the Field of a Strong Electromagnetic Wave'' he derived his famous Bessel function formulae for ionization rates. F.H.M. Faisal \cite{Faisal1973} and H.R. Reiss \cite{Reiss1980}, building also on earlier work by Perelomov, Popov and Terent'ev~\cite{Perelomov1966, Perelomov1967, Perelomov1967b}, extended this theory to calculate electron spectra in the process to be named Above-Threshold ionization (ATI).
These results were the first instances of the Strong Field Approximation, named ``KFR theory'' after the authors. 
Note that in this initial phase, SFA was exclusively applied to ionization problems. 
The first experiments on ATI date from 1979~\cite{Agostini1979}; in fact, the name of
the observed phenomenon slowly shifted from ``continuum-continuum transitions'' to ATI~\cite{Kruit1981, Fabre1982, Kruit1983}. The first observations of HHG date from 1988~\cite{Ferray1988, McPherson1987}, and of non-sequential two-electron ionization originate from 1983 [\citealp{Anne1983a,Anne1983b}; for more recent reviews see~\citealp{Mainfray1991, Krausz2007}].
Keldysh theory was further developed to calculate total ionization rates for various atomic species and states by M.V.\ Ammosov, N.B.\ Delone and V.P.\ Krainov~\cite{Ammosov1986}---the resulting expressions are known as ADK rates. 
More broadly, Keldysh theory has been an inspiration for many years for many scholars. M.L.\ first learned the Keldysh theory from a preprint by L.~Davidovich et al., ICTP Trieste. In fact, Davidovich later published several interesting papers on Keldysh theory~\cite{Brandi1982, Neto1990, Becker1994}. In the beginning of the 1980s several non-perturbative, quantum-optics-inspired models of ``continuum-continuum transitions'' and ATI were introduced by Z.~Bia{\l}ynicka-Birula~\cite{Bialynicka1983}, J.H.~Eberly\cite{Deng1984},  K.~Rz\k{a}\.zewski~\cite{Rzazewski1985} and others (for a review see Ref.~\citealp{EberlyRev1991}). These models stimulated M.L. to try to combine them with the Keldysh theory.

\item{\bf Kroll-Watson theory.}
Another inspiration for the contemporary SFA came from the seminal papers of N.M. Kroll and K.M. Watson on electron-atom \cite{Kroll1973} and atom-atom \cite{Kroll1976} scattering in the presence of a strong electromagnetic wave. The earlier paper clearly dealt with the problem of ``continuum-continuum transitions'', dressed by the laser field, leading to the expected Bessel-function dependence of the corresponding transition amplitudes. This observation led to the formulations of the ATI theory as a theory of multichannel decay and continuum-continuum transitions, dressed by the laser field~\cite{Lewenstein1985}. This approach, employing the relation between ATI and electron scattering in the intense laser field has been deepened and developed further in Ref.~\citealp{Grochmalicki1986}. In the contemporary language, the results of these studies described ATI as a combination of direct tunnelling processes, and rescattering processes occurring in the laser dressed continuum. At that time, however, the underlying quasi-classical theory and the simple man's model was yet not known. It is worth noting that this approach was also applied to two-electron ionization in Ref.~\citealp{Lewenstein1986a}, where the direct two-electron tunnelling processes were analysed.

\item{\bf Numerical studies of TDSE.} 
Numerical simulations always played, play and will play a fundamental role in our understanding of physics of matter in intense laser fields. A particularly important role was played here by 1D models of one- and two-electron systems, initiated by J.H. Eberly on ``Eberlonium'', also known as the Rochester atom model. This series of studies allowed the description of several qualitatively important results but, more importantly, it allowed---by appropriate tuning of the parameters---the finding of accurate quantitative predictions concerning ATI~\cite{Javanainen1988}, HHG~\cite{Eberly1989, Eberly1989a}, stabilization of a 1D atom in the presence of a strong field of high frequency~\cite{Su1990}---all of that  optimizing the ``model atom for multiphoton physics''~\cite{Su1991}. 
This approach was very successfully generalized to 1D two-electron models \cite{Grobe1992, Grobe1993}, which in turn stimulated the development of other quasi-1D approaches to the two-electron problems in intense laser fields. These developments were very important, especially in a view of the difficulties and computational cost of solving TDSE for helium in 3D~[\citealp{Dundas1999, Taylor2003, Parker2000, Emma2011}; see also \citealp{Feist2008,Pazourek2012}].
Analysis of the classical pathways for simultaneous escape of two electrons showed that there are two saddle points in the effective potential landscape located symmetrically with respect to the field polarization axis~\cite{Sacha2001}. This led to a modification of the 1D model, where electrons move along axes inclined symmetrically with respect to the polarization direction~\cite{ Eckhardt2006, Prauzner2007, Prauzner2008}. 
Within this model the ionization for three active electrons was also recently considered~\cite{Thiede2018}.
Another model, in which the center of mass movement was restricted to the polarization axis was introduced by Ruiz et al.~\cite{Ruiz2006}, and successfully applied to momentum distributions~\cite{Staudte2007}---for a comparison of various quasi-1D approaches, see Ref.~\citealp{Efimov2018}.

Nevertheless, the most important were investigations of the TDSE in 3D, led in those years by K. Kulander, who not only developed codes for solving the TDSE, but also propagation and phase matching in HHG, and collaborated intensively with the top experimental groups. Several seminal papers were written on ATI~\cite{Schafer1993}, double ionization of helium~\cite{Fittinghoff1992, Walker1994} and phase matching in HHG~\cite{Lhuillier1991}. The one that was the most important for the formulation of the simple man's models was the theory paper on HHG from atoms and ions in the high-intensity regime~\cite{Krause1992}, in which the famous cut-off law for HHG was discovered: the high harmonics cease to exist above the photon  energy $I_p+ 3.2U_p$, where $I_p$ is the ionization potential, and $U_p$ is the ponderomotive energy.

\item{\bf Classical phase-space methods.}
The key idea in these approaches was to mimic the evolution of the electronic system in terms of classical trajectories, governed by the completely classical Hamiltonians, but satisfying an initial phase-space distribution. 
In fact, that is why some versions of this approach were termed the ``truncated Wigner function'' approach. Initially, these methods were proposed for microwave perturbations of atomic systems by J. Leopold and I.C. Percival~\cite{Leopold,Percival}. They gained a lot of attention and popularity in the studies of quantum chaos and quantum dynamical localization~\cite{Casati1987}. 
Later, these methods were extended to the regime of mid-infrared, and even high laser frequencies and two-electron systems. There are two variants of these methods that can be distinguished: those where the initial distribution is calculated classically~\cite{ Groch1991, Gajda1992, Rzazewski1994, Wojcik1995, Ho2007}, and those which use the below-barrier tunneling approximations for the calculation of the initial distribution~\cite{Chen2000}. 
The quality of classical phase-space averaging methods can be checked by comparing their results with those from the corresponding quantum-mechanical models~\cite{Shvetsov2016}. Considerable progress was made with these methods on the study of the ionization yields~\cite{Ho2005, Mauger2009} and momentum distributions~\cite{Panfili2001}. 
Still, getting HHG spectra within classical methods exclusively seems to be a complicated task---for a recent discussion see Ref.~\citealp{Berman2018}.
These approaches are related to the so-called Initial Value Representation methods used in quantum chemistry~\cite{VanDeSand1999, Zagoya2012, Zagoya2014, Symonds2015}, though those methods are not purely classical, since they also account for phase shifts.

\item{\bf Simple man's models.} 
There were simple man's models before the simple man's model. An extended discussion of precursors is contained in a recent review in Ref.~\citealp{Ivanov2014}, where the earlier quantum formulation of ``Atomic Antennas'' of  M.Y.\ Kuchiev is discussed~\cite{Kuchiev1987}, as well as early attempts by F.\ Brunel and P.\ Corkum himself~\cite{Brunel1987, Brunel1990, Corkum1989}. 
Essentially the same formulae as the ones derived later in the framework of the SFA for HHG were obtained by F.\ Ehlotzky~\cite{Ehlotzky1992}, but without the underlying quasi-classical picture. The history of science chooses its own heroes. 
Nowadays, the simple man's model, also known as the ``three-step'' or ``recollision'' model, is usually attributed to P.\ Corkum~\cite{Corkum1993}, K.\ Kulander~\cite{Krause1992, Kulander1993}, and to a conference contribution of  H.\ Muller. 
These formulations were  done in the right place in the right time, and were truly seminal---they have revolutionized the whole area! 
M.L.\ learned about the simple man's model for HHG during and immediately after the famous NATO Workshop on Super Intense Laser Atom Physics (SILAP) in Han-sur-Lesse, Belgium, in January of 1993~\cite{footnote}. 
After the workshop M.\ Yu.\ Ivanov visited Saclay and stayed at M.L.'s house---that is how our first paper on the SFA for HHG, based on simple man's model, and  co-authored by  A. L'Huillier and other colleagues, was written~\cite{Anne1993}. 
After a long fight with Phys.\ Rev.\ Lett., this paper was published as a Rapid Communication in Phys.\ Rev.\ A, entitled ``High-order harmonic-generation cutoff''.  We termed the formula we used to evaluate the time-dependent dipole moment a {\it dynamical Landau-Dykhne formula}.

M.L. went to JILA in February 1993 and started to work on the long version of the theory, including a detailed discussion of the relation of the simple man's model with the quasi-classical (better termed quantum-orbit) saddle point approximations, along with concrete calculations for what we called Gaussian models, i.e.~models in which the ground state of the atom of interest was approximated by a Gaussian function. 
The paper on the theory of HHG by low-frequency laser fields appeared in Phys. Rev. A in 1994, and soon became a reference paper for theorists and experimentalists working in the field~\cite{Lewenstein1994}. 
During his stay in JILA M.L. worked with K. Kulander on the extension of the newly-formulated version of the SFA to ATI, stimulated by the observation of the intensity-dependent rings in the high order ATI~\cite{Yang1993}. The paper that extended a quasi-classical analysis (i.e.~one based on simple man's model) of rescattering processes in ATI appeared in Phys. Rev. A in 1995~\cite {Lewenstein1995}. 
On one hand, it explicitly demonstrated in which sense the SFA is a systematic perturbation theory in part of the Hamiltonian describing the continuum-continuum transitions. 
On the other hand, we introduced here for the first time the model atom involving a separable (non-local) potential. 
This kind of approach was a generalization of the so-called zero-range Becker's model~\cite{Becker1994a}. Recently, it turned out to be extremely useful in modelling
HHG and ATI in molecular dimers, trimers and quadrimers~\cite{Noslen1, Noslen2, Noslen3, Noslen4, Noslen-diss}. Other uses of separable potentials in the literature are discussed in Refs.~\citealp{Faisal1987, Faisal1987b, Faisal1989, Faisal1990, Faisal1991, Faisal1993, Faisal1993b, Radozycki1993, Kaminski1989, Klaiber2005, TetchouNganso2007, TetchouNganso2011, TetchouNganso2013, Galstyan2017, Galstyan2018}.

The quantum simple man's models, as the novel SFA was termed sometimes, proved very useful in explaining the relation of quantum orbits to phase matching in HHG. It allowed thus to realise coherence control in high-order harmonics~\cite{Pascal1995}, and  understand the behaviour of the phase of the atomic polarization in high-order harmonic generation~\cite{Maciek1995}, which in turn allowed the construction of schemes for generation of attosecond pulse trains using HHG~\cite{Antoine1996}. 
The first such trains were observed in by P. Agostini {\it et consortes} in 2001~\cite{Paul2001}. 
Equally well, the quantum simple man's model allowed for explanations, both intuitive and quantitatively accurate, of the generation of a single isolated attosecond pulse by an ultrashort, few-cycle laser pulse~\cite{Hentschel2001, Drescher2001, Kienberger2004, Sansone2006}.

The crowning of the SFA applications in the 1990s was perhaps the \textit{Science} paper~\cite{Pascal2001}, in which theory was confronted with the experimental results of the groups of the late B. Carr\'e and P. Sali\`eres at Saclay, on quantum orbits in HHG, and of G. Paulus and H. Walther at MPI Garching, on quantum orbits in ATI, driven by elliptically-polarized laser fields.

\end{itemize}

\subsection{The SFA today}
\label{subsec:1c-SFA-today}

Since the formulation of various versions of the SFA, starting from Keldysh in 1964-1965 until the approaches based on simple man's models, formulated in 1993-1994, the SFA kept being one of the most important theoretical tools of the physics of matter in intense laser fields. There are several important review articles and books that either describe or include description of developments and applications of the SFA~\cite{Krausz2009, Milosevic2006, Becker2012, Reiss2018}. 
There were various directions in which the theory of the SFA developed in the recent 20 years, with many authors working on technical issues of improving the accuracy of the theory, while others generalised the theory to novel regions, as we move forward into the future~\cite{Lindroth2019}. 
We summarize here some of the main trends in this body of work:

\begin{itemize}

\item {\bf Coulomb corrections.} 
A number of strong-field phenomena, particularly in ionization experiments, show features caused by the ion's Coulomb potential that evade the SFA, ranging from Coulomb focusing~\cite{Rudenko2005} and the asymmetric photoelectron spectra produced by elliptical polarizations~\cite{Bashkansky1988} to the more recent ``ionization surprise'' of the so-called Low-Energy Structures~\cite{Blaga2009}. 
Early work focused on including the Coulomb potential through a Born series, often with a single rescattering used very successfully for the ATI plateau and non-sequential double ionization (NSDI)~\cite{Milosevic1998}, but this is generally insufficient for Coulomb-dependent phenomena. Current approaches include the use of an oscillating Coulomb-wave basis for the continuum (the Coulomb-Volkov approximation~\cite{Arbo2008}), the eikonal inclusion of the Coulomb Hamiltonian to solve the continuum TDSE (the analytical $R$-matrix theory~\cite{Torlina2012}) and the direct modification of the SFA's trajectory language to include Coulomb potential influence on the action and the trajectories (the Coulomb-Corrected SFA~\cite{Popruzhenko2008}). 
In addition to more ``phenomenological'' approaches (cf. Refs.~\citealp{Yan2010, Smirnova2008}) there have been very elegant approaches based on the Feynman path integral formulation~\cite{Lai2015}. The most recent results based on the path-integral approach are discussed in Refs.~\citealp{Maxwell2017, Milosevic2017, Maxwell2018,  Maxwell2018a}.

\item{\bf Saddle point techniques.}
The simple man's model's classical trajectories are encoded in the SFA as the quantum orbits obtained as the saddle-point contributions to the oscillatory integrals. 
Understanding the nature of these saddle points in the complex time and momentum planes~\cite{Kopold2000, Kopold2002, Morisson2000} allows for a clear understanding of the coherent contribution of each pathway~\cite{Pascal2001}, and it also paves the way for experiments showing the contributions of other orbits~\cite{Sansone2006b, Zair2008}. 
Technical improvements include the regularization of discontinuities at the cutoff via uniform approximations \cite{Morisson2002, Milosevic2002}, and the extension to multi-electron configurations \cite{Ivanov2014}. On the other hand, some problems, such as the inclusion of field dressing of the ground state~\cite{PerezHernandez2009}, are less amenable to saddle-point analysis.

\item{\bf Applications to novel systems: Two-electron systems.}
In the last two decades the SFA has been successfully applied to two-electron systems.  
Most of these approaches used SFA formulations based on $S$-matrix theory \`a la W.~Becker~\cite{Kopold2000, Becker2012}, while the use of the physics of strong laser fields for imaging goes back to the seminal Refs.~\citealp{Itatani2004} (for HHG) and \citealp{Huismans2011} (for ATI). 
In fact, one could argue that two-electron experiments on cold target recoil ion momentum spectroscopy (COLTRIMS)~\cite{Weber2000} pioneered the imaging methods using strong-field physics. 
The phenomenon of interest here is NSDI, which occurs in accordance with P.~Corkum's idea of a recollision-driven model~\cite{Corkum1993}. 
Still, NSDI has two faces. 
If the ionization potential of the target ion is smaller than $3.17 U_p$, the recolliding electron may directly cause stripping of another electron, since there is enough energy for that to occur; this scenario is called  Electron Impact ionization (EII). On the other hand, if the recolliding electron does not have enough energy, it may still excite the target ion to an excited state, from which direct tunneling might easily take place; this scenario is known as Recollision Excitation with Subsequent ionization (RESI). 
In EII, electrons are typically ejected step by step, with most of the quantum interference effects getting washed out, and the standard SFA and  quasi-classical trajectory models work very well~\cite{Figueira2011, Becker2012, Staudte2007, Parker2006, Ye2008, Emmanouilidou2008}. The early studies of the RESI observed that time delay leads to back-to-back electron ejection, and it was assumed that interferences between different channels (different intermediate excited states, etc.) were irrelevant~\cite{Eremina2003, Haan2010}. 
Pretty soon, however, a myriad of shapes in the electron momentum distributions  were observed in RESI and, moreover, experiments were in clear contrast to the simplest SFA theories~\cite{Bergues2012, Sun2014, Kubel2014, Ye2010, Emmanouilidou2009}. 
It was then realised that the interference must be accounted for in RESI~\cite{ Shaaran2010, Chen2010, Hao2014}. 
A lot of insight was gained by the sophisticated analysis of the saddle-point approach~\cite{Shaaran2012}. A more complete understanding of the RESI phenomenon, taking into account interference effects, was only achieved recently~\cite{Maxwell2015, Kubel2014, Maxwell2016}. 
All these results allow, in principle, to work backwards toward the experimental data to reconstruct the states of the excited electron involved in RESI. Amazingly, the channel interference in RESI seems to have been observed recently~\cite{Quan2017}.

\item{\bf Applications to novel systems: Attosecond streaking.} 
The ideas of SFA have also been successfully applied to the single-photon ionization of atoms by high-frequency XUV pulses in the presence of a moderately strong infrared (IR) field, where the IR dresses the continuum and affects the final momentum. This configuration, known as attosecond streaking~\cite{Itatani2002, Yakovlev2005}, is extremely sensitive to the details of both fields and the ionization process. Because of this, it is routinely used as a benchmark for ultrafast light sources, providing direct measurements of optical waveforms~\cite{Goulielmakis2004}, as well as high-temporal-resolution observations of ionization processes~\cite{Schultze2010}.

\item{\bf Applications to novel systems: Large molecules.} 
One of the most natural arenas for application of the SFA and related theories is to the strong-field dynamics of molecules. 
The earliest approaches used the intense-field many-body $S$-matrix theory~(IMST) \cite{Faisal1999, Muth-Bohm2000, Muth-Bohm2001, Jaron-Becker2003, Jaron-Becker2004} later referred to as the molecular SFA~\cite{Kjeldsen2004, Milosevic2006b}. This is a Bessel-function expansion of SFA theory, which was used to model ATI ionization rates~\cite{Faisal1999, Muth-Bohm2000, Muth-Bohm2001} and photoelectron angular distributions~\cite{Jaron-Becker2003, Jaron-Becker2004, Kjeldsen2004} of molecular targets.
Further work used saddle-point methods to link back to the semi-classical dynamics, such as the recollision physics analysed in Refs.~\citealp{Lein2002},\,\citealp{Spanner2004}, for HHG~\cite{Chirila2006a, Chirila2006b, Faria2007}, including multi-electron effects~\cite{Smirnova2009}, photoelectron angular distributions~\cite{Milosevic2006b} and spectra~\cite{Hetzheim2007} for ATI, as well as NSDI~\cite{Faria2008}; for reviews on these methods see Refs.~\citealp{Lein2007, Augstein2012}.
Two related alternatives to SFA for molecules are the so-called molecular ADK (MO-ADK)~\cite{Tong2002} and quantitative rescattering (QRS)~\cite{Le2008} theories.
This theoretical work has also been mirrored and enhanced by substantial advances in the experimental control of strong-field processes for molecules, including work on molecular structure \cite{Krausz2009, Xu2016, Meckel2008, Blaga2012, Wolter2016, Burt2018} and dynamics \cite{Krausz2009, Lein2007, Baker2006, WenLi2008, Worner2010, Wolter2016, Amini2019}, electronic excitation in molecules in the XUV and X-ray energy domains \cite{Villeneuve2018, Chatterley2016}, and the ability to follow the dynamics of an attosecond electron wave packet ejected from molecules \cite{Itatani2004, Krausz2009}.

In  a series of recent papers  we revisited the SFA model on ATI for few-cycle infrared laser pulses~\cite{Noslen1, Noslen2, Noslen3, Noslen4, Noslen-diss}. 
We compared it first with the numerical solution of the TDSE in one (1D) and two (2D) spatial  dimensions for an atomic system~\cite{Lewenstein1995}. 
We developed and generalized there an analytical atomic model based on a non-local (short range) potential. 
In the first paper~\cite{Noslen1} we analysed ATI for an atom, followed by ATI~\cite{Noslen2} and HHG~\cite{Noslen3} for diatomic molecules. Here we paid special attention to non-physical terms which arise in the theory if plane waves are used instead of the exact continuum scattering states of the system.
Finally, in Refs.~\citealp{Noslen4} and~\citealp{Noslen-diss},  we generalized our approach to molecular trimers and quadrimers, and attempted to describe Laser-Induced Electron Diffraction for such targets.
The ultimate goal of this theory is to characterize the time evolution of the target (its size, configuration, its molecular orbital, its dynamical configuration, etc.) by looking at the ATI spectrum and angular distributions, especially in the region of high energies, corresponding to rescattering processes.

\item{\bf Applications to novel systems: Solid state.}
In the past 5 years, the strong field community has increasingly turned its attention to condensed matter systems, following the observation of high-order harmonics from bulk crystals subjected to strong laser fields~\cite{Ghimire2011, Ghimire2019}.  
Our understanding of dynamics in gases, based essentially on the SFA, has recently been extended to yield crucial insights into microscopic attosecond phenomena in condensed matter \cite{Vampa2015, Vampa2014, Osika2017}. 
The merger of strong field physics with solids has the potential to revolutionize contemporary electronics~\cite{Krausz2014}, as well as yield crucial fundamental insights into long-standing problems in condensed matter physics. 
For instance, the first direct measurements of the Berry phase~\cite{Liu2017, Luu2018} were accomplished using HHG. 
These recent measurements promise to be of great interest to the condensed-matter community, due to the important role played by Berry phase in the anomalous Hall effect and in topological insulators, among other fields.

\item{\bf Applications to novel systems: Atto-nanophysics.}
In the last decade the SFA has been successfully applied to situations in which HHG, ATI or NSMI come directly from nano-structures, for instance through plasmonic excitations, or from atomic/molecular targets located close to nano-structures. In the latter case, plasmonic enhanced electromagnetic fields close to the structures serve to excite the targets. Recent developments of this area, termed atto-nanophysics, are extensively described in the review~\cite{Ciappina2017}.

\item {\bf Applications to novel systems: Quantum simulators.} 
The strong-field dynamics described by the SFA have very close analogues in the motion of cold atoms in optical traps~\cite{Dum1998}, particularly via the Kramers-Henneberger correspondence between a dipole coupling and a `shaken' atomic potential~\cite{Morales2011}. 
In the decades since, cold trapped atoms have become one of the primary platforms for quantum simulation~\cite{LewensteinQuantumSimulators}, and several works have explored the possibility of using cold-atom simulators to probe the strong-field dynamics described by SFA~\cite{Arlinghaus2010, Luhman2015, Sala2017, Rajagopal2017}, thereby allowing a complementary look at observables (like e.g. the instantaneous wavefunction, or a full quantum state tomography on the outgoing particles) that are inaccessible to conventional experiments. 
While some dedicated experimental efforts to implement this are still at the tool-building stage~\cite{Ramos2018}, there are already functioning quantum simulation platforms for ultrafast physics~\cite{Senaratne2018, Ramos2019}, which should provide growing opportunities to test SFA physics in new ways. 
Similar complementary views on strong-field dynamics should also be available via photonic simulation, using the natural Schr\"odinger-equation correspondence for optical fibers~\cite{Kahn2017}.

\end{itemize}

\subsection{The present paper}

The present article is organized as follows. 
Section~\ref{sec:1-introduction}, Introduction, covers the motivations for the paper (Subsection~\ref{subsec:1a-motivation}), and the past (Subsection~\ref{subsec:1b-history}) and present (Subsection~\ref{subsec:1c-SFA-today}) of the SFA. 
In Section~\ref{sec-2-strong-field-processes} we present a short explainer of the basic phenomena and processes: HHG, ATI and NSMI, including representative and explanatory figures. 
In Section~\ref{sec:3-background} we review in more detail the theory, which describes the HHG and  ATI processes within the version of Strong Field Approximation following Ref.~\citealp{Lewenstein1994}. 
In particular, we present the derivation of the transition amplitudes for both the direct and the rescattered electrons, as well as for the time-dependent dipole moment.  We develop in detail the mathematical foundations towards the final results by starting from the Hamiltonian, which describes the atomic system and the TDSE associated to it.
Section~\ref{sec:4-two-electrons} is devoted entirely to two-electron processes. We derive the SFA for this case in an explicit time-dependent approach, using exact eigenfunctions for the states in the one- and two-electron continua and perturbation theory with respect to the ``less-singular'' parts of the continuum-continuum matrix elements. We then analyze here the interplay between the EII and RESI processes.

In Section~\ref{sec:5-molecules}, we formulate our theory for the case when the single active electron (SAE) approximation is applied to a molecule undergoing temporal evolution of its nuclear configuration.
This is done using the Born-Oppenheimer approximation and classical equations of motion for the nuclei. We consider first the simplest case in which the molecule's dynamics (vibrations, dissociations) do not affect the SAE electronic dynamics. 
Even in this simple case, novel effects appear in the SFA dynamics, such as the appearance of a temporal Berry phase. 
In Subsection~\ref{subsec:5b-large-targets} we go beyond this approximation and consider the self-consistent dynamics, in which SAE dynamics affect nuclei and vice versa. 
We conclude in Subsection~\ref{subsec:5c-molecular-dynamics} where we present an outlook on extending our quasi-analytical model to more complex atomic and molecular systems. 
Finally, in Section~\ref{sec:6-solids} we briefly review the recent application of SFA theory to the generation of harmonics in solid-state systems.

Appendix~\ref{app:A-TD-ADK} discusses time dependent ADK rates. 
In Appendix~\ref{app:B-ground-state} we sketch calculations of $a(t)$---the amplitude of the ground state within our generalized SFA theory. 
In Appendix~\ref{app:C-atom-molecule-model} we introduce the model for our atomic and molecular systems that uses a particular form of a non-local short-range separable potential. The matrix elements to describe the direct ionization and the re-scattering processes are then computed analytically.

We then offer additional material regarding the two-electron theory:
in Appendix~\ref{app:D-dipoles} we discuss the properties of the dipole matrix elements involved, and 
in Appendix~\ref{app:E-two-electron} we pose full forms for the two-electron integro-differential equations derived from the TDSE.
In Appendix~\ref{app:F-toy-model} we analyze a single-electron toy model for HHG from a quenched molecule.
Finally, in Appendix~\ref{app:G-RESI-EII}, we present solutions of the RESI and EII equations using additional approximations for the dipole matrix elements, neglecting electron-electron interaction effects for those elements.

\section{Strong-field processes in an atomic gas}
\label{sec-2-strong-field-processes}
\begin{figure}[htbp]
  \includegraphics[width=0.85\textwidth]{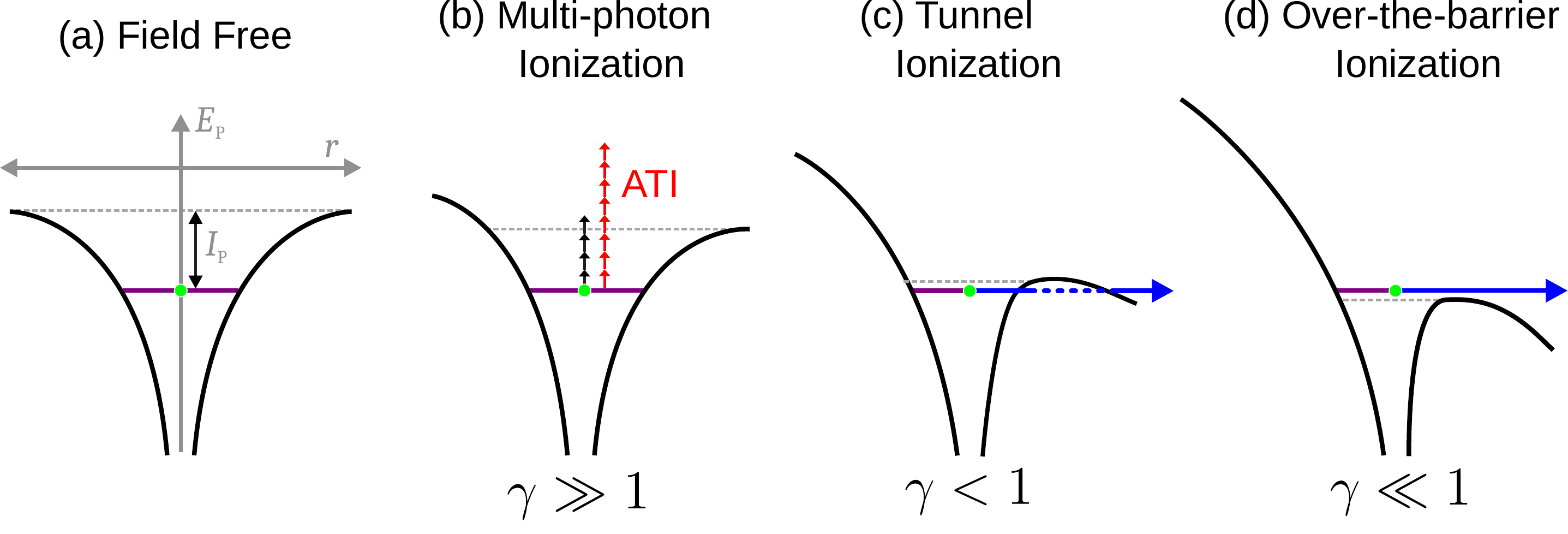}
  \caption{The atomic Coulomb potentials under the influence of intense laser fields are shown for the (a) field free, (b) multi-photon ionization, (c) tunnel ionization, and (d) over-the-barrier ionization cases. Above-threshold ionization (ATI) is also presented (red vertical arrows) as compared to multi-photon ionization (black vertical arrows). Here, $I_{p}$ is the ionization potential, and $\gamma$ is the Keldysh parameter. Figure adapted from Ref.~\citealp{Hansen2012}.}
  \label{fig_SF}
\end{figure}

Over the lifetime of strong-field physics, SFA theory has accounted for a broad variety of physical phenomena which could not be explained by traditional perturbation theory. These phenomena involve light-matter interaction using lasers whose field strengths are comparable to the Coulomb force of attraction between electrons and protons. Consequently, this can lead to the distortion of the Coulomb potential and the subsequent lowering of the barrier to ionization in the strong-field regime. Many nonlinear ionization processes can be initiated in this regime, such as multi-photon ionization (MPI), above-threshold ionization (ATI), tunnel ionization (TI), and over-the-barrier (OTB) ionization. These processes are shown in Fig.~\ref{fig_SF} with their operating conditions summarized in Table~\ref{tbl_SF}.
The ionization regime of operation can be identified by the Keldysh parameter, $\gamma$, given by
\begin{equation}
\gamma = \sqrt{\frac{I_{p}}{2U_{p}}},%
\label{eqn_Keldysh}
\end{equation}
where $I_{p}$ is the ionization potential (i.e.\ the energy required to eject an electron from the ground state to the ionization continuum), and $U_{p}$ is the ponderomotive energy (i.e.\ the average kinetic energy of the oscillations of a free electron in a laser field) given by
\begin{equation}
U{\rm _p} 
= \frac{e^2\mathcal E_0^2}{4m_e\omega_0^2} 
= \frac{I_0e^2\lambda_0^2}{8\pi^2m_{\rm e}\epsilon_0c^3}  
= 9.337\times10^{-20}  I_0\lambda_0^2  \ \frac{ \mathrm{eV} }{\mathrm{W} \: \rm cm^{-2} \: nm^2},
\end{equation}
where $e$ is the elementary charge, $\mathcal E_0$ is the electric field amplitude, $m_{\rm e}$ is the mass of an electron, $\omega_0$, $\lambda_0$ and $I_0$ are the central laser frequency, wavelength and intensity, respectively, $\epsilon_0$ is the vacuum permittivity, and $c$ is the speed of light. We give, in the rest of this section, a brief summary of strong-field processes that are based on ATI and TI.

\begin{table}[ht]
\centering
\medskip
\begin{tabular}{rcl}
	\hline
	\textbf{Ionization regime} & $\ $ & \textbf{Operating condition} $\ $ \\
	\hline
	Single-photon ionization (SPI) & & $\hbar\omega>I_{p} \gg U_{p}$ \\
	Multi-photon ionization (MPI) & & $I_{p}>\hbar\omega \gg U_{p}$\\
	Above-threshold ionization (ATI) & & $I_{p}>U_{p}>\hbar\omega$\\
	Tunnel ionization (TI) & & $U_{p}>I_{p}>\hbar\omega$\\
\end{tabular}
\caption{The operating conditions in terms of photon energy, $\hbar\omega$, ionization potential, $I_{p}$, and ponderomotive energy, $U_{p}$ are presented for single-photon ionization (SPI), multi-photon ionization (MPI), above-threshold ionization (ATI) and tunnel ionization (TI).}
 \label{tbl_SF}
\end{table}

\subsection{Above-Threshold Ionization (ATI)}
Above-threshold ionization (ATI) is an extension of multi-photon ionization where multiple photons are absorbed to not only access the ionization continuum but to surpass the $I_{p}$ by more than one photon, $\hbar\omega$~\cite{Agostini1979}. 
In the initial studies of ATI there was substantial interest in determining the nature of this process, and discerning the relative contributions of perturbative multi-photon ionization and tunnel ionization. Nowadays, ATI is considered holistically as a process where non-perturbative phenomena coexist with perturbative ones (see, for instance, Ref.~\citealp{Paulus1994}).

\begin{figure}[ht]
	\includegraphics[width=0.5\textwidth]{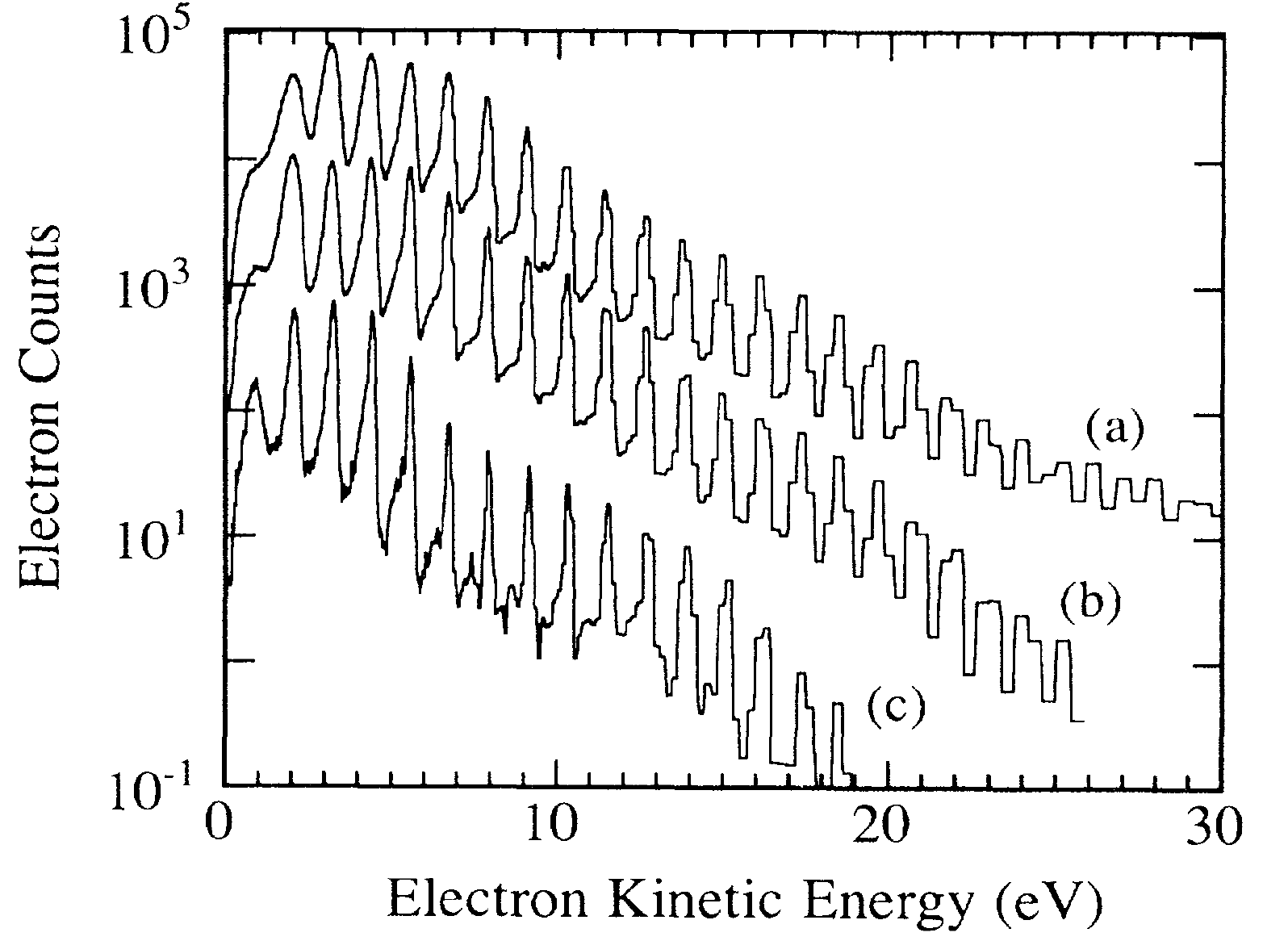}
	\caption{Above-threshold ionization (ATI) spectrum of xenon illuminated with a $\SI{50}{ps}$, $\SI{1.05}{\micro m}$ laser pulse at three intensities: (a) $\SI{2.0e13}{W/cm^2}$, (b)$\SI{1.5e13}{W/cm^2}$, and (c) $\SI{1.0e13}{W/cm^2}$. Figure taken from Ref.~\citealp{Schafer1993}.}
	\label{fig_ATI}
\end{figure}

In a typical ATI photoelectron spectrum, as shown in Fig.~\ref{fig_ATI}, a series of peaks are observed that correspond to each photon absorbed above the $I_{p}$, each of which is separated by a single photon energy, $\hbar\omega$.
More strongly, ATI can be observed in the high energy range of the photoelectron spectrum ($2U_{p} \le E_r \le 10U_{p}$), referred to as high-order ATI (HATI), where recollision-based strong-field physics can appear, giving rise to elastic and inelastic scattering.

\subsection{High-Harmonic Generation (HHG)}
\begin{figure}[t!]
  \includegraphics[width=0.5\textwidth]{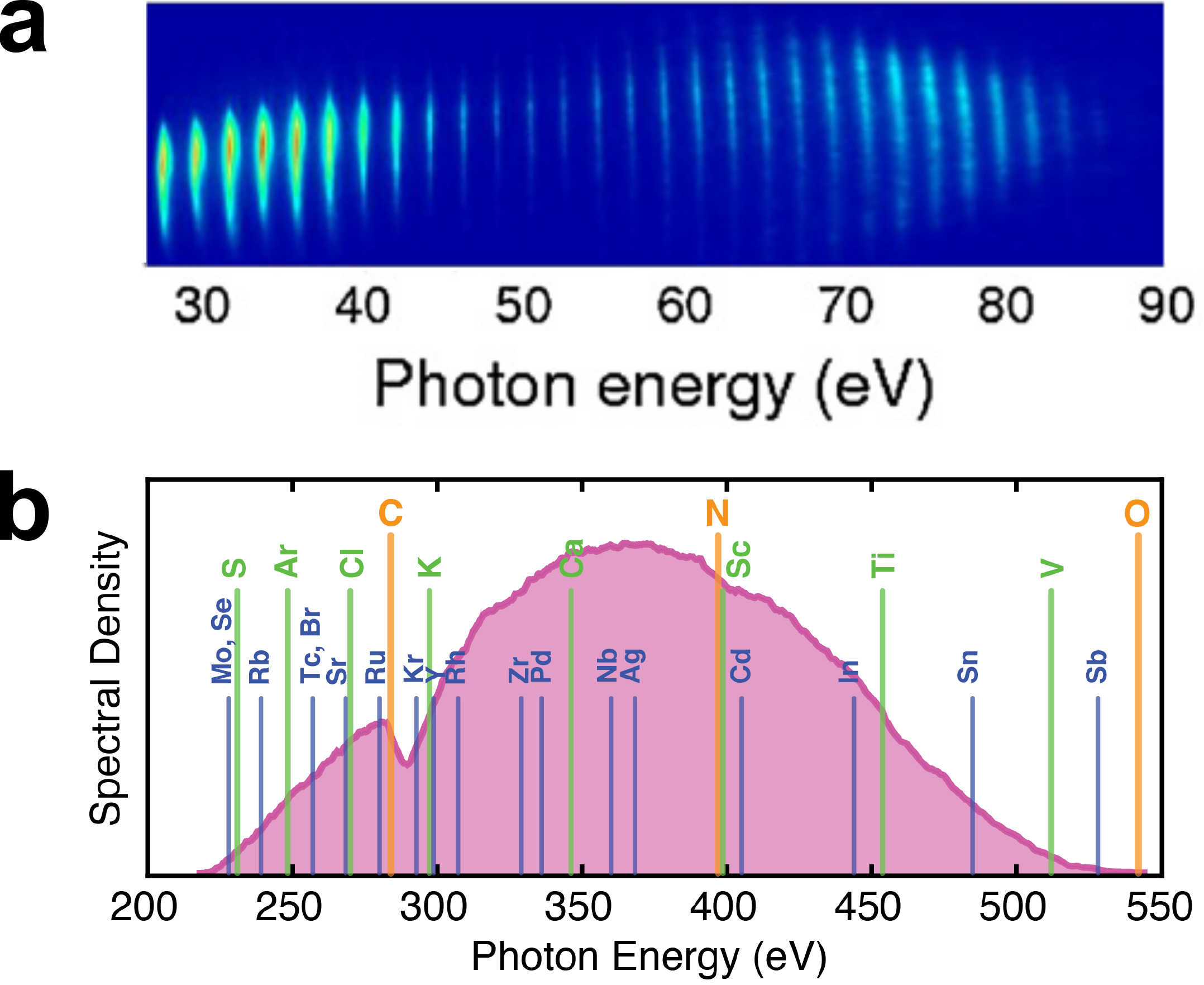}
  \caption{ HHG spectrum for (a) a train of attosecond pulses and (b) an isolated attosecond pulse. In panel (a), argon was illuminated with a $\SI{40}{fs}$ (10-cycle), $\SI{1200}{nm}$ driving laser pulse to generate a spectrum over the $30\text{-}\SI{90}{eV}$ range. In panel (b), helium was ionized by a $\SI{12}{fs}$ (1.8-cycle), $\SI{1850}{nm}$ laser pulse to generate a broadband spectrum over the entire water window range of $284\text{-}\SI{543}{eV}$, with the K- (orange), L- (green), and M-shell (blue) absorption edges indicated by vertical lines. Figures in panels (a) and (b) were adapted from Refs.~\citealp{Jin2011} and~\citealp{Buades2018}, respectively.}
  \label{fig_HHG}
\end{figure}

Attosecond laser pulses of high-photon energies in the extreme ultraviolet (XUV) and X-ray energy region can be produced by high-harmonic generation~\cite{Krause1992, Corkum1993, Krausz2009}. Pulse trains of attosecond radiation are generated using a multi-cycle femtosecond driving laser pulse, as presented in Fig.~\ref{fig_HHG}a with the $27^{\rm th} - 85^{\rm th}$ harmonics shown~\cite{Jin2011}. Similarly, a single attosecond pulse with a broadband spectrum can be generated using a near-single cycle driving laser pulse with a continuous broadband spectrum, as shown in Fig.~\ref{fig_HHG}b~\cite{Buades2018}, though other so-called `gating' schemes are also possible~\cite{CorkumBurnett1994, Kovacev2003, Mashiko2008, Pfeifer2007, Vincenti2012}. The maximum HHG cut-off energy, $E_{\rm max}$, that can be generated is given by the simple-man's-model-like $E_{\rm max} = 3.17U_{p} + I_{p}$, and is observed as the abrupt end to the HHG plateau. 

These microscopic aspects aside, it is also important to remark that HHG is a macroscopic nonlinear optical process that requires the coherent combination of a large number of emitters to be observed experimentally, and this requires that specific attention be paid to the phase-matching conditions~\cite{Heyl2016}, which are often the determining limitation in the production of harmonics.

\subsection{Inelastic Scattering: Non-Sequential Double Ionization (NSDI)}
\begin{figure}[t!]
  \includegraphics[width=0.5\textwidth]{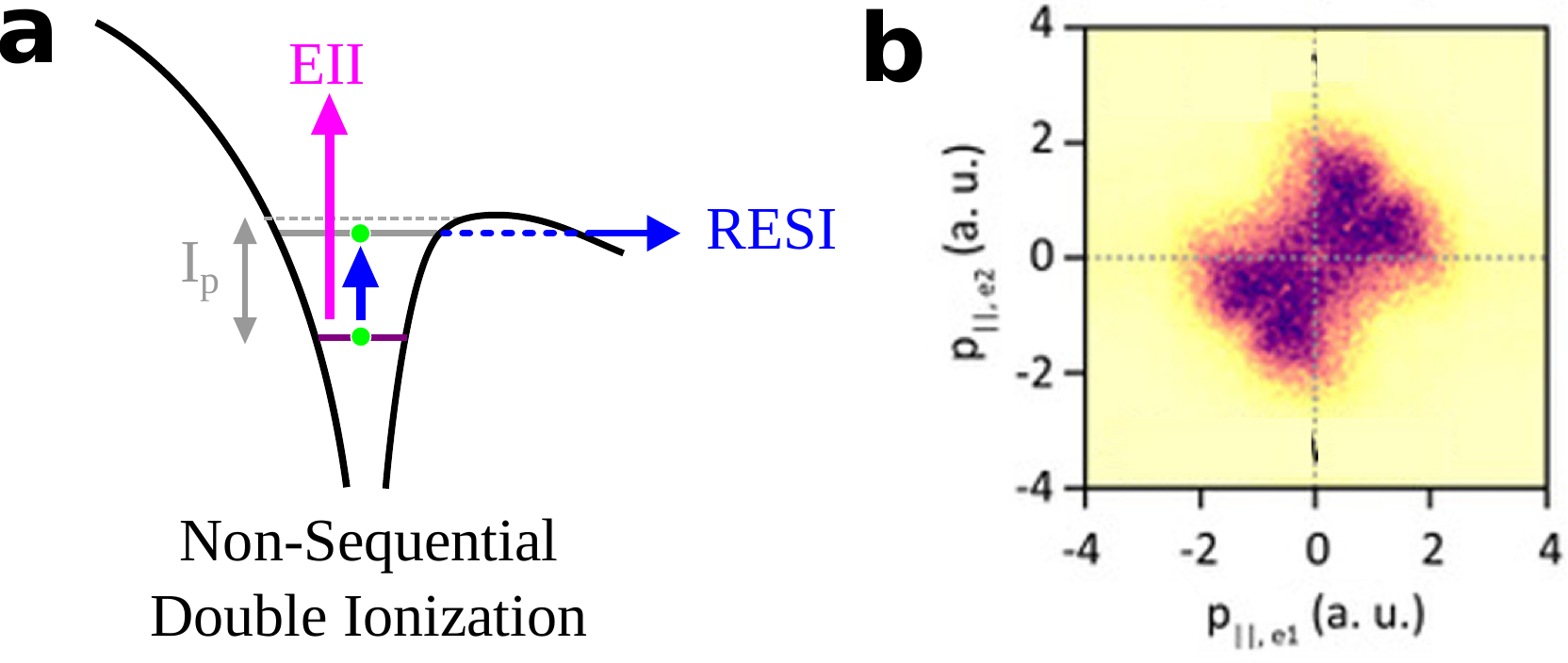}
  \caption{(a) Non-sequential double ionization (NSDI) illustrated after the recollision of $e_1$ with the target ion using an atomic Coulomb potential. After inelastic recollision, either the second electron is ejected through: (i) an electron-impact process (pink arrow) or (ii) excited and subsequently ionized through recollision-excitation with subsequent ionization (RESI; blue arrow). (b) The typical EII signature in a two-dimensional momentum map $(p_{\parallel,1},p_{\parallel,2})$ of the longitudinal momenta of the two electrons in atomic units (a.u.) from strong-field ionized Xe$^{2+}$ ions. 
  (In contrast, RESI may lead to a myriad of shapes in correlated electron-electron distributions, and in principle occupies all four quadrants.)
  Panel (b) was adapted from Ref.~\cite{Pullen2017}.}
  \label{fig_NSDI}
\end{figure}

If the recollision of the returning electron ($e_1$) with the parent ion is inelastic, then it can transfer enough energy to eject a second electron ($e_2$). This process is known as non-sequential double ionization (NSDI)~\cite{Walker1994}, and it can proceed through two ionization pathways upon the recollision of $e_1$~\cite{Pullen2017}, as shown in Fig.~\ref{fig_NSDI}a: (i) direct ionization of $e_2$, known as electron-impact ionization (EII)~\cite{Corkum1993}; or (ii) impact excitation of $e_2$ subsequently followed by its delayed tunnel ionization, known as recollision-excitation with subsequent ionization (RESI)~\cite{Feuerstein2001}. A typical signature of NSDI, particularly in the EII regime, is the correlated detection of two electrons ($e_1$ and $e_2$) in the same emission direction within the two-dimensional momentum map $(p_{\parallel,1},p_{\parallel,2})$ of the longitudinal momenta of the two electrons, as shown in Fig.~\ref{fig_NSDI}b.

\begin{figure}[b!]
	\includegraphics[width=0.5\textwidth]{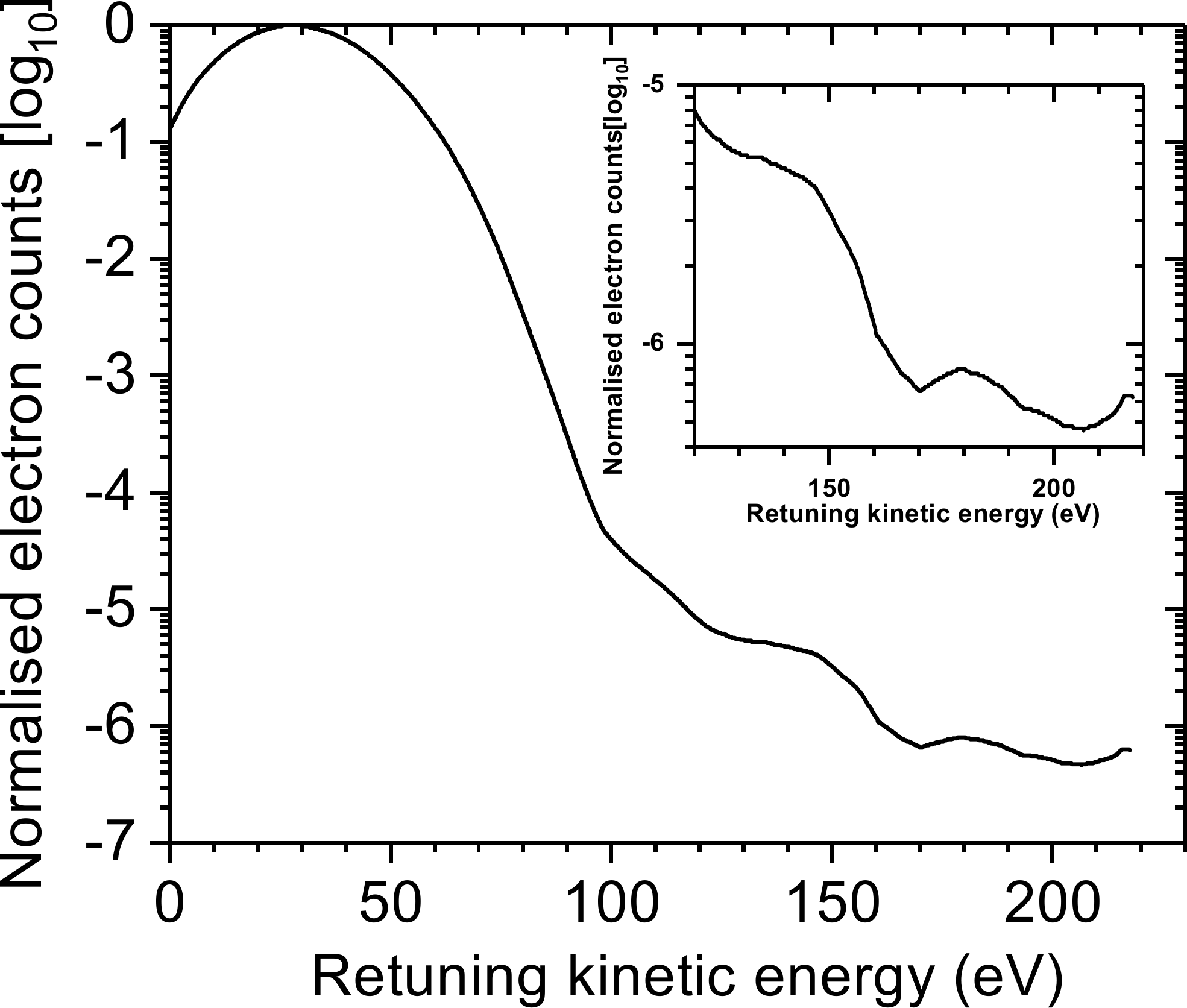}
	\caption{Typical photoelectron spectrum recorded with LIED of C$_2$H$_2^+$, with the direct photoelectrons ($<3.5\:$a.u.) and rescattered electrons ($>3.5\:$a.u.) present. Oscillations are clearly seen in the rescattered energy range of the electron signal that is a result of the coherent molecular interference signal which is dependent on the target's geometric structure. The inset shows a zoomed-in view of the oscillations in the differential cross-section of the scattering energy range. Figure adapted from Ref.~\citealp{Pullen2016}.}
	\label{fig_LIED}
\end{figure}

\subsection{Elastic Scattering: Laser-Induced Electron Diffraction (LIED)}
The highly-energetic returning electron can collide elastically and scatter on the target ion, leading to a momentum transfer between the electron and parent ion. This is known as laser-induced electron diffraction (LIED)~\cite{Meckel2008, Blaga2012, Wolter2016, Pullen2016, Ito2016, Amini2019} and it can be explained in the framework of laser-driven electron recollision~\cite{Krause1992,Corkum1993}. Structural information is embedded in the photoelectron momentum distribution, appearing as oscillations in the high-energy part of spectrum corresponding to recollision-based physics ($2U_{p} \le E_r \le 10U_{p}$) as a function of the emission angle, as shown in Fig.~\ref{fig_LIED} with a zoomed-in view of these oscillations given in the inset. It should be noted that, as opposed to HHG, phase-matching is irrelevant for inelastic and elastic scattering processes (i.e.\ NSDI and LIED), since the macroscopic observable is an incoherent combination of the emission from the different atoms in the laser focus.

\section{Strong field approximation for single-electron processes}
\label{sec:3-background}

Strictly speaking, neglecting nuclear motion, an atomic or molecular system interacting with  a strong electric field pulse is described by the time-dependent Schr\"odinger equation (TDSE) that  captures  both the evolution of the (electronic) wave function and the time evolution of the physical observables.  
The numerical solution of the TDSE offers a full quantum mechanical description  of the laser-matter interaction processes; it has been used extensively to study HHG~\cite{KulanderGaarde1998, TateDiMauro2007, Marcelo2012, Scrinzi2014} and ATI~\cite{Kulander1987, Muller1999, Bauer2005classical, Blaga2009, RufKeitel2009} in atomic and molecular systems. 
However, the full numerical integration of the TDSE in all the  degrees of freedom of the system is computationally very demanding, when it is at all possible. 
Moreover, a physical interpretation of the numerical results is highly nontrivial, as always  for an {\it ab initio} technique. 
Within this framework, then, approximate methods are welcome, and the SFA has consistently been shown over the years to be the workhorse tool for that role.

\subsection{Hamiltonian and TDSE}
Let us consider an atom or molecule under the influence of an intense laser field in the so-called single active electron (SAE) approximation. In the limit when the wavelength of the laser $\lambda_0$ is large compared with the Bohr radius, $a_0$ ($\SI{5.29e-11}{m}$), the electric field of the laser beam around the interaction region can be considered spatially homogeneous. 
Consequently, the interacting atoms will not experience the spatial  dependence of the laser electric field and, hence, only its time variation is taken into account---this is the so-called dipole approximation. Note, on the other hand, that certain dynamical effects, even in the long-wavelength limit, can break this approximation~\cite{Reiss2018}. Within this framework, the laser electric field can be written as:
\begin{equation}
 \textbf{E}(t) = \mathcal{E}_0\:f(t) \sin(\omega_0 \:t + \phi_0)\,{\bf e}_{z}.
 \label{Eq:Efield}
\end{equation}
The field of Eq.~(\ref{Eq:Efield}) has a carrier frequency $\omega_0 = \frac{2\pi c}{\lambda_0}$, where $c$ is the speed of light, and a peak amplitude $\mathcal{E}_0$. 
We consider here that the laser field is linearly polarized along the $z$ direction, with a pulse envelope ${f(t)}$ and a carrier-envelope phase $\phi_0$. 
More generally, we could consider time-dependent polarization, i.e.\ replace $\mathcal{E}_0{\bf e}_{z} \to {\bf E}_0(t)$.

The TDSE  reads:
\begin{equation}
i\hbar\frac{ \partial}{ \partial t} | \Psi(t) \rangle=\hat{H} | \Psi(t) \rangle, \label{Eq:SE}
\end{equation}
where the Hamiltonian, $\hat{H}$, describes the laser-target system in SAE approximation, and is the sum of two terms, i.e.
\begin{eqnarray}
\hat{H} &= &\hat{H_0}+\hat{U}, \label{Eq:H}
\end{eqnarray}
where $\hat{H}_0$ is the laser-free Hamiltonian of the atomic or molecular system
\begin{eqnarray}
\hat{H}_0&=&- \frac{\hbar^2\nabla^2}{2m} + V(\hat{\textbf{r}}),
\end{eqnarray}
with $V(\hat{\textbf{r}})$ the effective SAE atomic or molecular potential, $m$ the electron mass, and $\hat{U}=-e{\bf E}(t)\cdot \hat{\bf r}$ the  dipole coupling, which describes the interaction of the atomic or molecular system with the laser radiation, written in the length gauge~\cite{Bauer2005, Galstyan2016} and under the dipole approximation.  
Note that in atomic units, the electron charge, denoted  by $e$, is $e=-1\:$a.u., and the Planck constant and electron mass are both set to unity, $\hbar = m = 1\:$a.u. In this work, however, we keep the explicit constants.

\subsection{``Standard'' SFA \`a la Lewenstein}
We shall restrict  ourselves to  the regime of low laser frequency and relatively high intensity,
where the SFA is expected to be valid~\cite{Keldysh1965, Faisal1973,
Reiss1980, Lewenstein1994,Lewenstein1995, Grochmalicki1986} and to  describe well the laser-matter interactions.
This corresponds to the tunnelling regime, where
the Keldysh parameter $\gamma$ is less than one, $\gamma<1$. 
In this regime the effects of atomic effective potential on the dynamics of electrons in the continuum are assumed to be small, and they can be treated using perturbation theory.
These observations suggest to formulate the ``standard SFA'' as follows:

\begin{enumerate}%[(i)]

\item[(i)] The strong field laser does not couple with any bound  state beyond the ground state, $|0 \rangle$, so that only it and the continuum  (scattering) states, $|\textbf{p}\rangle$, are taken into account in the dynamics;

\item[(ii)] 
The amplitude of the ground state, $a(t)$, is  considered to be known.

\item[(iii)] 
The continuum states are taken from the basis of {\it exact} scattering  states, which are eigenstates via
\begin{equation}
\hat{H}_0 |\textbf{p}\rangle = \frac{1}{2m}\textbf{p}^2 |\textbf{p}\rangle
\end{equation}
of the atomic Hamiltonian with a fixed outgoing (kinetic) momentum  ${\bf p}$. The continuum-continuum matrix element from ${\bf p}$ to ${\bf p}'$ are then decomposed into their most singular part, proportional to $i \hbar {\nabla}_{\bf p}\delta({\bf p}-{\bf p}')$, and the ``rest''~\cite{Grochmalicki1986, Lewenstein1994, Lewenstein1995}. The ``rest'' is then treated in a perturbative manner~\cite{Lewenstein1995}.
\end{enumerate}
The following comments are necessary in order to specify more precisely the above points.
\begin{enumerate}[align=left]

\item[{\bf Ad (i)}] 
Based on the statement (i), the electronic state $|\Psi(t)\rangle$ that represents the time evolution of the system is a coherent superposition of the ground $|0 \rangle$ and the continuum $|\textbf{p}\rangle$ states~\cite{Lewenstein1994,Lewenstein1995}:
\begin{equation}
| \Psi(t) \rangle= e^{\textit{i}I_p\textit{t}/\hbar}\bigg(a(t) |0 \rangle + \: \int{\textit{d}^3 \textbf{p} \:  \textit{b}( \textbf{p},t) |\textbf{p}\rangle} \bigg).
\label{PWavef}
\end{equation}
The factor $a(t)$, representing the amplitude of the ground state, is assumed to be known (see below for the ways to evaluate or estimate it). 
The prefactor $e^{\textit{i}I_p\textit{t}/\hbar}$ represents the phase oscillations which describe the accumulated electron energy in the ground state ($I_p=-E_0$ is the ionization potential, with $E_0$ the ground-state energy of the target system). Furthermore, the transition amplitude to the continuum states is denoted by  $\textit{b}(\textbf{p},t)$, and it depends both on the kinetic momentum  of the outgoing electron and the laser pulse.  Note that, if needed, other (relevant) bound states may be taken into account in the expression \eqref{PWavef} (cf.\ Refs.~\citealp{Ivanov2014, PerezHernandez2009, Sanpera1996}).

\item[{\bf Ad (ii)}]
There are several ways of evaluating or estimating $a(t)$, depending on the regime of parameters.
\begin{itemize}
\item 
First, one can use {\it ab initio} TDSE of the target system to determine the amplitude $a(t)$. This is obviously quite costly numerically, but it is much less costly than a full solution of the TDSE which is also required to calculate photoelectron momentum spectra or angular distributions which would need much higher precision, memory and disk storage, and higher computation times.

\item 
Second, one can use any ``cheap'' approximate method to calculate $a(t)$, such as phase-space averaging or the truncated Wigner approximation~\cite{Berman2018}.

\item 
Third, a broadly-used method is to calculate $a(t)$ analytically using the  ionization rates according to the Ammosov-Delone-Krainov theory (ADK rates~\cite{Ammosov1986}). To this end, one generalizes these rates to depend on time locally through the time dependence of the laser electric field (also known as the quasi-static approximation), which is generally a rather straightforward task (see Appendix~\ref{app:A-TD-ADK}). This approach is valid in the quasi-static regime, when not only the laser frequency, but also the rate of change of the pulse envelope function $f(t)$ are small---meaning that the laser pulse is longer, so that it includes several optical periods.

\item 
Fourth, when the pulse is very short, or it is long but not too strong, there is practically no depletion of the ground state, i.e.\ $a(t)\simeq 1$. This happens, for instance, for moderately long pulses when the ponderomotive  energy is lower than the saturation energy of the system  $(U_p < U_\mathrm{sat})$.

\item 
Fifth, one can calculate $a(t)$ within our SFA self-consistently. This approach was already discussed in Ref.~\citealp{Lewenstein1994},  but it turned out not to be very precise for the longer pulses---the ADK rates were giving much better agreement with the exact solutions of the TDSE and with the experimental data. This approach seems to be, however, much more adequate and precise for ultrashort, few-cycle pulses. We describe it in detail in Appendix~\ref{app:B-ground-state}.
\end{itemize}

\item[{\bf Ad (iii)}]
The continuum-continuum matrix element, independently of the fact whether the effective SAE potential is short-range (as it is for model atoms and negative ions) or Coulomb-like, has the general form:
\begin{equation}
e\langle {\bf p}|\hat{\bf r}| {\bf p}' \rangle=  ie \hbar {\nabla}_{\bf p}\delta({\bf p}-{\bf p}') + \hbar {\bf g}({\bf p},{\bf p}'),
\label{c-c}
\end{equation}
where the part $\hbar{\bf g}({\bf p},{\bf p}')$ is less singular---typically the strongest singularity it contains corresponds to the on-energy-shell gradient of the Dirac delta of ${\bf p}^2-({\bf p}')^2$. 
This part is responsible for rescattering effects in ATI and recollisions in NSMI. 
Note that since we insist on using  the {\it exact} scattering states, the dipole matrix element $\langle {\bf p}|\hat{\bf r}|0\rangle$ (together with the rescattering continuum-continuum matrix elements) does include the full effects of the effective SAE potential, comprising both the short-range effects as well as any long-range Coulomb effects (if present). 

Note also that the SFA in the present formulation (actually equivalent to that of Ref.~\citealp{Lewenstein1995}) does not involve plane waves or Volkov solutions! The majority of authors, including ourselves, ``erroneously'' (in the view of the present formulation) claim that SFA corresponds to the use of Volkov states in the continuum. 
This is, in principle, false and dangerous. One can use additional approximations, and approximate the {\it exact} scattering states by plane waves in the calculations, but {\it this is an additional approximation!} It does simplify life and allows one to calculate many things more easily, but it also leads to problems, especially in the case of molecules and other extended targets. 

These problems are due to the fact that a plane wave $|{\bf p}\rangle$ is not orthogonal to the ground state $|0\rangle$, so that the matrix element $\langle {\bf p}|{\bf R}_0|0\rangle \ne 0$, where ${\bf R}_0$ is the typical internuclear distance, which is just a constant vector. 
The lack of orthogonality of $\langle {\bf p}|0\rangle \ne 0$ leads to various non-physical and misleading results in applications of, say, ``primitive'' SFA to molecules (for remedies see Refs.~\citealp{Noslen2, Noslen3}).
We stress: no remedies are needed, on the other hand, if the {\it exact} scattering states are used, since then the orthogonality is assured by construction.

Why, then, do the plane waves and Volkov solutions appear at all? Clearly, this is due to the fact that in the zeroth approximation of SFA we neglect the contribution of $ \hbar {\bf g}({\bf p},{\bf p}')$. In this case, the full continuum-continuum matrix element becomes $\langle {\bf p}|\hat{\bf r}| {\bf p}' \rangle=  i \hbar {\nabla}_{\bf p}\delta({\bf p}-{\bf p}')$, and is exactly equal to that obtained for plane waves and Volkov solutions. 
That means that the quasi-classical action, describing the propagation of electrons in the continuum, does indeed have a free electron form. 
For short-range effective potentials this is acceptable, but not for the Coulomb-like ones. That is why the so-called Coulomb corrections are easily included in $\langle {\bf p}|\hat{\bf r}|0\rangle$  or $\hbar {\bf g}({\bf p},{\bf p}')$, but much effort has been devoted to find Coulomb corrections to the action---see the Introduction for the relevant references.

\end{enumerate}

\subsection{Solutions of the SFA equations}

Our main task in this subsection will be to derive a
general expression for the amplitude  ${b}({\bf p},t)$, which then will be used to calculate ATI spectra and angular distributions, as well as HHG spectra.  After some algebra, the time variation
of the ground state amplitude, $a(t)$, and the transition amplitude $\textit{b}(\textbf{p},t)$ read:
\begin{align}
\dot{a}(t) 
& = 
\frac{i}{\hbar}
\int d^3{\bf p} \: 
\textbf{E}(t)\cdot \textbf{d}^*\mathopen{}\left( \textbf{p}\right)\mathclose{} \:
b(\textbf{p},t) 
\\
\dot{b}( \textbf{p},t) 
& =
-\frac{i}{\hbar}\left(\frac{\textbf{p}^2}{2m}
+ I_p\right)\textit{b}( \textbf{p},t) 
+\frac{i}{\hbar}\textbf{E}(t) \cdot {\bf d}({\bf p})a(t)
 \nonumber \\ & \qquad
-\,e{\bf E}(t)\cdot\nabla_{\bf p} b({\bf p},t) 
+ i \textbf{E}(t)\cdot \int{\textit{d}^3 \textbf{p}^{\prime}\: \textit{b}( \textbf{p}^{\prime},t){\bf g}( \textbf{p}, \textbf{p}^{\prime})}.
\label{Eq:New5}
\end{align}
The first term on the right-hand side of Eq.~(\ref{Eq:New5}) represents
the free phase evolution of the electron in the absence if the  oscillating laser field. 
In the second term we have defined the bound-free transition dipole matrix element as
\begin{equation}
e \langle \textbf{p} |\hat{\textbf{r}}|0 \rangle
=
\textbf{d}( \textbf{p}).
\label{Eq:dv}
\end{equation}
Finally, the last two terms describe the continuum-continuum transition, $\nabla_{\bf p} b({\bf p},t)$, without  the influence of the scattering center, and by considering the core potential, 
$\int{\textit{d}^3 e\textbf{p}^{\prime}\: \textit{b}( \textbf{p}^{\prime},t){\bf g}({\bf p},{\bf p}')}$. 
Here ${\bf g}({\bf p},{\bf p}')$ denotes the  rescattering  transition matrix  element, where the potential core plays an essential role:
\begin{equation}
e\langle {\bf p}|\hat{\bf r}| {\bf p}' \rangle
=
ie \hbar {\nabla}_{\bf p}\delta({\bf p}-{\bf p}') 
+ \hbar {\bf g}({\bf p},{\bf p}'),
\label{Eq:CCM1}
\end{equation}
Note that (\ref{Eq:New5}) is a linear integro-diffential equation for $\textit{b}( \textbf{p},t)$. 
In the following, we shall describe how it is possible to compute the transition amplitude, $ b({\bf p},t)$, by applying zeroth- and first-order perturbation theory to the solution of the partial differential equation (\ref{Eq:New5}). 
We will split the solution of the transition amplitude, $b({\bf p},t)$, into two parts: $b_0({\bf p},t)$ and $b_1({\bf p},t)$, i.e.~$b({\bf p},t)=b_0({\bf p},t)+b_1({\bf p},t)$.  
The zeroth order of our perturbation theory $b_0({\bf p},t)$ will be called the direct term. It describes the transition amplitude for a laser-ionized electron that will never rescatter with the remaining ion.
On  the  other  hand, the first-order term, which we call the rescattered term, $b_1({\bf p},t)$, refers to the electrons that, once ionized, will have a certain probability of rescattering with the potential of the parent ion.

\subsection{Direct-ionization amplitude}
Let us consider the  process in which the electron is ionized and does not return to its parent ion. 
This process is modelled by the direct photoelectron transition amplitude $b_0({\bf p},t)$. 
As the direct ionization process should have a larger probability  compared  with the rescattering one~\cite{Lewenstein1995}, one can neglect the last term in Eq.~(\ref{Eq:New5}). 
This is what we refer to as the zeroth order solution:
\begin{equation}
{\partial }_tb_0( \textbf{p},t) 
=
-\frac{i}{\hbar}\left(\frac{\textbf{p}^2}{2m}+ I_p \right){b}_0( \textbf{p},t)  
+ \frac{\textit{i}}{\hbar}\: \textbf{E}(t) \cdot  \textbf{d}( \textbf{p})a(t) +\textbf{E}(t) \cdot \nabla_{\bf p}\textit{b}_0( \textbf{p},t).
\end{equation}

The above equation is a first-order inhomogeneous differential equation, which is easily solved by conventional integration methods (see e.g.\ Ref.~\citealp{Lev}). Therefore, the solution can be written as
\begin{equation}
\begin{split}
b_0( \textbf{p},t) 
= & 
\frac{i}{\hbar} \int_0^t{\textit{d} \textit{t}^{\prime}\:\textbf{E}(t^{\prime})}\:\cdot \textbf{d}\left( \textbf{p}+e\textbf{A}(t)/c-e\textbf{A}(t^{\prime})/c\right)
\\
& \qquad \times 
\exp\mathopen{}\left(
  -\textit{i} \:
  \int_{t^{\prime}}^{t} d\tilde{t}
  \left[
 \frac{1}{2m} (\textbf{p}+e\textbf{A}(t)/c-e\textbf{A}(\tilde{t})/c)^2
 +I_p
 \right]/\hbar
 \right)
a({t}^{\prime}). 
\label{Eq:IntDirecTerm}
\end{split}
\end{equation}

Here, we have considered that the electron appears in the
continuum with kinetic momentum ${\bf p}(t')={\bf p}+e{\bf
A}(t)/c-e{\bf A}(t')/c$ at the time $t'$, where {\bf p}~is the final
kinetic momentum, and $\textbf{A}(t) =-c\int^{t}{
\textbf{E}(t^{\prime})dt^{\prime}}$ is the vector potential of the
electromagnetic field, with $c$ the speed of light. In particular, the vector potential at the time when the electron appears at the continuum $t'$ is ${\bf A}(t')$, and at a certain detection time $t$, the vector potential reads ${\bf A}(t)$. 
In addition, it is possible to write Eq.~(\ref{Eq:IntDirecTerm}) as a function of the canonical momentum $\textbf{p}_{c}$, defined by $\textbf{p}_{c} = {\bf p} + e{\bf A}(t)/c$, and therefore the probability transition amplitude for the direct electrons simplifies to that from Ref.~\citealp{Lewenstein1994}, where we have eliminated the subscript~${c}$:
\begin{equation}
b_0( \textbf{p},t) 
= 
\frac{i}{\hbar} \int_0^t{\textit{d} \textit{t}^{\prime}\:
\textbf{E}(t^{\prime})}\cdot \textbf{d}\left( \textbf{p}-e\textbf{A}(t^{\prime})/c\right)
a(t^{\prime})
\:
\exp\mathopen{}\left(
  -\textit{i} \:
  \int_{t^{\prime}}^t{d{\tilde t}\,
  \left[
  \frac{1}{2m}(\textbf{p}-e\textbf{A}({\tilde t})/c)^2
  +I_p
  \right]/\hbar}
  \right)
.
\label{Eq:b_0}
\end{equation}
This expression is understood as the sum of all the ionization events which occur from the time $t'$ to $t$~\cite{Pascal2001}. 
Then, the instantaneous transition probability amplitude of an electron at a time $t'$, at which it appears into the continuum with momentum  ${\bf p}(t')= \textbf{p}-e\textbf{A}(t^{\prime})/c$, is defined by the argument of the integral in Eq.~(\ref{Eq:b_0}).
Furthermore, the exponent phase factor in Eq.~(\ref{Eq:b_0}) denotes the ``semi-classical action'', ${S}({\bf p},t,t^{\prime})$, that defines a possible electron trajectory  from the birth time $t'$ until the ``detection'' time $t$~\cite{Lewenstein1995}:
 \begin{equation}
{S}({\bf p},t,t^{\prime}) 
= 
\int_{t^{\prime}}^{t}d{\tilde t}
\left[
\frac{1}{2m}({\bf p}-e\textbf{A}({\tilde t})/c)^2
+I_p
\right].
\end{equation}
As our purpose is to obtain the final transition amplitude $b_0({\bf p},t)$, the time $t$ will be fixed at the end of the laser field, $t=t_{\rm F}$. 
For our calculations, we thus define the integration time window as $t \in [0,t_{\rm F}]$.
Therefore, we set ${\bf E}(0) = {\bf E}(t_{\rm F}) = {\bf 0}$, in such a way to make sure that the electromagnetic field is a time oscillating wave and  does not  have static  components. 
The same arguments are applied to the vector potential ${\bf A}(t)$. 
In concrete calculations we have defined the laser pulse envelope as $f(t)=\sin^2(\frac{\omega_0t}{2N_c})$ where $N_c$ denotes the number of total cycles.

Note that, for an arbitrary electromagnetic field, it is possible for ${\bf A}_{\rm F}(t)\ne  0$, i.e.\ the vector potential does not necessarily vanish at the end of the pulse, implying that the kinetic momentum at $t_{\rm F}$ is ${\bf p}_\mathrm{kin}={\bf p}-e{\bf A}(t_{\rm F})/c$; 
if that is the case then it should be considered carefully, since it is ${\bf p}_\mathrm{kin}$ which is detected in experiments. 
However, for laser pulses that are focused away from their source and in the paraxial approximation, nonzero-area pulses of this form are not possible, and the vector potential can be taken as zero on both sides of the pulse.

\subsection{Rescattering transition amplitude}

In order to find a solution for the transition amplitude of the rescattered photoelectrons, $b_1({\bf p},t)$, we have considered the rescattering core matrix element $\textbf{g}(\textbf{p},\textbf{p}^{\prime})$ term of Eq.~(\ref{Eq:New5}) different than zero, i.e.\ $\textbf{g}(\textbf{p},\textbf{p}^{\prime}) \not= \textbf{0}$. 
In addition, the first-order perturbation theory is applied to obtain $b_1({\bf p},t)$ by inserting the zeroth-order solution $b_0({\bf p},t)$ in the right-hand side of Eq.~(\ref{Eq:New5}). 
Then, we obtain $b_1({\bf p},t)$ as a function of the canonical momentum ${\bf p}$ (neglecting the subscript $c$) as follows:
\begin{equation}
\begin{split}
b_1( \textbf{p},t) 
=& 
\left(\frac{i}{\hbar}\right)^2 
\int_0^t{\textit{d}t^{\prime}\exp{\left[-\textit{i} S({\bf p},t,t')/\hbar\right]}\,{\textbf{E}(t^{\prime})\cdot}} \int_0^{t^\prime}{\textit{d} \textit{t}^{{\prime}{\prime}}}\int{\textit{d}^3\textbf{p}^{\prime} }  \textbf{g}\left(\textbf{p}-e\textbf{A}(t^{\prime})/c,\textbf{p}^{\prime}-e\textbf{A}(t^{\prime})/c\right)
\\
& \qquad \times 
\textbf{E}(t^{{\prime}{\prime}}) \cdot \textbf{d}\left( \textbf{p}^{\prime} -e\textbf{A}(t^{{\prime}{\prime}})/c\right) a(t^{{\prime}{\prime}})\: \exp{\left[-\textit{i}  S({\bf p}',t',t'')/\hbar\right]}
 .
\end{split}
\label{Eq:b_1}
\end{equation}
This last equation contains all the information about the rescattering  process. In particular, it refers to the probability amplitude of an emitted electron at the time $t^{\prime\prime}$, with an amplitude given by
\begin{equation}
\textbf{E}(t^{{\prime}{\prime}}) \cdot \textbf{d}\left(
\textbf{p}^{\prime} -e\textbf{A}(t^{{\prime}{\prime}})/c\right) a(t^{{\prime}{\prime}}).
\end{equation}
In this step the electron has a kinetic momentum of ${\bf v}^{\prime}(t'')=\textbf{p}^{\prime}-e\textbf{A}(t^{{\prime}{\prime}})/c$.
The last factor, $\exp{\left[-\textit{i} S({\bf p}^{\prime},t{'},t'')\right]}$, is the accumulated phase of an electron born at the time $t^{\prime\prime}$ until it rescatters at time $t^\prime$. 
The intervening term, ${\bf g}({\bf p}-e{\bf A}(t')/c,{\bf p}'-e{\bf A}(t')/c)$, contains the structural matrix element of the continuum-continuum transition at the re-scattering time $t'$. 
At this particular moment in time, the electron changes its kinetic momentum from ${\bf p}'-e{\bf A}(t')/c$ to ${\bf p}-e{\bf A}(t')/c$. 
We stress, however, that  the term  ${\bf g}({\bf v},{\bf v}')$ does not necessarily imply that the electron returns to the ion core.

In addition to this, the phase term $\exp\left[-\textit{i}S({\bf p},t,t')\right]$ defines the accumulated phase of the electron after the rescattering from the time $t'$ to the ``final'' one $t$ when the electron is ``measured'' at the detector with momentum \textbf{p}.
In particular, note that the photoelectron spectrum, $|b({\bf p},t_{\rm F})|^2 $, is a coherent superposition of both solutions, $b_0({\bf p},t_{\rm F})$ and~$b_1({\bf p},t_{\rm F})$, together with an interference term:
\begin{align}
|b({\bf p},t_{\rm F})|^2 
& =
|b_0({\bf p},t_{\rm F})+b_1({\bf p},t_{\rm F})|^2,
\nonumber \\
& =
|b_0({\bf p},t_{\rm F})|^2 + |b_1({\bf p},t_{\rm F})|^2 + b_0({\bf p},t_{\rm F}){b_1^*}({\bf p},t_{\rm F}) 
+ \mathrm{c.c.} 
\label{Eq:ATIS}
\end{align}

So far we have formulated  a model, which describes the photoionization process leading to two main terms, namely, a direct $b_0({\bf p},t_{\rm F})$ and a rescattering $b_1({\bf p},t_{\rm F})$ one. 
As the complex transition amplitude, Eq.~(\ref{Eq:b_0}), is a single time integral, it can be integrated numerically without major problems.  
However, the multiple  time (``2D'') and momentum (``3D'') integrals of the re-scattering term, Eq.~(\ref{Eq:b_1}), present an increasingly difficult and demanding task from a computational perspective. 
In order to reduce the computational difficulties, and to obtain a physical meaning of the ATI process, one may employ saddle-point methods  to evaluate these highly-oscillatory integrals (see Section~\ref{subsec:saddle-point} below for details).

The main challenge  to calculate the ATI spectrum is then the computation of the bound-free transition dipole matrix  element, ${\bf d}({\bf p})$, and  the  continuum-continuum transition re-scattering  matrix  element ${\bf g}({\bf p},{\bf p}')$ for a given atomic or molecular system. 
In the Appendices, we discuss how to do this analytically for a model atom or molecule with a short-range separable potential.

\subsection{Time-dependent dipole moment}

Finally, to analyse the HHG we need to know the electron acceleration, or at least the  time dependent electron dipole moment. This is dominantly given by the zeroth order solution of the SFA equations. It is then  given by the dynamical version of the celebrated {\it Landau-Dykhne} formula,
\begin{align}
\begin{split}
\langle {\bf r}(t)\rangle 
& =
\mathrm{Re}\mathopen{}\left[
\frac{i}{\hbar} 
\int_0^t\textit{d} 
\textit{t}^{\prime} 
\int d^3{\bf p} \:
a^*(t) \,
\textbf{d}^*\mathopen{}\left( \textbf{p}-e\textbf{A}(t)/c\right)\mathclose{}
\ 
\textbf{E}(t^{\prime})\cdot \textbf{d}\left( \textbf{p}-e\textbf{A}(t^{\prime})/c\right)
a(t^{\prime})
\right. \\ & \qquad \qquad \left. \times 
\exp\mathopen{}\left(
  -\textit{i} \:\int_{t^{\prime}}^t d{\tilde t}
  \left[
  \frac{1}{2m}(\textbf{p}-e\textbf{A}({\tilde t})/c)^2 +I_p
  \right]/\hbar
  \right)
\right]
,
\end{split}
\label{Eq:HHG-td-dipole}
\end{align}
which is then generally compared to experiment via its frequency-domain version, the Fourier transform
\begin{align}
\label{Eq:HHG-fd-dipole}
\tilde{\bf r}(\Omega)
&= 
\int_{-\infty}^\infty \langle {\bf r}(t)\rangle e^{+i\Omega t}  dt
\nonumber \\
&=
\mathrm{Re}\mathopen{} \left[
\frac{i}{\hbar}
\int_{-\infty}^\infty dt 
\int_0^t\textit{d} \textit{t}^{\prime} 
\int d^3{\bf p} \:
a^*(t) 
\textbf{d}^*\mathopen{}\left( \textbf{p}-e\textbf{A}(t)/c\right)\mathclose{}
\textbf{E}(t^{\prime})\cdot \textbf{d}\left( \textbf{p}-e\textbf{A}(t^{\prime})/c\right)a(t^{\prime})
\right. \\ & \qquad \qquad \left. \times 
\nonumber
\exp\mathopen{}\left(
  -\textit{i} \:\int_{t^{\prime}}^td{\tilde t}
  \left[
  \frac{1}{2m}(\textbf{p}-e\textbf{A}({\tilde t})/c)^2 +I_p
  \right]/\hbar +i\Omega t 
  \right)
\right]
,
\end{align}
where $\Omega$ is the frequency of the emitted harmonic.

\subsection{Saddle-point methods and quantum orbits}
\label{subsec:saddle-point}
The SFA results as we have obtained them thus far, i.e. Eqs.~\eqref{Eq:b_0} and \eqref{Eq:b_1} for the direct- and rescattered-electron momentum wavefunctions and Eq.~\eqref{Eq:HHG-fd-dipole} for the frequency-domain harmonic dipole, in what is known as their time-integrated versions with explicit integrals over the times of ionization and recollision or recombination, are often perfectly sufficient for the evaluation of the relevant physical observables via a direct numerical integration. However, they generally involve the integration of highly oscillatory terms, such as those contained in the phase factor of the harmonic dipole in \eqref{Eq:HHG-fd-dipole},
\begin{equation}
\exp\mathopen{}\left(
-i S_\Omega (\textbf{p},t,t') / \hbar
\right)
=
\exp\mathopen{}\left(
  -\frac{\textit{i}}{\hbar}
  \int_{t^{\prime}}^t
    \left[ 
      \frac{1}{2m} (\textbf{p}-e\textbf{A}({\tilde t})/c)^2 
      +I_p 
      \right]
    d{\tilde t} 
  +i\Omega t 
  \right)
,
\label{Eq:HHG-phase}
\end{equation}
where the phase of the exponential can vary wildly, introducing extreme cancellations in the integrand that require increased precision in the numerical integration to calculate correctly. Generally speaking, the phase in this factor can be estimated by considering its scaling once the pulse amplitude $\mathcal{E}_0$ and frequency $\omega$ are factored out by de-dimensionalizing the integral in the exponent, giving contributions which scale with the so-called strong-field parameter $z=U_p/\omega$ and with $I_p/\omega$. For experiments with a strong low-frequency field, both of these parameters are large, and the exponent in Eq.~\eqref{Eq:HHG-phase} will quickly cover many radians without giving the rest of the integrand time to change, giving rise to cancellations in the integral.

As mentioned earlier, this problem can be overcome by approximating any relevant oscillatory integrals using the method of steepest descents, which approximates the integrals using the values of the integrand at stationary points of the action -- in exactly the same way as the classical trajectories emerge as the stationary-action points of the Feynman path integral~\cite{Pascal2001}. Using the paradigmatic case of HHG as an example, this requires us to find solutions to the stationary-point equations over the three integration variables,
\begin{align}
\frac{\partial}{\partial t'} S_\Omega (\textbf{p},t,t') = \frac{1}{2m} (\textbf{p}-e\textbf{A}(t')/c)^2 +I_p & = 0
\label{Eq:tunnelling-eq}
\\
\frac{\partial}{\partial t} S_\Omega (\textbf{p},t,t') + \Omega = \frac{1}{2m} (\textbf{p}-e\textbf{A}(t)/c)^2 +I_p & = \hbar \Omega
\label{Eq:recollision-eq}
\\
\nabla_\textbf{p} S_\Omega (\textbf{p},t,t') = \frac{1}{m} \int_{t^{\prime}}^t
\left[ \textbf{p}-e\textbf{A}({\tilde t})/c \right]
d{\tilde t} & = 0 
\label{Eq:return-eq}
\end{align}
which are often termed the tunnelling, recollision, and return equations, respectively. (For other phenomena, these should be adjusted accordingly, by e.g.\ dropping (\ref{Eq:recollision-eq}, \ref{Eq:return-eq}) for the direct-electron ionization amplitude.) The tunnelling equation here, Eq.~\eqref{Eq:tunnelling-eq}, is the central, determining structure, both because of its prevalence over all SFA applications and because both of the terms on its left-hand side, $\frac12 (\textbf{p}-e\textbf{A}(t')/c)^2$ and $I_p$, are ostensibly positive (for real $t'$), which means that solutions will only be possible if $t'$ (and, with it, all the other variables) are complex-valued.

\begin{figure}[htb]
\begin{tabular}{c}
\includegraphics[scale=1]{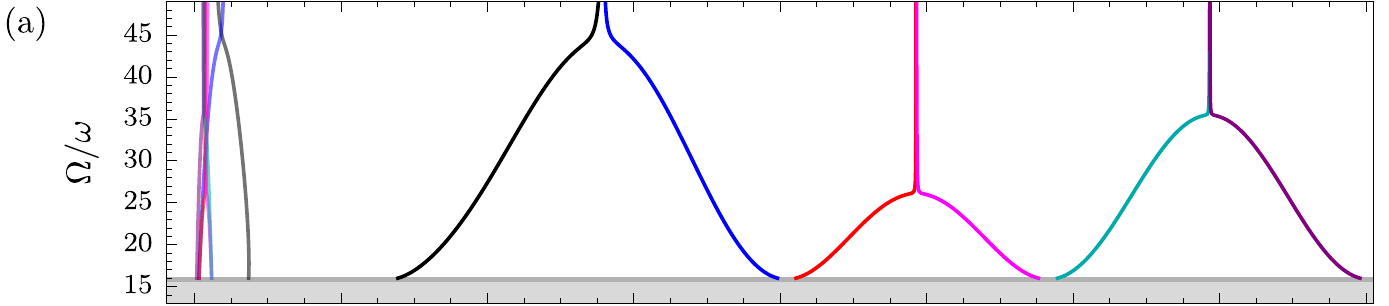} \\
\includegraphics[scale=1]{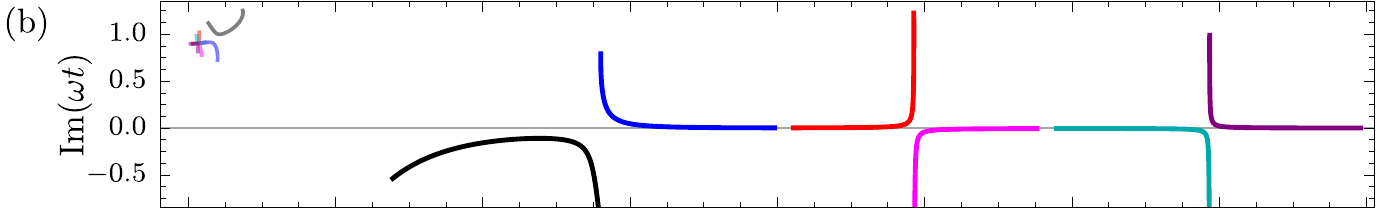} \\
\includegraphics[scale=1]{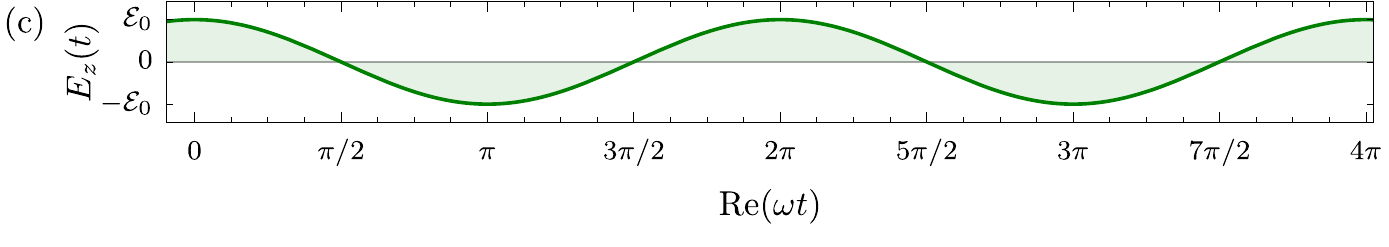}
\end{tabular}
\caption{Saddle-point trajectories for HHG produced by helium in a monochromatic field of wavelength $\SI{800}{nm}$ and intensity $\SI{2e14}{W/cm^2}$, showing (a) the energy-time relationship, (b) the saddle-point solutions in the complex time plane, and (c) the electric field, for reference. 
The recollision and ionization saddle points (shown in solid and faint lines in panel panel (b)) form a series of curves in complex time, with close-to-real recollision times in the harmonic plateau that then veer off into imaginary time at the cutoff. 
When plotted as an energy-time relationship (or, more precisely, as the harmonic order $\Omega/\omega$ versus the recollision time, shown in panel (a)), the saddle-point curves `wrap' around the simple-man model's purely classical relationships. 
 The below-threshold region, which is not well described by the SFA, is shown shaded in panel (a).
}
\label{fig:saddle-point-curves}
\end{figure}

Within that steepest-descent approximation, then, SFA amplitudes are given by a sum over all the relevant saddle-point roots that contribute to the deformed integration contour,
\begin{align}
\tilde{\bf r}(\Omega)
& =
\mathrm{Re}\mathopen{} \left[ \textit{i} \: 
\sum_s 
H(t_s,t'_s,\textbf{p}_s)
a^*(t_s) \mathbf{d}\mathopen{}\left( \textbf{p}_s-e\textbf{A}(t_s)/c\right)
\:\mathbf{E}(t'_s)\cdot \mathbf{d}\mathopen{}\left( \textbf{p}_s-e\textbf{A}(t'_s)/c\right)a(t'_s)
e^{ -i S_\Omega (\textbf{p}_s,t_s,t'_s) / \hbar }
\right]
.
%\label{Eq:HHG-fd-dipole}
\end{align}
with an additional Hessian factor $H(t_s,t'_s,\textbf{p}_s)$ that accounts for the width of the complex-integration Gaussians being approximated~\cite{Bleistein1975}. (Alternatively, it is also possible to perform a partial saddle-point approximation over momentum only, keeping the unique root of the return equation \eqref{Eq:return-eq} $\textbf{p}_s=\textbf{p}_s(t,t')$ as a function of the ionization and recollision times, and then integrate numerically.) For the full saddle-point method, the ionization time typically has a large imaginary part and it is confined to a small window shortly after the peak of the field, while the recollision time comes in a series of so-called quantum orbits that span the following periods, as shown in Fig.~\ref{fig:saddle-point-curves}.

Typically, the only quantum orbits that contribute significantly to harmonic generation are the so-called short and long trajectories, shown in black and blue (resp.) in Fig.~\ref{fig:saddle-point-curves}(b). The long trajectories, which ionize closer to the peak of the field, have a higher single-atom harmonic yield, but the phase-matching conditions are typically chosen to select the contribution of the short trajectories, which are easier to phase-match. The higher-order returns (shown in red, pink, green and purple), which recollide more than one period after the ionization, typically spend too much time in the continuum accumulating an intensity-dependent phase for them to form a macroscopically-coherent emission, but under dedicated circumstances it is still possible to observe signatures of their presence~\cite{Sansone2006b,Zair2008}.

%Saddle-point methods, quantum orbits, trajectories, long-and-short trajectories. For HHG and ATI.
%
%Emilio.

\subsection{Polarization effects}

The above analysis of quantum orbits becomes, obviously, more complex when the laser fields have more complicated patterns of polarization beyond the simple linear one. Elliptical polarization was considered in the context of HHG already in the pioneering papers by P. Corkum~\cite{Corkum1993}: the electron trajectories in such situation form ellipses, and essentially miss the parent ion, leading to a rapid decrease of the HHG efficiency with increasing ellipticity. 
These trajectory-based predictions were first confirmed in experiments by Budil et al.~\cite{Budil1993}, 
and they can be used to produce `gating' schemes~\cite{CorkumBurnett1994, Kovacev2003, Mashiko2008} to produce isolated attosecond pulses by using a time-dependent polarization that changes across the pulse from circular to linear and back.

The late Bertrand Carr\'e was also among the pioneers of polarization studies in HHG. 
The first experimental results of ellipticity dependence of the harmonic yield were published in Ref.~\citealp{Antoine1996elliptical}, in which the detailed SFA theory of HHG  by an elliptically polarized laser field was investigated. 
The following seminal paper~\cite{Antoine1997} was the first one to discuss measurements of the (partial) polarization of high harmonics generated by elliptically polarized laser fields, with careful comparison to SFA-based theory including propagation. 
This paper stimulated many researchers to search for ways to control polarization of harmonics. The Holy Grail was to generate high-order harmonics with left- and right-circular polarization to be able to use them to study circular dichroism in absorption---to distinguish, for instance, chiral molecules---, or to study chiral effects in magnetism. 

The rapid decrease of HHG efficiency with ellipticity suggested looking for scenarios based around a linearly-polarized IR driver. 
Pioneering ideas were formulated by P.-M. Paul in his doctoral thesis~\cite{Paul2001}, and developed further in the group of B.\ Carr\'e by Y.\ Mairesse, first employing two-photon absorption of one XUV and one IR photon, and later resonant HHG~\cite{Ferre2015a, Ferre2015b}, as well as HHG generated by linearly-polarized light pulses applied to aligned molecules~\cite{Levesque2009, Zhou2012, Shafir2012}.

A breakthrough method was proposed by D.\,B.\ Milo\v{s}evi\'c~\cite{Milosevic2000} and later implemented by O.\ Cohen, using two circularly-polarized beams with a frequency ratio of $1{:}2$ and opposite helicity~\cite{Fleischer2014}, which permits the generation of bright phase-matched circularly-polarized extreme ultraviolet high harmonics~[\citealp{Kfir2015}; for a review see \citealp{PisantyThesis2016}]. 
The original method and results of Ref.~\citealp{Antoine1997} was developed further to completely characterize the state of elliptically-polarized light by electron-ion vector correlations~\cite{Veyrinas2013}, and finally to realise the complete polarimetry of high harmonics~\cite{Veyrinas2016}. 
These methods have recently been applied to HHG generated under O.\ Cohen's `bicircular' fields~\cite{Barreau2018}, providing a clear evidence for depolarization of high harmonics.  

It is worth mentioning that more laser fields with ``exotic'' polarization (spin) and orbital angular momentum have been proposed recently (see Ref.~\citealp{Pisanty2018conservation} and references therein). These so-called polarization torus knots, proposed in Ref.~\citealp{Pisanty2018knotting}, when applied to atoms in a form of ultrashort and ultraintense pulse, generate ``exotic'' harmonics that conserve  torus-knot angular momentum, a topologically-nontrivial mixture between spin and orbital angular momentum.

One should also say that ellipticity and polarization effects play an important role in ATI and multielectron ionization. 
A nice example of these ATI results is included in the Science paper of 2001~\cite{Pascal2001}, where G. Paulus was able to characterize a whole plethora of trajectories corresponding to rescattering of electrons in elliptically polarized laser fields. 
In the same paper, Carr\'e and Sali\`eres present spectra of high harmonics that allow one to identify directly the contribution of the ``short'' and ``long'' electronic trajectories.

\section{Two-Electron Processes}
\label{sec:4-two-electrons}

\subsection{Hamiltonian and states}

In order to describe higher order processes we must extend this formalism to include more active electrons. We start by formulating the SFA equations for two active electrons, which allows us to model higher-order strong field ionization process such as non-sequential double ionization (NSDI). We also include excited states in the wave function to allow for the recollision excitation with subsequent ionization (RESI) pathway of NSDI; otherwise only the direct electron impact ionization (EII) pathway would be present. Following a similar procedure to the one electron case,  the ansatz for the wavefunction can be written as
\begin{align}
	\ket{\psi(t)}=\exp(iE_0 t/\hbar)\left(a(t)\ket{0}+\int d^3\mathbf{p}\; b(\mathbf{p},t)\ket{\mathbf{p},0}
	+\sum_{\eta}\int d^3 \mathbf{p}\; c(\mathbf{p},\eta,t)\ket{\mathbf{p},\eta}
	+\iint d^3\mathbf{p}d^3\mathbf{p}'\; d(\mathbf{p},\mathbf{p}',t)\ket{\mathbf{p},\mathbf{p}'}	
	\right),
	\label{Eq:2e-PWavef}
\end{align}
where $\ket{0}$ is the two-electron ground state, $\ket{\mathbf{p},0}$ gives the two-electron state, where the first electron has been promoted to a continuum scattering state with momentum $\mathbf{p}$, $\ket{\mathbf{p},\eta}$ is similar but the second electron is in an excited state with a principal quantum number $\eta$, while $\ket{\mathbf{p},\mathbf{p}'}$ denotes both electrons in continuum scattering states. The state $\ket{\eta,0}$ could be included to allow for some additional effects such as single electron frustrated tunnelling, but we will neglect it for now as we are interested in two-electron effects, where this state will play almost no role. These states are all eigenstates of the two-particle Hamiltonian
\begin{equation}
\hat{H}_{0}
=
\sum_{i=1}^{2}
\left(
  \frac{\hat{\mathbf{p}}_i^2}{2 m}
  +V(\hat{\mathbf{r}}_i)
  \right)
+V(\hat{\mathbf{r}}_1-\hat{\mathbf{r}}_2),
\label{Eq:2e-H0}
\end{equation} 
where $\hat{\mathbf{p}}_i$ are single particle momentum operators, $V(\mathbf{r}_i)$ gives the interaction of each particle with the atomic or molecular core and $V(\mathbf{r}_1-\mathbf{r}_2)$ gives the interaction between the two electrons. Note that including the interaction between electrons means that none of the two-particle states introduced above can be written as products of one-particle states, e.g. $\ket{\mathbf{p},0}\ne\ket{\mathbf{p}}\ket{0}$ . However, we can write the energy eigenvalue equations for each sector:
\begin{align}
\hat{H}_{0}\ket{0} & = -E_{0}\ket{0},
& 
E_0 & = I_{2p}
\ \text{(two-electron ionization potential)} ;
\\
\hat{H}_{0}\ket{\mathbf{p},0} 
& = 
\left(\frac{p^2}{2m}-E_{10}\right) \ket{\mathbf{p},0},
&
E_{10} & = I_{1p}
\ \text{(one-electron ionization potential)};
\\
\hat{H}_{0}\ket{\mathbf{p},\eta}
& = 
\left(\frac{p^2}{2m}-E_{1\eta}\right) \ket{\mathbf{p},\eta},
&
E_{1\eta} & = I_{1\eta,p}
\ \text{(one-electron excited-state ionization potential)};
\\
\hat{H}_{0}\ket{\mathbf{p},\mathbf{p}'}
& =
\left(\frac{p^2}{2m}+\frac{p'^2}{2m}\right) \ket{\mathbf{p},\mathbf{p}'}
.
\end{align}
Given that we are accounting for electron correlation, $E_0$ will generally be different from $2E_{10}$; however, often the correlation is weak and then this is a good approximation to make. 
Note that here the eigenstates $|\mathbf{p},0\rangle$, $|\mathbf{p},\eta\rangle$ and $|\mathbf{p},\mathbf{p}'\rangle$ do not denote plane-wave states in the continuum, but instead the full one- and two-electron scattering eigenstates with asymptotic outgoing momenta $\mathbf{p}$ and $\mathbf{p}'$.
This explains the particular form of the eigenenergies.

If one assumes non-interacting electrons, this amounts to dropping the last term in Eq.~\eqref{Eq:2e-H0}, then the following substitution can be made for each of the two particle states
\begin{align}
%|0>
	\ket{0}&\rightarrow\ket{\psi_0}
	\quad \quad \ \ =\ket{0}\ket{0},\hspace{4cm}\\
%|k,0>
	\ket{\mathbf{p},0}&\rightarrow\ket{\psi_0(\mathbf{p})}
	\quad \; =\frac{1}{\sqrt{2}}\left(
	\ket{\mathbf{p}}\ket{0}+\ket{0}\ket{\mathbf{p}}
	\right),\hspace{4cm}\\
%|k,\eta>
	\ket{\mathbf{p},\eta}&\rightarrow\ket{\psi_0(\mathbf{p},\eta)}
	\ =\frac{1}{\sqrt{2}}\left(
	\ket{\mathbf{p}}\ket{\eta}+\ket{\eta}\ket{\mathbf{p}}
	\right),\hspace{4cm}\\
%|k,k'>
	\ket{\mathbf{p},\mathbf{p}'}&\rightarrow\ket{\psi_0(\mathbf{p},\mathbf{p}')}
	=\frac{1}{\sqrt{2}}\left(
	\ket{\mathbf{p}}\ket{\mathbf{p}'}+\ket{\mathbf{p}'}\ket{\mathbf{p}}
	\right).\hspace{4cm}
\end{align}
Note we do not use anti-symmetric superpositions as we consider two electrons from a singlet spin state, so that the spins will already be anti-symmetric. These will be eigenstates of the Hamiltonian $\hat{H}_0$ \emph{without} the electron-electron interaction term. 

As in the one-particle case the full Hamiltonian is given by laser-free and dipole coupling Hamiltonians as in Eq.~\eqref{Eq:H} but now in the two particle case the dipole coupling is given by
\begin{equation}
	\hat{U}=-\sum_{i=1}^{2}e\mathbf{E}(t)\cdot\hat{\mathbf{r}}_i.=-\mathbf{E}(t)\cdot e(\hat{\mathbf{r}}_1+\hat{\mathbf{r}}_2).
\end{equation}

We will proceed as before and derive the integro-differential equations for $a(t)$, $b(\mathbf{p},t)$, $c(\mathbf{p},\eta,t)$ and $d(\mathbf{p}, \mathbf{p}', t)$. However, first we will introduce the dipole matrix elements required for each possible kind of transition between the two particle states. The matrix elements will follow the convention that the left-hand state will have a lower or equal energy to the right-hand state. Then the dipole matrix elements can be defined in the following way
\begin{align}
%d1
	\mathbf{d}(\mathbf{p})&:=\braket{0|e(\hat{\mathbf{r}}_1+\hat{\mathbf{r}}_2)|\mathbf{p},0},&
%g1
	\mathbf{g}(\mathbf{p},\mathbf{p}')&:=\braket{\mathbf{p},0|e(\hat{\mathbf{r}}_1+\hat{\mathbf{r}}_2)|\mathbf{p}',0},
	\nonumber\\
%d2
	\mathbf{d}(\mathbf{p},\eta)&:=\braket{0|e(\hat{\mathbf{r}}_1+\hat{\mathbf{r}}_2)|\mathbf{p},\eta},&
%g2
	\mathbf{g}(\mathbf{p},\mathbf{p}',\eta)&:=\braket{\mathbf{p},0|e(\hat{\mathbf{r}}_1+\hat{\mathbf{r}}_2)|\mathbf{p}',\eta},
	\nonumber\\
%d3
	\mathbf{d}(\mathbf{p},\mathbf{p}')&:=\braket{0|e(\hat{\mathbf{r}}_1+\hat{\mathbf{r}}_2)|\mathbf{p},\mathbf{p}'}&
%g3
	\mathbf{g}(\mathbf{p},\mathbf{p}',\mathbf{p}'')&:=\braket{\mathbf{p},0|e(\hat{\mathbf{r}}_1+\hat{\mathbf{r}}_2)|\mathbf{p}',\mathbf{p}''},
	%\nonumber
	\\
	\mathbf{d}(\eta) &:= \braket{0|e\hat{\mathbf{r}}|\eta}_1
	& \nonumber \\
%	
%h1
	\mathbf{h}(\mathbf{p},\eta,\mathbf{p}',\eta')&:=\braket{\mathbf{p},\eta|e(\hat{\mathbf{r}}_1+\hat{\mathbf{r}}_2)|\mathbf{p}',\eta'},
	&\nonumber\\
%h2
	\mathbf{h}(\mathbf{p},\eta,\mathbf{p}',\mathbf{p}'')&:=\braket{\mathbf{p},\eta|e(\hat{\mathbf{r}}_1+\hat{\mathbf{r}}_2)|\mathbf{p}',\mathbf{p}''},
	&
%i
	\mathbf{i}(\mathbf{p},\mathbf{p}',\mathbf{p}'',\mathbf{p}''')&:=\braket{\mathbf{p},\mathbf{p}'|e(\hat{\mathbf{r}}_1+\hat{\mathbf{r}}_2)|\mathbf{p}'',\mathbf{p}'''}
	,
	\nonumber
\end{align}
where the subindices refer to the two electrons. Note that, due to symmetry, the dipole matrix element from $\ket{0}$ to $\ket{0}$ will be zero. Each matrix element has an important physical meaning, which we will discuss in some detail for both the interacting and non-interacting cases in the Appendix~\ref{app:D-dipoles}.

\subsubsection{Example for RESI}
\begin{figure}
	\includegraphics[width=0.45\textwidth]{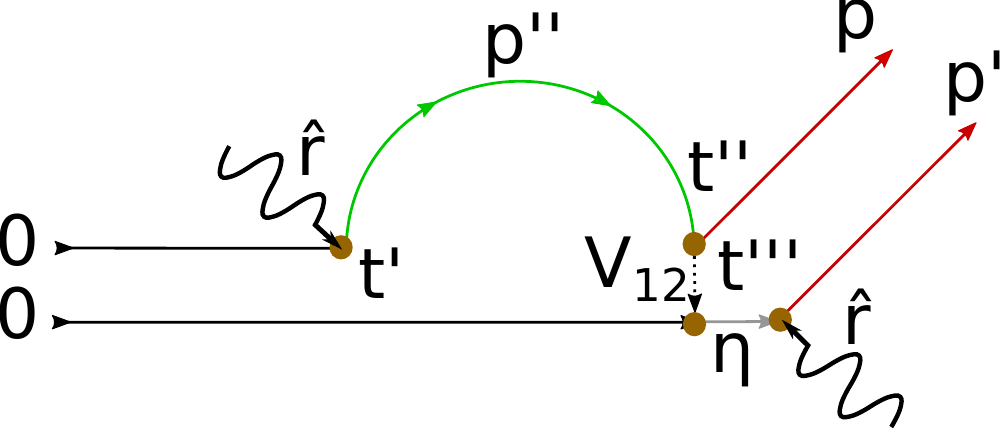}
	\caption{The complete RESI process showing all three transitions. The three nodes marked at the times $t'$, $t''$ and $t'''$ show the three transitions identified in the chain in Eq.~\eqref{Eq:2e_RESI_Chain}.}
	\label{Fig:2e-FeynmannRESI}
\end{figure}

Here we use this formulation to recover the equations for the RESI mechanism of NSDI.
%\textcolor{red}{I have only done the simplest case, where electron-electron interaction is only considered in the pivotal step of excitation.}
The process of RESI goes through each states in our two-electron wavefunction ansatz given by Eq.~\eqref{Eq:2e-PWavef}. Hence, it goes through the `chain'
\begin{equation}
\ket{0}
\underset{\mathbf{d}(\mathbf{p}'')}{\xrightarrow{\hspace*{6mm}}}
\ket{\mathbf{p}'',0}
\underset{\mathbf{g}(\mathbf{p}'',\mathbf{p},\eta)}{\xrightarrow{\hspace*{12mm}}}
\ket{\mathbf{p},\eta}
\underset{\mathbf{h}(\mathbf{p},\eta,\mathbf{p},\mathbf{p}')}{\xrightarrow{\hspace*{12mm}}}
\ket{\mathbf{p},\mathbf{p}'}
,
\label{Eq:2e_RESI_Chain}
\end{equation}
where the dipole matrix elements underneath are essential for the  transitions between states, so must be included to describe RESI. In Fig.~\ref{Fig:2e-FeynmannRESI}, the complete Feynman diagram for RESI is shown. In addition to these dipole matrix elements we include those responsible for self propagating the states
\begin{align}
\ket{\mathbf{p},0}&\rightarrow\ket{\mathbf{p}',0}
& \text{implemented by} \quad &
\mathbf{g}(\mathbf{p},\mathbf{p}'),\\
\ket{\mathbf{p},\eta}&\rightarrow\ket{\mathbf{p}',\eta'}
& \text{implemented by} \quad &
\mathbf{h}(\mathbf{p},\eta,\mathbf{p}',\eta'),\\
\ket{\mathbf{p},\mathbf{p}'}&\rightarrow\ket{\mathbf{p}'',\mathbf{p}'''}
& \text{implemented by} \quad &
\mathbf{i}(\mathbf{p},\mathbf{p}',\mathbf{p}'',\mathbf{p}''').
\end{align}
The remaining dipole matrix elements, $\mathbf{d}(\mathbf{p},\eta),\ \mathbf{d}(\mathbf{p},\mathbf{p}')$, and $\mathbf{g}(\mathbf{p},\mathbf{p}',\mathbf{p}'')$, will not contribute significantly to RESI and thus we can neglect their contributions to the corresponding part of the time-dependent Schr\"odinger equation.
In this example we will take the simplest case where electron interaction is only considered in the necessary step of the excitation of the second electron. As in the case of SFA in the SAE approximation, the crucial point is to determine the most singular parts of the relevant matrix elements and the less singular ``rest''. The SFA will then correspond to the systematic expansion of the ``rest''.

Below we provide the necessary decompositions of the relevant matrix at two levels: (a) at the level of exact matrix elements, calculated for two-electron Hamiltonian; (b) approximated matrix elements, calculated, neglecting the electron-electron interactions. In this way we will be  able to compare with previous SFA results for RESI. We list both expressions (a) and (b)  in their most explicit form,
\begin{align}
	\mathbf{d}(\mathbf{p})&= \mathbf{d}(\mathbf{p})\simeq\frac{2 e}{\sqrt{2}}\braket{\mathbf{p}|\hat{\mathbf{r}}|0_1},\\
	\mathbf{g}(\mathbf{p},\mathbf{p}')&=ie \hbar \nabla_{\mathbf{p}}\delta(\mathbf{p}-\mathbf{p}')+\hbar \tilde{\mathbf{g}}(\mathbf{p},\mathbf{p}')\simeq ie \hbar \nabla_{\mathbf{p}}\delta(\mathbf{p}-\mathbf{p}')+\hbar {\mathbf{g}}_1(\mathbf{p},\mathbf{p}'),\\
	\mathbf{h}(\mathbf{p},\eta,\mathbf{p}',\eta')&=\delta_{\eta \eta'}\left(
	ie \hbar \nabla_{\mathbf{p}}\delta(\mathbf{p}-\mathbf{p}')+\hbar\tilde{\mathbf{h}}(\mathbf{p},\mathbf{p}',\eta)
	\right)
	+e \delta(\mathbf{p}-\mathbf{p}')\tilde{\mathbf{d}}(\eta,\eta')\\ \nonumber
&\simeq     \delta_{\eta \eta'}\left(
	ie \hbar \nabla_{\mathbf{p}}\delta(\mathbf{p}-\mathbf{p}')+\hbar{\mathbf{g}}_1(\mathbf{p},\mathbf{p}')
	\right)
	+e \delta(\mathbf{p}-\mathbf{p}')\braket{\eta|\hat{\mathbf{r}}|\eta'},\\
	\mathbf{h}(\mathbf{p},\eta,\mathbf{p}',\mathbf{p}'')&\simeq
	e\delta(\mathbf{p}-\mathbf{p}'')\tilde{\mathbf{h}}(\mathbf{p}, \mathbf{p}', \eta)
	+e\delta(\mathbf{p}-\mathbf{p}')\tilde{\mathbf{h}}(\mathbf{p}, \mathbf{p}'', \eta) \nonumber \\
&\simeq
	e\delta(\mathbf{p}-\mathbf{p}'')\braket{\eta|\hat{\mathbf{r}}|\mathbf{p}'}
	+e\delta(\mathbf{p}-\mathbf{p}')\braket{\eta|\hat{\mathbf{r}}|\mathbf{p}''}, \\
	\mathbf{i}(\mathbf{p},\mathbf{p}',\mathbf{p}'',\mathbf{p}''')&\simeq
	\delta(\mathbf{p}'-\mathbf{p}''') (ie\hbar\nabla_{\mathbf{p}}\delta(\mathbf{p}-\mathbf{p}'')+\hbar\tilde{\mathbf{g}}(\mathbf{p},\mathbf{p}',
\mathbf{p}'', \mathbf{p}'''))\notag\\
	& \quad +
	\delta(\mathbf{p}'-\mathbf{p}'') (ie\hbar\nabla_{\mathbf{p}}\delta(\mathbf{p}-\mathbf{p}''')+\hbar\tilde{\mathbf{g}}(\mathbf{p},\mathbf{p}',
\mathbf{p}'', \mathbf{p}'''))\notag\\
	& \quad +\delta(\mathbf{p}-\mathbf{p}''') (ie\hbar\nabla_{\mathbf{p}'}\delta(\mathbf{p}'-\mathbf{p}'')+\hbar\tilde{\mathbf{g}}(\mathbf{p},\mathbf{p}',
\mathbf{p}'', \mathbf{p}'''))\notag\\
	& \quad +
	\delta(\mathbf{p}-\mathbf{p}'') (ie\hbar\nabla_{\mathbf{p}'}\delta(\mathbf{p}'-\mathbf{p}''')+\hbar\tilde{\mathbf{g}}(\mathbf{p},\mathbf{p}',
\mathbf{p}'', \mathbf{p}'''))
\notag \\
    &\simeq
    \delta(\mathbf{p}'-\mathbf{p}''') (ie\hbar\nabla_{\mathbf{p}}\delta(\mathbf{p}-\mathbf{p}'')+\hbar\mathbf{g}_1(\mathbf{p},
\mathbf{p}''))\notag\\
	& \quad +
	\delta(\mathbf{p}'-\mathbf{p}'') (ie\hbar\nabla_{\mathbf{p}}\delta(\mathbf{p}-\mathbf{p}''')+\hbar\mathbf{g}_1(\mathbf{p},\mathbf{p}'''))\notag\\
	& \quad +
	\delta(\mathbf{p}-\mathbf{p}''') (ie\hbar\nabla_{\mathbf{p}'}\delta(\mathbf{p}'-\mathbf{p}'')+\hbar\mathbf{g}_1(\mathbf{p}',\mathbf{p}''))\notag\\
	& \quad +
	\delta(\mathbf{p}-\mathbf{p}'') (ie\hbar\nabla_{\mathbf{p}'}\delta(\mathbf{p}'-\mathbf{p}''')+\hbar\mathbf{g}_1(\mathbf{p}',\mathbf{p}'''))
\end{align}
The convention that we use above is that the less singular parts of the matrix elements with tilde are calculated ``exactly'', taking into account electron-electron interaction, while the matrix  with subscript 1 stem from an approximate calculation, in which we neglect the electron-electron interactions, so that these matrix elements can be obtained from the corresponding single electron dipole moments, calculated in the SAE approximation. Thus
$ \hbar {\mathbf{g}}_1(\mathbf{p},\mathbf{p}')=\hbar {\mathbf{g}}(\mathbf{p},\mathbf{p}')$ from the previous chapters.

In addition we have perhaps the most important matrix element that describes re-scattering of the electron accompanied by the excitation of the remaining electron,
\begin{equation}
\mathbf{g}(\mathbf{p},\mathbf{p}',\eta)
=
\delta(\mathbf{p}-\mathbf{p}')\mathbf{d}(\eta)
+\hbar \tilde{\mathbf{g}}(\mathbf{p},\mathbf{p}',\eta)
= 
e \delta(\mathbf{p}-\mathbf{p}')\braket{0|\hat{\mathbf{r}}|\eta}_1
+\hbar {\mathbf{g}}_1(\mathbf{p},\mathbf{p}', \eta)
.
\end{equation}
We will keep this matrix element in its entirety, but we will treat it as a perturbation.

Since we treat $\mathbf{g}(\mathbf{p},\mathbf{p}',\eta)$ as a first order perturbation, we can then keep only the most singular parts of the remaining matrix elements, neglecting the less singular parts such as $\tilde{\mathbf{g}}(\mathbf{p},\mathbf{p}')$.
These less singular contributions can be very interesting,  leading to Coulomb effects such as distortion of interference structures in ATI \cite{Maharjan2006, Meckel2008}, and should certainly receive some attention, but in the present instance we will take the most basic form of RESI and neglect them. Thus, we will take, for instance, the single electron re-scattering $\tilde{\mathbf{g}}(\mathbf{p},\mathbf{p}')=0$ in the above equations. Substituting these into the integro-differential equations (see Eq.~\eqref{Eq:2e-New5} in Appendix~\ref{app:E-two-electron}) leads to a much simplified form:
\begin{align}
	%a
	\dot{a}(t)&=\frac{i}{\hbar}\mathbf{E}(t)\cdot\int d^3\mathbf{p}\; \mathbf{d}^*\mathopen{}(\mathbf{p})b(\mathbf{p},t)\mathclose{}\notag\\
	%b
	\dot{b}(\mathbf{p},t)&=-\frac{i}{\hbar}\Bigg[
	\left(\frac{\hbar^2p^2}{2m}+E_0-E_{10}\right)b(\mathbf{p},t)
	%\notag\\& \hspace{1cm}
	- \mathbf{E}(t)\mathbf{d}(\mathbf{p})a(t)-ie \hbar\mathbf{E}(t)\cdot\mathbf{\nabla}_{\mathbf{p}}b(\mathbf{p},t)\Bigg]
	+ \cdots
	\notag\\ \hspace{1cm}
	%c
	\dot{c}(\mathbf{p},\eta,t)&=-\frac{i}{\hbar}\Bigg[
	\left(\frac{\hbar^2p^2}{2m}+E_0-E_{1\eta}\right)c(\mathbf{p},\eta,t)
	-\mathbf{E}(t)\cdot\int d^3\mathbf{p}'\; \mathbf{g}(\mathbf{p}',\mathbf{p},\eta)b(\mathbf{p}',t)
	-ie\hbar \mathbf{E}(t)\mathbf{\nabla}_{\mathbf{p}}c(\mathbf{p},\eta,t)\Bigg]
	+ \cdots \notag
	\\ \hspace{1cm}
	%d
	\dot{d}(\mathbf{p},\mathbf{p}',t)&=-\frac{i}{\hbar}\Bigg[
	\left(\frac{\hbar^2p^2}{2m}+\frac{\hbar^2p'^2}{2m}+E_0\right)d(\mathbf{p},\mathbf{p}',t)
	-e\mathbf{E}(t)\cdot\sum_{\eta \ne 0}(c(\mathbf{p}',\eta,t)\tilde{\mathbf{g}}(\mathbf{p},\mathbf{p}',\eta)+c(\mathbf{p},\eta,t)\tilde{\mathbf{g}}(\mathbf{p}',\mathbf{p},\eta))
	\notag\\& \qquad \qquad
	-2ie\hbar \mathbf{E}(t)\cdot(\mathbf{\nabla}_{\mathbf{p}}+\mathbf{\nabla}_{\mathbf{p}'})d(\mathbf{p},\mathbf{p}',t)
	\Bigg]
	+ \cdots
	\label{Eq:2e-New5-RESI1}
\end{align}
The above equations contain only the terms relevant for the perturbative solution in the first order in $\tilde{\mathbf{g}}(\mathbf{p},\mathbf{p}',\eta)$ -- they have thus reduced to a very simple form.
Now integral solutions of each of these equations can be formulated, where $\dot{d}(\mathbf{p},\mathbf{p}',t)$ is expressed in terms of $\dot{c}(\mathbf{p},\eta,t)$, while $\dot{c}(\mathbf{p},\eta,t)$ is in terms of $\dot{b}(\mathbf{p},t)$, and $\dot{b}(\mathbf{p},t)$ is in terms of $a(t)$, which we assume to know (or we set to unity for the not-too-strong and not-too-long driving pulses). The solutions are as follows:
\begin{align}
%b
	b(\mathbf{p}'',t'')
	&=
	\frac{i}{h}\int_0^{t''}dt'\; 
	\exp\mathopen{}\left[
		-\frac{i}{\hbar} S_b\left(\mathbf{p}'',t',t''\right)
	\right] 
	\mathbf{E}(t') \cdot \mathbf{d}(\mathbf{p}''-e\mathbf{A}(t')/c) a(t'),\\
%c
	c(\mathbf{p},\eta,t''')
	&=
	\frac{i}{\hbar}\int_0^{t'''}dt''\;\int d^3\mathbf{p}''\; 
	\exp\mathopen{}\left[
		-\frac{i}{\hbar} S_c\left(\mathbf{p},t'',t'''\right)
	\right]
%	\notag
%	\\ &
	\mathbf{E}(t'')\cdot\mathbf{g}(\mathbf{p}-e\mathbf{A}(t'')/c,\mathbf{p}''-e\mathbf{A}(t'')/c,\eta)b(\mathbf{p}'',t'') ,\\
%d
	d(\mathbf{p},\mathbf{p}',t)
	&=
	\frac{i}{\hbar}\int_0^{t}dt'''\;
	\exp\mathopen{}\left[
		-\frac{i}{\hbar} S_d\left(\mathbf{p},\mathbf{p}',t''',t\right)
	\right]	
	\mathbf{E}(t''')\cdot\sum_{\eta \ne 0}
	\bigg(
	\tilde{\mathbf{g}}(\mathbf{p}-e\mathbf{A}(t''')/c,\mathbf{p}'-e\mathbf{A}(t''')/c,\eta) \: c(\mathbf{p}',\eta,t''')
	\\ \notag & \qquad \qquad \qquad \qquad \qquad \qquad \qquad \qquad \qquad \qquad \qquad 
	+
	\tilde{\mathbf{g}}(\mathbf{p}'-e\mathbf{A}(t''')/c,\mathbf{p}-e\mathbf{A}(t''')/c,\eta) \: c(\mathbf{p},\eta,t''')
	\bigg),
\end{align}
where
\begin{align}
	S_b\left(\mathbf{p}'',t',t''\right)
	&=
	\int_{t'}^{t''}d \tau \; \left[\frac{1}{2m}\left(\mathbf{p}''-e\mathbf{A}(\tau)/c\right)^2
	+E_0-E_{10}\right]\\
	S_c\left(\mathbf{p},t'',t'''\right)
	&=
	\int_{t''}^{t'''}d \tau \; \left[\frac{1}{2m}\left(\mathbf{p}-e\mathbf{A}(\tau)/c\right)^2
	+E_0-E_{1\eta}\right]\\
	S_d\left(\mathbf{p},\mathbf{p}',t''',t\right)
	&=
	\int_{t'''}^{t}d \tau \; \left[\frac{1}{2m}\left(\mathbf{p}-e\mathbf{A}(\tau)/c\right)^2
	+\frac{1}{2m}\left(\mathbf{p}'-e\mathbf{A}(\tau)/c\right)^2
	\right]
\end{align}
The S-Matrix  transition amplitude of this process can be related to the above expression in the following way
\begin{align}
	M(\mathbf{p},\mathbf{p}')&=\lim\limits_{t\rightarrow \infty}\braket{\mathbf{p},\mathbf{p}'|\psi(t)}\\
	&=\lim\limits_{t\rightarrow \infty}d(\mathbf{p},\mathbf{p}',t)
\end{align}

\begin{figure}[b!]
	\includegraphics[width=0.6\textwidth]{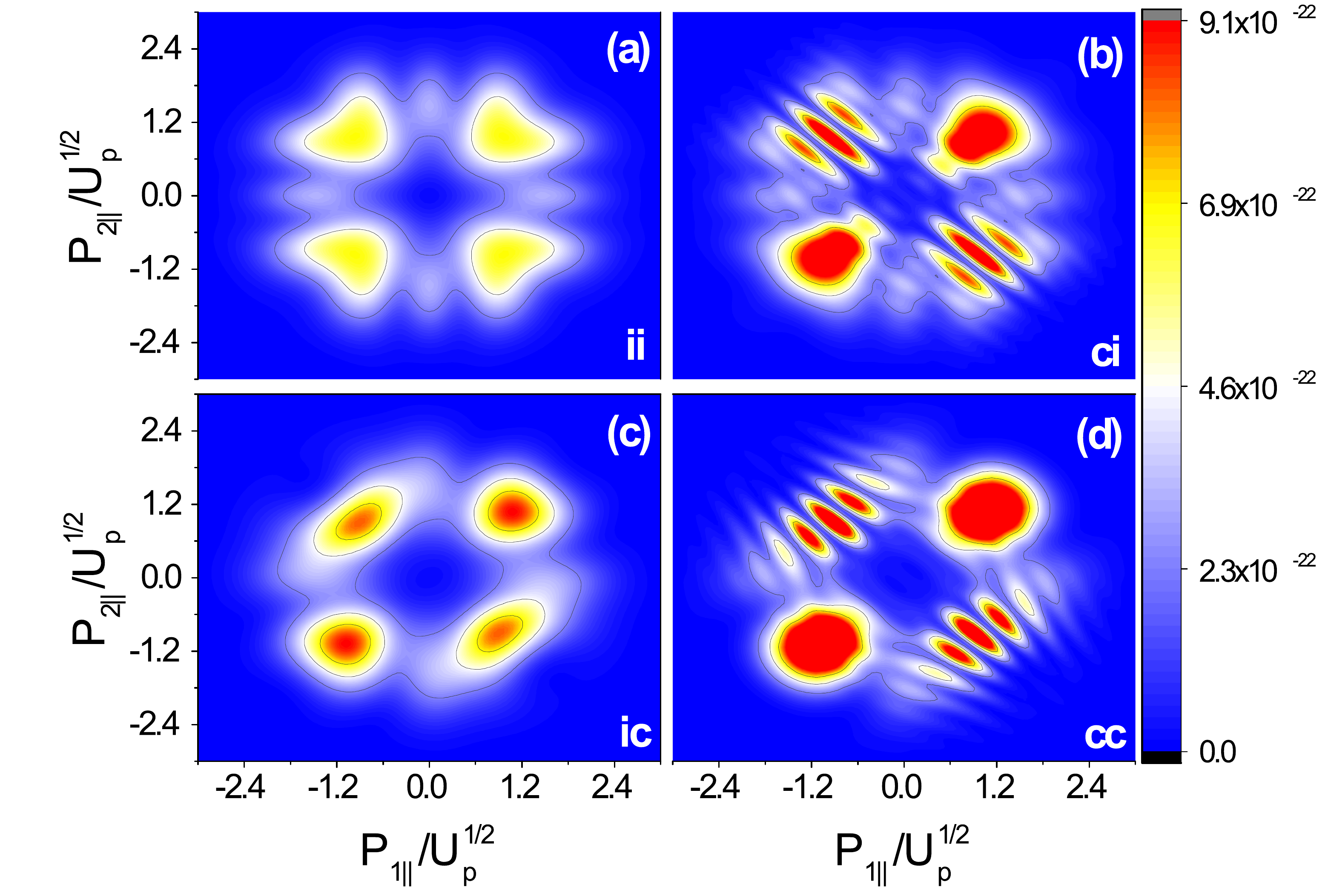}
	\caption{Momentum distribution of RESI for argon showing coherent and incoherent sums of pathways relating to different intermediate excited states and pathways related by symmetries. Whether the sum is coherent or incoherent (denoted c and i) is given in the bottom right for both the pathways relating to symmetries and excited states respectively.
	The ponderomotive energy is given by $U_{\mathrm{p}}=0.1$~a.u.\ ($I=4.56 \times 10^{13} \mathrm{~W/cm}^2$) corresponding to an angular frequency $\omega=0.057$~a.u. or wavelength $\lambda=800$~nm.
	Figure adapted from Ref.~\citealp{Maxwell2015}.
	}
	\label{Fig:FullChanSum}
\end{figure}

We get thus the final result for the RESI amplitude
\begin{align}
d(\mathbf{p},\mathbf{p}',t)
&=
\left(\frac{i}{\hbar}\right)^3
\sum_{\eta \ne 0}
\int_0^{t}dt'''\;
\int_0^{t'''}dt''\;
\int_0^{t''}dt'\;
\int d^3\mathbf{p}''\;
\exp\mathopen{}\left[
  \frac{i}{\hbar} S_d\left(\mathbf{p},\mathbf{p}',t''',t\right)
  \right]
\notag \\ & \quad \times
\left(
\mathbf{E}(t''')\cdot\tilde{\mathbf{g}}(\mathbf{p}'-e\mathbf{A}(t''')/c,\mathbf{p}-e\mathbf{A}(t''')/c,\eta)
\ 
\exp\mathopen{}\left[
  -\frac{i}{\hbar} S_c\left(\mathbf{p},t'',t'''\right)
  \right]
\notag \right. \\ & \quad \qquad \qquad \qquad \left. \times
\mathbf{E}(t'')\cdot\mathbf{g}(\mathbf{p}-e\mathbf{A}(t'')/c,\mathbf{p}''-e
\mathbf{A}(t'')/c,\eta)
\ 
\exp\mathopen{}\left[
  -\frac{i}{\hbar} S_b\left(\mathbf{p}'',t',t''\right)
  \right] 
\mathbf{d}(\mathbf{p}''-e\mathbf{A}(t')/c) 
a(t') 
\right. \notag \\ & \qquad \qquad \qquad 
+\left\{\mathbf{p}' \to \mathbf{p}\right\}\bigg)
\end{align}

In this expression, the three actions will be combined to give the SFA action for the RESI processes. The three integrals with the times $t'$, $t''$ and $t'''$ can be associated with ionization, recollision excitation, and final ionization, respectively. The integral over momentum can be related to the intermediate momentum. These integrals can be solved by the saddle point approximation, which makes this problem computationally tractable, as was done in \cite{Maxwell2015, Maxwell2016}. In that work, the probability distributions for monochromatic fields were calculated for the momentum components parallel to the laser field polarisation, where the components perpendicular to the laser field polarisation are integrated over, as shown in Fig.~\ref{Fig:FullChanSum}. This shows that different pathways for the RESI process will interfere, by plotting coherent and incoherent sums. These are pathways related to ionization via different excited states; for argon there are six pathways that contribute, as well as pathways related to those via symmetries such as the indistinguishability of electrons.

\subsubsection{Example for EII}
\begin{figure}[b]
	\includegraphics[width=0.45\textwidth]{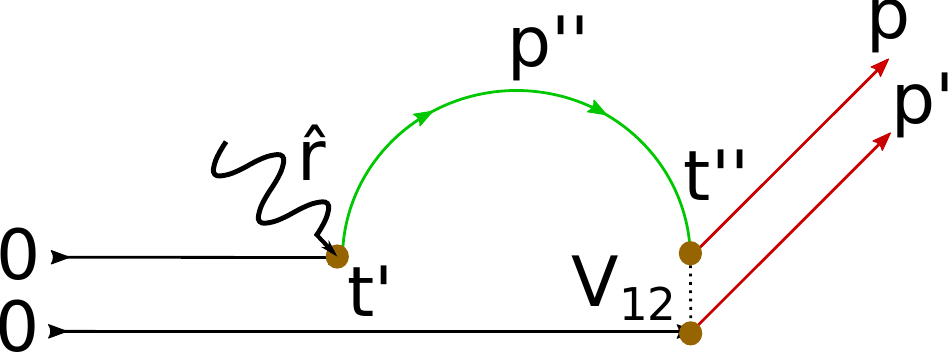}
	\caption{The complete EI process showing the two transitions. The two nodes marked at $t'$ and $t''$ show the two transitions identified in the chain in Eq.~\eqref{Eq:2e_EI_Chain}.}
	\label{fig:EII-chain}
\end{figure}
Using the same logic it is easy to do the same for the EII of NSDI. Considering similar restrictions as before only the integro-differential equation for  $d(\mathbf{p},\mathbf{p}',t)$ needs to be changed. This can easily be seen in the EII ``chain'' (see Fig.~\ref{fig:EII-chain})
\begin{equation}
\ket{0}
\underset{\mathbf{d}(\mathbf{p}'')}{\xrightarrow{\hspace*{6mm}}}
\ket{\mathbf{p}'',0}
\underset{\mathbf{g}(\mathbf{p}'',\mathbf{p},\mathbf{p}')}{\xrightarrow{\hspace*{12mm}}}
\ket{\mathbf{p},\mathbf{p}'}
,
\label{Eq:2e_EI_Chain}
\end{equation}
Here, we neglect the dipole matrix element for recollision excitation RESI contribution given by $\mathbf{g}(\mathbf{p},\mathbf{p}',\eta)$, and instead include in a similar way the matrix element $\mathbf{g}(\mathbf{p},\mathbf{p}',\mathbf{p}'')$. We can proceed as before, and
now the integro-differential equation for $d(\mathbf{p},\mathbf{p}',t)$ can be written out, this time it only depends on 	$b(\mathbf{p}'',t'')$ and is given by,
\begin{align}
	\dot{d}(\mathbf{p},\mathbf{p}',t)&=-\frac{i}{\hbar}\Bigg[
	\left(\frac{\hbar^2p^2}{2m}+\frac{\hbar^2p'^2}{2m}+E_0\right)d(\mathbf{p},\mathbf{p}',t)
	-\mathbf{E}(t)\cdot \int d^3 \mathbf{p}''\; \mathbf{g}(\mathbf{p},\mathbf{p}',\mathbf{p}'')
	b(\mathbf{p}'',t'')
	\notag\\&\hspace{3cm} 
	-2\mathbf{E}(t) \cdot (\mathbf{\nabla}_{\mathbf{p}} + \mathbf{\nabla}_{\mathbf{p}'}) d(\mathbf{p},\mathbf{p}',t)
	\Bigg].
\end{align}
Then the solution can be written as
\begin{align}
d(\mathbf{p},\mathbf{p}',t)
& =
\frac{i}{\hbar}\int_0^{t}dt''\;
\int_0^{t''}dt'\; 
\int d^3 \mathbf{p}''\; 
\exp\mathopen{}\left[
  \frac{i}{\hbar} S_d\left(\mathbf{p},\mathbf{p}',t'',t\right)
  \right]
\mathbf{E}(t'') \cdot \mathbf{g}(\mathbf{p}-e\mathbf{A}(t'')/c,\mathbf{p}'-e\mathbf{A}(t'')/c,\mathbf{p}'')
\notag \\ & \qquad \times 
\exp\mathopen{}\left[
  -\frac{i}{\hbar} S_b\left(\mathbf{p}'',t',t''\right)
  \right] 
\mathbf{E}(t') \cdot \mathbf{d}(\mathbf{p}''-e\mathbf{A}(t')/c)
\: a(t').
\end{align}
Ultimately, both the EII and RESI mechanism can be included in $d(\mathbf{p},\mathbf{p}',t)$ and it will still be integrable and, as expected, will simply be equal to the sum of these two solutions.
Note that EII is much more ``classical'' than RESI, i.e., that interference effects get washed out upon integration over momentum components perpendicular to the driving field polarization~\cite{Figueira2004, Figueira2004b}.

\section{Large molecules and targets in strong laser fields}
\label{sec:5-molecules}
In this section we formulate the general problem we want to attack with the help of SFA. The Hamiltonian describing a multi-atomic molecule or atomic cluster has the following general form:
\begin{equation}
\hat H= \hat H_\mathrm{nuc} + \hat H_\mathrm{el}
, 
\label{Htot}
\end{equation}
where the nuclear hamiltonian reads
 \begin{equation}
\hat H_\mathrm{nuc} 
= 
\sum_{i=1}^N \frac{\hat{\bf P}^2_i}{2M_i} 
+ V(\hat{\bf R}_1,\ldots,\hat{\bf R}_N),
\label{Hnuc}
\end{equation}
with the inter-nuclear potential
\begin{equation}
V(\hat{\bf R}_1, \ldots, \hat{\bf R}_N)= \frac{1}{2}\sum_{i\ne j}^N \frac{Z_iZ_je^2} {|\hat{\bf R}_i - \hat{\bf R}_j|},
\end{equation}
where $N$ is the number of nuclei involved, $i$ enumerates the nuclei, and ${\bf R}_i$, ${\bf P}_i$ are their positions and momenta, respectively, $Z_i$ are the nuclear charges, and $M_i$ the nuclear masses. 
In principle, we could include more complex nucleus-nucleus interactions, taking into account deeply-bound electrons via effective potentials and similar methods. 
We neglect here the influence of the laser electric field on the nucleus---they are simply too heavy to be affected by the short laser pulses.

The electronic Hamiltonian depends parametrically on the positions of the nuclei, via
\begin{equation}
\hat H_\mathrm{el}
= 
\sum_{i=1}^ M \frac{\hat{\bf p}^2_i}{2m} 
+ \frac{1}{2}\sum_{i\ne j}^M \frac{e^2} {|\hat{\bf r}_i - \hat{\bf r}_j|}
- \sum_{i, j}^{M,N} \frac{Z_je^2} {|\hat{\bf r}_i - {\bf R}_j|} - \sum_{i=1}^M e{\bf E}(t)\hat{\bf r}_i.
\end{equation}
Here $M$ is the number of electrons involved, $i$ enumerates them, and ${\bf r}_i$, ${\bf p}_i$ are their positions and momenta, respectively.
Again, we could replace bare Coulomb potentials by the dressed effective ones. 
Also, we assume that the target is large, but still smaller than the wavelength, so that a global dipole approximation holds.

\subsection*{Born-Oppenheimer Approximation}

In the following we assume that the nuclear motion is slower than that of the electrons, so we use the Born-Oppenheimer approach. To this is end, we first determine the electronic wave function, $\Psi_\mathrm{el}(\{{\bf r}_i\}_{i=1}^M, t; \{{\bf R}_i\}_{i=1}^N)$, in Dirac's notation  denoted as $|\Psi_\mathrm{el}(t)\rangle$, that fulfils TDSE with fixed nuclear positions
\begin{equation}
i\hbar\frac{d}{dt}|\Psi_\mathrm{el}(t)\rangle =   \hat H_\mathrm{el}|\Psi_\mathrm{el}(t)\rangle.
\end{equation}
We define then instantaneous electronic potential for the nuclei
$$ E_\mathrm{el}(\{{\bf R}_i\}_{i=1}^N, t) = \langle \Psi_\mathrm{el}(t)| \hat H_\mathrm{el}|\Psi_\mathrm{el}(t)\rangle,
$$
and treat the motion of the nuclei classically by solving the resulting Newton equations
\begin{eqnarray}
\frac{d{\bf R}_i}{dt} &=& \frac{{\bf P}_i}{M_i}, \\
\frac{d{\bf P}_i}{dt} &=& -{\bf\nabla}_{{\bf R}_i} V({\bf R}_1,\ldots,
{\bf R}_N) -{\bf\nabla}_{{\bf R}_i} E_\mathrm{el}({\bf R}_1,\ldots,
{\bf R}_N).
\end{eqnarray}
The solutions of these classical equations are then inserted into the electronic TDSE, and so on. In general, it has to be done self-consistently. We will discuss below a couple of cases when some simplifications are possible.

\subsection*{Single Active Electron  Approximation}
The theory formulated above can be reduced to a single-electron TDSE using the SAE approximation.  
In principle, it can be done in the same way as it is done in the static case for atoms or molecules. The only difference is that we now have to consider the fact that the effective potential must now be time-dependent through the parametric dependence on the nuclear coordinates,
\begin{equation}
\hat H_\mathrm{eff}=\frac{\hat{\bf p}^2}{2m} +  \hat V_\mathrm{eff}({\bf R}_1(t),\ldots,
{\bf R}_N(t), t) -  e{\bf E}(t) \cdot \hat{\bf r}.
\end{equation}
While calculations of $V_\mathrm{eff}$ for atoms belong still to the domain of the atomic physics, calculations of the effective potential for molecules, especially  in dynamical situations, clearly require the use of methods from molecular physics and quantum theoretical chemistry.

\subsection{Strong Field Approximation for quenched molecules}

The equations of the above section are very complex. There are some situations, however, when they can be radically simplified. 
One example of such a situation is the instant quench, in which the molecule is suddenly stripped of, say, one of the electrons, or photoexcited to a certain well-defined state. 
This can be achieved, for instance, applying an ultrashort attosecond XUV or soft X-ray pulse to the molecule. 
Right after the pulse, the molecule will find itself in the ground state corresponding to one missing electron, or in the well-defined excited state. 
In both situations, the nuclei configuration will be by no means stable. The molecule will start to vibrate, rotate, and maybe even dissociate.

If the excitation occurs to a weakly bound molecular state, the following vibrations or dissociation will occur on a rather slow time scale of $\SI{100}{fs}$ to $\SI{1}{ps}$. 
In that case, the HHG or ATI caused by an intense few-femtosecond pulse  may be used for an instant imaging of the dynamically changing molecular structure (for seminal experiments see Refs.~\citealp{WenLi2008,Worner2010}). 
If the electron removal or excitation occurs to a strongly-bound state, the resulting dynamics might be much faster: stripping of electrons, for instance, might lead to dissociation completely controlled by the Coulomb forces, and occurring on the timescales of an atomic unit (fractions of a femtosecond). 
These are the situations we want to consider in this section.

\subsection*{SFA and molecular dynamics}

If we then apply a short femtosecond laser pulse in the mid-infrared range, we may expect that, similarly to the standard HHG or ATI processes, the femtosecond laser induced electronic dynamics will not affect the intrinsic molecular dynamics. That means that, from the point of view of nuclei, we can replace the electronic Hamiltonian
\begin{equation}
\hat H_\mathrm{eff}
=
\frac{\hat{\bf p}^2}{2m} 
+ \hat V_\mathrm{eff}({\bf R}_1(t),\ldots,{\bf R}_N(t), t) 
- e{\bf E}(t)\hat{\bf r}
=
\hat H_{0}
- e{\bf E}(t) \cdot \hat{\bf r},
\end{equation}
with
\begin{equation}
\hat H_{0}
=
\frac{\hat{\bf p}^2}{2m} 
+  \hat V_\mathrm{eff}({\bf R}_1(t),\ldots, {\bf R}_N(t), t).
\end{equation}
The Born-Oppenheimer Newton equations for the nuclei can then be solved self-consistently, as
\begin{align}
\frac{d{\bf R}_i}{dt} 
&= 
\frac{{\bf P}_i}{M_i},
 \label{Newt1}\\
\frac{d{\bf P}_i}{dt} 
&= 
-{\bf\nabla}_{{\bf R}_i} V({\bf R}_1,\ldots,{\bf R}_N) 
-{\bf\nabla}_{{\bf R}_i} E_\mathrm{el}({\bf R}_1,\ldots, {\bf R}_N),
\label{Newt2}
\end{align}
where
$$ E_\mathrm{el}(\{{\bf R}_i\}_{i=1}^N, t) = \langle \Psi_\mathrm{el}(t)| \hat H_{0}(\{{\bf R}_i\}_{i=1}^N, t)|  \Psi_\mathrm{el}(t)\rangle.
$$
Assuming the that the ionization during the process is weak, the contribution of the continuum part of the electronic wavefunction will give negligible contribution to the electronic energy, so that
\begin{equation}
 E_\mathrm{el}(\{{\bf R}_i\}_{i=1}^N, t) \simeq \langle \Psi_{0}(t)| \hat H_{0}(\{{\bf R}_i\}_{i=1}^N, t)|  \Psi_{0}(t)\rangle |a(t)|^2,
\label{elecEn}
\end{equation}
where $|\Psi_{0}(t)\rangle$ is the time-dependent ground electronic state, and $a(t)$ is the probability amplitude of being in this state.
If we know $|a(t)|^2$ (for instance, if we can assume that $|a(t)|^2\simeq 1$), then the solutions can be simply introduced into Eq.~(\ref{elecEn}), and one can then calculate explicitly---i.e.\ without self-consistency conditions---both $|\Psi_0(t)\rangle$ and the corresponding continuum functions that fulfill
\begin{align}
E_0(\{{\bf R}_i\}_{i=1}^N, t) |\Psi_{0}(t)\rangle 
&=
\hat H_{0}(\{{\bf R}_i\}_{i=1}^N, t)|  \Psi_{0}(t)\rangle
\label{groundt}\\
\frac{1}{2m}{\bf p}^2 |{\bf p}(t)\rangle
&=
\hat H_{0}(\{{\bf R}_i\}_{i=1}^N, t)|{\bf p}(t)\rangle
.
\label{contit}
\end{align}
Both of these functions depend explicitly on time through the time dependence of the positions of the nuclei. Note, that the equations can be even more simplified if we can simplify the effects of $E_0$ in Eq. (\ref{Newt2})---the equations will not even require self-consistency!
For instance, in the case of stripping of, say, $K$ electrons, for large internuclei distance, the only effect of $E_0$ in Eq.~(\ref{Newt2}) will be to screen the nuclei charges, that is replace $Z_i$s by $\tilde Z_i$s, where $\sum_i Z_i= \sum_i \tilde Z_i - K$.

\subsection*{SFA for a quenched molecule}
The expression derived above implicitly assumes that  we proceed in fact as in Section~\ref{sec:3-background}. 
That is, we write the full electronic wave function as
\begin{equation}
|\Psi(t)\rangle
= 
e^{\textit{i}I_p(t)\textit{t}/\hbar + i\phi_B(t)}\bigg(a(t) |\Psi_0(t) \rangle + \: \int{\textit{d}^3 \textbf{p} \:  \textit{b}( \textbf{p},t) |\textbf{p}(t)\rangle} \bigg).
\label{PWavefq}
\end{equation}
where we set $I_p(t)= -E_0(t)$. The new effect here is $\phi_B(t)= \langle \Psi_0(t)|\partial_t \Psi_0(t)\rangle$---the Berry phase arising from projecting/expanding  the electronic wave function in  the time dependent basis. The equations still have practically the same form as before; for instance the direct transition amplitude fulfills:
\begin{equation}
{\partial }_tb_0( \textbf{p},t) =-\frac{i}{\hbar}\left(\frac{\textbf{p}^2}{2m}+ \tilde I_p(t) \right){b}_0( \textbf{p},t)  + \frac{\textit{i}}{\hbar}\: \textbf{E}(t) \cdot  \textbf{d}( \textbf{p},t)a(t) +\textbf{E}(t) \cdot \nabla_{\bf v}\textit{b}_0( \textbf{p},t),
\end{equation}
where we have included now the Berry phase in $\tilde I_p(t)=I_p(t) +\hbar \phi_B(t)$. Note that the Berry phase is nonzero if and only if the ground-state wavefunction is complex. 
This typically happens if the time-reversal symmetry is broken, i.e.\ for instance in the presence of a magnetic field or a so-called ``artificial'' gauge field. 
Also, the matrix element now depends explicitly on time, through the time dependence of the positions of the nuclei.

\subsection*{Generalized SFA expression for a quenched molecule}

The above equations for the electronic dynamics, together with the Newton equations (\ref{Newt1}) and (\ref{Newt2}), as well as the expressions for the  electronic energy (\ref{elecEn}-\ref{contit}), allow us to derive thus:
\begin{itemize}
\item The direct ATI amplitude:
\begin{equation}
\begin{split}
b_0( \textbf{p},t) 
= &
\frac{\textit{i}}{\hbar} \:
\int_0^t{\textit{d} \textit{t}^{\prime}\:
\textbf{E}(t^{\prime})} \cdot \textbf{d}\left( \textbf{p}-e\textbf{A}(t^{\prime})/c, t\right)a(t^{\prime})\\
& \qquad \times 
\exp\mathopen{}\left(
-\frac{\textit{i}}{\hbar} \:
\int_{t^{\prime}}^t d{\tilde t}
\left[
\frac{1}{2m}(\textbf{p}-e\textbf{A}({\tilde t})/c)^2 +\tilde I_p(t)
\right]
\right).
 \end{split}
 \label{Eq:b_0g}
\end{equation}
\item The re-scattering amplitude:
\begin{equation}
\begin{split}
b_1( \textbf{p},t) 
=&
 -\int_0^t{\textit{d}t^{\prime}
 \exp\mathopen{}\left[-\textit{i} S({\bf p},t,t')/\hbar\right]
 \,{\textbf{E}(t^{\prime})\cdot}} 
 \int_0^{t^\prime}{\textit{d} \textit{t}^{{\prime}{\prime}}}
 \int{\textit{d}^3\textbf{p}^{\prime} } \:
 \textbf{g}\left(\textbf{p}-e\textbf{A}(t^{\prime})/c,\textbf{p}^{\prime}-e\textbf{A}(t^{\prime})/c, t\right)
 \\  &\times 
 \textbf{E}(t^{{\prime}{\prime}}) \cdot \textbf{d}\left( \textbf{p}^{\prime} -e\textbf{A}(t^{{\prime}{\prime}})/c,t\right) \:
 a(t^{{\prime}{\prime}})\: 
 \exp\mathopen{}\left[-\textit{i}  S({\bf p}',t',t'')/\hbar \right],
  \end{split}
    \label{Eq:b_1g}
\end{equation}
where
\begin{equation}
{S}({\bf p},t,t^{\prime}) 
= 
\int_{t^{\prime}}^{t}d{\tilde t}
\left[
\frac{1}{2m}({\bf p}-e\textbf{A}({\tilde t})/c)^2 +\tilde I_p(\tilde t)
\right].
\end{equation}

\item The time-dependent dipole moment
\begin{align}
\langle {\bf x}(t)\rangle 
& =
\mathrm{Re}\mathopen{}\left[
\frac{i}{\hbar} \:\int_0^t{\textit{d} \textit{t}^{\prime} \int d^3{\bf p} a^*(t) \textbf{d}\left( \textbf{p}-e\textbf{A}(t)/c\right)\:\textbf{E}(t^{\prime})}\:\cdot \textbf{d}\left( \textbf{p}-e\textbf{A}(t^{\prime})/c\right)a(t^{\prime})
\nonumber \right. \\ & \qquad \qquad \left. \times 
\exp\mathopen{}\left(
  -\textit{i} \:\int_{t^{\prime}}^t d{\tilde t}
  \left[
  \frac{1}{2m}(\textbf{p}-e\textbf{A}({\tilde t})/c)^2 +I_p
  \right]/\hbar
  \right)
\right]
.
\end{align}

\end{itemize}
Note that if $|a(t)|^2$ is ``known'', then the solutions do not require self-consistency. 
Otherwise, they have to be obtained in the manner discussed below.
In the Appendix~\ref{app:F-toy-model} we discuss a toy model of 1D quenched $\rm H_2^+$ molecule, looking at qualitative and even semi-quantitative effects in HHG.

\subsection{SFA for large targets}
\label{subsec:5b-large-targets}
Here we consider another situation: the molecule (a large target) is initially in the ground state, and is impinged by an intense, short (few-cycle) laser pulse in the mid-infrared range.
This pulse causes the ionization of the single active electron, and induces thus structural dynamics of the target, i.e.\ the motion of the nuclei.
Amazingly, the expressions describing the quantities of interest are {\it exactly the same} as in the previous section. 
The way to obtain them, however, is much more complex: now we have to determine the evolution of ${\bf R}_i(t)$ and ${\bf P}_i(t)$ simultaneously and self-consistently with the dynamics of the electronic wave function, $|\Psi_\mathrm{el}(t)\rangle$.

The protocol to follow is thus:
\begin{enumerate}
\item Calculate the  electronic state (the ground state of $\hat H_\mathrm{eff}$),
\begin{equation}
E_\mathrm{el}(\{{\bf R}_i(0)\}_{i=1}^N, t) |\Psi_\mathrm{el}(0)\rangle 
=
\hat H_\mathrm{eff}(\{{\bf R}_i(0)\}_{i=1}^N, 0)|  \Psi_\mathrm{el}(0)\rangle
,
\end{equation}
for the initial positions of the nuclei $\pm \Delta {\bf R}_i$ (to be able to calculate gradients).
\item Propagate the equations for nuclei,
\begin{align}
\frac{d{\bf R}_i}{dt} 
&= 
\frac{{\bf P}_i}{M_i}, 
\label{Newt1g}\\
\frac{d{\bf P}_i}{dt} 
&=
-{\bf\nabla}_{{\bf R}_i} V({\bf R}_1,\ldots, {\bf R}_N) 
-{\bf\nabla}_{{\bf R}_i} E_\mathrm{el}({\bf R}_1,\ldots,
{\bf R}_N)
,
\label{Newt2g}
\end{align}
to the next time instant, $t$. Calculate the new  $\{{\bf R}_i(t)\}_{i=1}^N$.

\item Calculate the new  electronic state $|\Psi_\mathrm{el}(t)\rangle$. This is the  state propagated using $\hat H_\mathrm{eff}$,
\begin{equation}
i\hbar\frac{d}{dt}|\Psi_\mathrm{el}(t)\rangle 
= 
\hat H_\mathrm{el}(\{{\bf R}_i(t)\}_{i=1}^N, t) |\Psi_\mathrm{el}(t)\rangle
,
\label{timedepPsi}
\end{equation}
for the actual positions of the nuclei $\pm \Delta {\bf R}_i$ (to be able to calculate gradients). 
Note that this propagation should be done using the SFA ansatz (\ref{PWavefq})).
Calculate then
\begin{equation}
E_\mathrm{el}(\{{\bf R}_i(t)\}_{i=1}^N, t) =
\langle \Psi_\mathrm{el}(t) | \hat H_\mathrm{eff}(\{{\bf R}_i(t)\}_{i=1}^N, 0)| \Psi_\mathrm{el}(t)\rangle
\end{equation}
for the actual positions of the nuclei $\pm \Delta {\bf R}_i$ (to be able to calculate gradients).

\item Calculate
\begin{equation}
 E_\mathrm{el}(\{{\bf R}_i(t)\}_{i=1}^N, t) = \langle \Psi_\mathrm{el}(t)| \hat H_\mathrm{el}(\{{\bf R}_i(t)\}_{i=1}^N, 0)|  \Psi_\mathrm{el}(t)\rangle.
\end{equation}

\item Go to 2.
\end{enumerate}

Obviously, the above procedure is quite complex, but it does not present giant numerical challenges, and it is relatively straightforward to implement.
Evidently, it is much easier than solving the TDSE involving classical (Born-Oppenheimer) dynamics of the nuclei, the feasibility of which is not entirely obvious.

\subsection{SFA and quantum molecular dynamics}
\label{subsec:5c-molecular-dynamics}
The use of classical Newton equations for molecules in dissociation or vibration processes might be questionable.
There is a simple method of including certain aspects of the quantum motion of molecules that we describe now.
Our starting assumption is the generalization of the SFA ansatz to the full wave function:
\begin{equation}
|\Psi(t)\rangle= \xi_0({\bf R}_1, ..., {\bf R}_N) a(t) |\Psi_0\rangle + \xi_1({\bf R}_1, \ldots, {\bf R}_N) \int{\textit{d}^3 \textbf{p} \:  \textit{b}( \textbf{p},t) |\textbf{p}\rangle},
\label{PWavefq-full}
\end{equation}
where $\xi_0$, $\xi_1$ are the normalized wave functions of the nuclei for the
molecule with $M$, $M-1$ electrons correspondingly.

As before $\hat H_\mathrm{eff} = \hat H_0 -  e{\bf E}(t){\bf r}$, whereas
\begin{equation}
\hat H_0=
\frac{{\bf p}^2}{2m} +  V_\mathrm{eff}({\bf R}_1,\ldots,
{\bf R}_N(t)).
\end{equation}
The electronic ground state is now time-independent, but it does explicitly depend on the nuclear positions via
\begin{equation}
E_0({\bf R}_1, ..., {\bf R}_N)|\Psi_0\rangle = \hat H_0 |\Psi_0\rangle.
\end{equation}
Similarly, the states in the continuum do not depend on time, but on the nuclear positions, entering via $\hat H_0$ as
\begin{equation}
\frac{{\bf p}^2}{2m}|{\bf p}\rangle = \hat H_0 |{\bf p}\rangle.
\end{equation}

We still use the Born-Oppenheimer approximation, but in the quantum version. Also, we use different Hamiltonians for the non-ionized and ionized part of the molecular electronic dynamics. 
Thus, for $\xi_0({\bf R}_1, \ldots, {\bf R}_N)$ we use
\begin{equation}
\hat H_{\mathrm{nuc},0} 
=
\sum_{i=1}^N \frac{\hat{\bf P}^2_i}{2M_i} 
+ \hat V({\bf R}_1,\ldots,{\bf R}_N) 
+|a(t)|^2\langle \Psi_{0}(t)| \hat H_{0}|\Psi_{0}(t)\rangle.
\label{Hnuc0}
\end{equation}
while for $\xi_1({\bf R}_1, \ldots, {\bf R}_N)$ we use simply
\begin{equation}
\hat H_{\mathrm{nuc},1} 
= 
\sum_{i=1}^N \frac{\hat{\bf P}^2_i}{2M_i} 
+ \hat V({\bf R}_1,\ldots,{\bf R}_N).
\label{Hnuc1}
\end{equation}
We neglect here the laser part of the electronic energy, as well as the kinetic energy of electrons in the continuum. 
Note that the equation (\ref{Hnuc1}) can be solved without any self-consistency conditions. 
As in the previous sections, equation (\ref{Hnuc0}) can also be solved that way, provided that the time dependence of $|a(t)|^2$ is known.

The last point is the derivation of the SFA equation. 
To this end we assume that the quantum fluctuations of the nuclear positions are small, and replace the ${\bf R}_i$ dependence in $|\Psi_{0}(t)\rangle$ by the average $\bar{\bf R}_i(t) = \int d^3{\bf R} \: {\bf R}_i|\xi_0({\bf R},t)|^2$. 
Similarly, we replace  the ${\bf R}_i$ dependence in the continuum part by the average 
$\bar{\bf R}_i(t) = \int d^3{\bf R} \: {\bf R}_i|\xi_1({\bf R},t)|^2$;
after that trick, the SFA equations can be projected on the normalized functions $\xi_{0,1}(t)$.
This leads to the following modified equations:
\begin{equation}
{\partial }_tb_0( \textbf{p},t) 
=
-\frac{i}{\hbar}\left(\frac{\textbf{p}^2}{2m} + I_p(t) \right) b_0(\textbf{p},t)  
- e \textbf{E}(t) \cdot \nabla_{\bf p}\textit{b}_0( \textbf{p},t)
+ \frac{\textit{i}}{\hbar}\: \textbf{E}(t) \cdot  \textbf{d}( \textbf{p},t)\langle \xi_0(t)|\xi_1(t)\rangle a(t),
\end{equation}
and
\begin{equation}
{\partial }_t a(t)
=
\frac{\textit{i}}{\hbar}\: \langle \xi_1(t)|\xi_0(t)\rangle
\int d^3 \textbf{p} \textbf{E}(t) \cdot  \textbf{d}( \textbf{p},t)b_0( \textbf{p},t)
.
\end{equation}
As we see, the final equation depends only on the overlap $\langle \xi_1(t)|\xi_0(t)\rangle$, generally called the nuclear autocorrelation function, which, despite the fact that the positions of the nuclei in each branch of the process are quite ``classical'', might become very small as the positions of the nuclei in the two channels change. 
This can then seriously limit the HHG and LIED signals from the process, both for the direct and re-scattering parts~\cite{Lein2005, Baker2006, Patchkovskii2009}.

\section{SFA for solids}
\label{sec:6-solids}
The most important challenges of strong laser field physics concern using strong laser fields, be it the ones used in atto-science, or in free electron lasers (FELs), or in X-ray lasers, to characterize the statics and dynamics of the targets. The complexity of those targets grows with their ``size'': in a sense, atoms are the simplest, molecules are much more complex, and solid state and condensed matter systems are perhaps the most challenging, especially if we consider strongly-correlated systems.

The question of the optical response of solids has been a long-lasting subject of investigation, from the basics of nonlinear optics \cite{Boyd2003}, through studies of femtosecond X-ray sicence and the optical properties of semiconductors~\cite{Pfeifer2006, Haug2008}; it is directly related to the fundamentals of attosecond optics~\cite{Chang2011} and contemporary photoelectron spectroscopy~\cite{Hufner2013}. Before we focus on solids, let us summarize shortly the visions with atoms and molecules over the past decade.

\begin{itemize}[itemsep=-1mm, topsep=1.5mm]

\item {\bf Visions with atoms.} In a sense, the basis for all of these applications is Ref.~\citealp{Lewenstein1994}. However, to use this theory in practice, one has to include important physical and quantitative effects, such as Coulomb focusing~\cite{Brabec1996}.
In the last ten years or so, the primary focus of the work on atoms was on dynamics, in particular the dynamics of the tunnelling processes, including the birth time of HHG photons~\cite{Dudovich2006}, collective many-electron dynamics~\cite{Shiner2011}, explicit probing of the atomic wave function~\cite{Shafir2009}, or time resolved auto-ionization~\cite{Wang2010}; the Lund group proposed to study the birth time of attosecond pulses using two-color fields~\cite{He2010, Dahlstrom2011}.  Recent highlights of attacking atomic challenges with ultra-strong and ultrashort laser pulses include studies of the time when an electron exits a tunnelling barrier~\cite{Shafir2012b}, observations of electron propagation and screening on the atomic length scale~\cite{Neppl2015}, and attosecond tunnelling interferometry~\cite{Pedatzur2015}. For a very recent discussion of these areas, see the beautiful review on ultrafast holographic photoelectron imaging in Ref.~\citealp{Figueira2019}.

\item {\bf Visions with molecules.} These visions, obviously, come back to the seminal work of Itatani et al.~\citealp{Itatani2004}, which was focused on the reconstruction of the static molecular wave functions from measured HHG spectra. In recent years these visions moved very strongly forward. A few important examples (though by no means totally representative) concern interferometry of multi-electron dynamics in
molecules with HHG~\cite{Smirnova2009}, following time-resolved chemical reactions~ \cite{Worner2010}, performing generalized molecular orbital tomography~\cite{Vozzi2011}, or observing the bending of molecules in real time -- the Renner-Teller effect using laser-induced electron diffraction~\cite{Amini2019}.

\item {\bf  Visions with solids.} This is probably the most challenging area, especially if one thinks about strongly correlated systems or systems with topological order.

\end{itemize}

\begin{figure}[b]
  \includegraphics{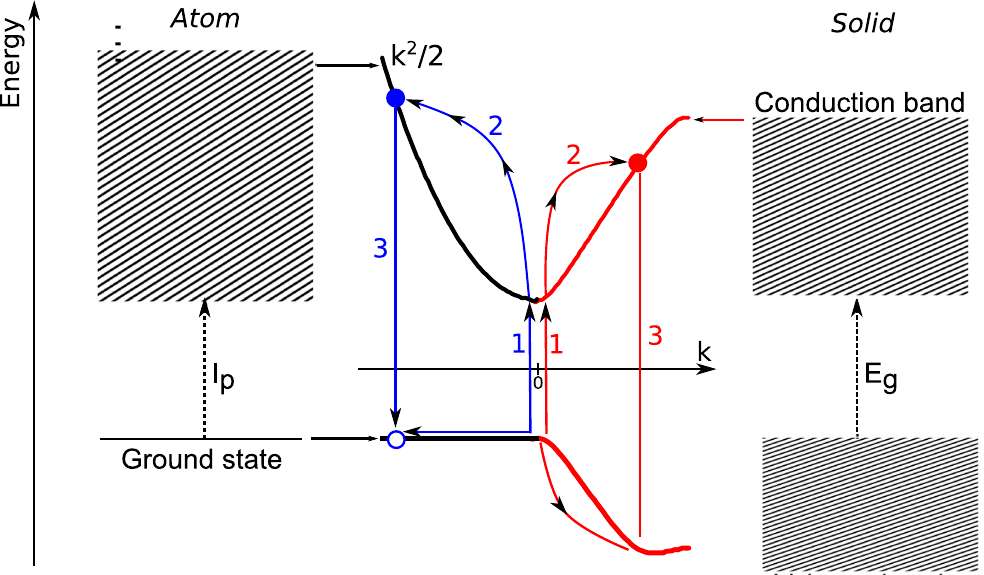}
  \caption{
    Sketch of the differences between the harmonic-emission processes in atoms and solids.
    In atoms (left side), the continuum band is parabolic, so the electron's motion in the continuum does not emit harmonics, and the hole it leaves behind remains stationary in a flat band.
    In a solid (right side), on the other hand, the hole can also move, and both holes and electrons experience dispersive forces in their continuum motion, which leads to the emission of intraband harmonics.
    However, when electrons and holes meet, they can also emit interband harmonics analogous to the gas-phase harmonics.
    Adapted from Ref.~\citealp{Vampa2017}.
  }
  \label{Fig:atom_vs_solid}
\end{figure}

The pioneering works on theory and experiments concerning  HHG in solids go back to the works of P.C.\ Becker et al.\ and, independently,  G.\ Petite and Ph.\ Martin in CEA Saclay~\cite{Becker1988, Martin2000, Pisani2001}, who developed and analyzed a ``happy electron model''. A more rigorous approach to SFA in solids was followed by F.\ Faisal's group few years later~\cite{Faisal2005a, Faisal2006}. Of course, the basis to analyze HHG or LIED in solids must involve precise and detailed static structure calculations~\cite{Goano2007}

The whole area started to grow exponentially roughly ten years ago. Important early theory developments were done by the S.\ Koch's group~\cite{Golde2008}. Around 2010, the ``volcano'' erupts: Goulielmakis and coworkers report real-time observation of valence electron motion \cite{Goulielmakis2010}, while S.\ Ghimire et al.\ publish their seminal paper on observation of high-order harmonic generation in a bulk crystal in 2011~\cite{Ghimire2011}. This is followed by a true explosion of interest in this area: HHG generation and propagation in crystals~\cite{Ghimire2012}, controlling dielectrics with light~\cite{Schultze2013}, attosecond band gap dynamics in silicon~\cite{Schultze2014}, electron propagation and dielectric screening on the atomic length scale~\cite{Neppl2015}, and efficient HHG in liquids~\cite{Heissler2014, Luu2018-liquids}.

While the disorder of the liquid phase makes spectroscopic studies of its structure more complicated, the rigid structure of solids allows for a much deeper understanding of the emission mechanisms as well as, potentially, broad and detailed high-harmonic spectroscopy~\cite{Luu2015}, real-time observation of interfering crystal electrons in HHG~\cite{Hohenleutner2015},  HHG in silicon \cite{Vampa2016}, and, more recently, interferometric measurements of the dipole phase in high harmonics from solids~\cite{Lu2019}. 
This is because, when an electron is released in a solid by a strong-field excitation from the valence to the conduction band, it explores a continuum which contains much more structure than the quadratic band of a free electron (as shown in Fig.~\ref{Fig:atom_vs_solid} and explained in depth in the review in Ref.~\citealp{Vampa2017}), and the dispersion induced by this structured continuum produces non-harmonic motion which leads to the emission of so-called ``intraband'' emission (cf.\ Ref.~\citealp{Schubert2014}).

Moreover, in atoms, the ionized electron leaves behind a stationary hole which is bound to the parent ion and cannot be displaced in space, but in solids this is no longer the case, and the motion of the hole in the valence band also needs to be considered. Nevertheless, when the hole and electron meet, they can recombine and emit so-called interband harmonics, exactly as in the atomic case (cf. Refs.~\citealp{McDonald2015, Wu2015}). However, despite that similarity, there are important differences, since that recombination can happen away from the origin, the electron and hole trajectories are subject to more complicated dynamics in their dispersive bands, and the bands themselves contain nontrivial parallel-transport effects that produce, through a nonzero geometrical phase, additional `anomalous' velocity terms that also contribute to the harmonic emission.

The theory of HHG in solids is based on the ``classical'' work of Blount~\cite{Blount1962}. To a good approximation, the electronic dynamics in a solid driven by a strong low-frequency laser pulse is governed by the semiconductor Bloch equations (SBE)~\cite{HaugKoch, Picon2019}, as derived for instance in Ref.~\citealp{Chacon2018}:
\begin{align}
\dot{n}_m({\bf K},t)
& =
\frac{i}{\hbar}\, s_m\:
e{\bf E}(t)\cdot{\bf d}_{cv}^*({\bf K} - e{\bf A}(t)/c)
\: {\pi}({\bf K},t)
+ \mathrm{c.c.},
\label{eqn:SBEs1}\\
\dot{\pi}({\bf K},t)
& =
-\frac{i}{\hbar}
\left[
  \varepsilon_g({\bf K} - e{\bf A}(t)/c)
  + e{\bf E}(t)\cdot{\bm \xi}_g({\bf K} - e{\bf A}(t)/c)
  - i \frac{\hbar}{T_2}
  \right]
{\pi}({\bf K},t)
\label{eqn:SBEs2}
\\ \nonumber & \quad \
-\frac{i}{\hbar} e{\bf E}(t)\cdot{\bf d}_{cv}({\bf K} - e{\bf A}(t)/c)
\, w({\bf K},t)
,
\end{align}
where $n_m$ is the population in band $m$, $\pi$ is the inter-band coherence (where we assume for simplicity a two-band model), $w=n_c-n_v$ is the population difference between the valence and conduction bands, $s_c=-s_v=1$, $\mathbf d_{mm'}(\mathbf k)$~is the inter-band dipole moment, and $\boldsymbol\xi_g(\mathbf k) = \boldsymbol\xi_c(\mathbf k) - \boldsymbol\xi_v(\mathbf k)$ is the difference in the Berry connections of the two bands~\cite[supplemental material]{Chacon2018}.
These variables give rise to the harmonic emission via the intra and inter-band components of the total current,
\begin{align}
{\bf J}_\mathrm{ra}(t)
&=
e\sum_m \int_{\rm \overline{BZ}} d^3 {\bf K}  \:
{\bf v}_m\left({\bf K} - e{\bf A}(t)/c\right)
n_m({\bf K},t),
\label{eqn:intra}\\
{\bf J}_\mathrm{er}(t)
& =
e{\frac{d}{dt}} \int_{\rm  \overline{BZ}} d^3 {\bf K} \:
{\bf d}_{cv}^{*}\left({\bf K} - e{\bf A}(t)/c\right)
\pi({\bf K},t)
+ \mathrm{c.c.}
\label{eqn:inter}
\end{align}

Let us discuss with a little more detail the theoretical description of HHG in solids, and its similarities to the SFA for atoms.  The basis for the SFA is the Keldysh theory of tunnelling ionization~\cite{Keldysh1965}. Keldysh considered the ionization of atoms in low frequency fields, and thus could use a quasi-static approximation to describe the tunnelling processes. It is worth noting that in the orginal paper from 1965 Keldysh already could generalise his theory to the solid state, and estimate tunneling ionization in solids. As detailed in Section~\ref{sec:1-introduction}, Keldysh theory was generalized, on one hand, by Ammosov, Delone and Krainov~\cite{Ammosov1986} to describe ionization rates from complex atoms, and, on the other hand, by Faisal and Reiss to describe ATI~\cite{Faisal1990, Reiss1980}; it was also a stimulus for the present formulation of the SFA~\cite{Lewenstein1994}.

The optical response of solids (insulators and semiconductors, and to some extent even metals) to strong laser fields is described by the semiconductor Bloch equations~(\ref{eqn:SBEs1},\,\ref{eqn:SBEs2}). In general, these equations  are too complex to solve directly other than numerically, but, as in the atomic case, in some situations they allow us to calculate an approximate expression for the electronic current.

In the atomic case, approximate analytic solutions of the TDSE are possible in the SFA, where the dynamics is reduced to the ground state and continuum states (with bound states other than the ground state getting neglected), and in the simplest approach Coulomb interactions are ignored. For solids, on the other hand, the situation is different from the very beginning, since we start with a many-body system of electrons in the valence band of the crystal. In some cases, it is possible to use effective-electron descriptions of the system (and even then we deal with effectively non-interacting quasi-particles) and we have to include all of the particles that fill the valence band. This is analogous to considering many filled bound states in an atom. Perhaps the best analogy to the atomic case is achieved by considering the solid-state dynamics using the Wannier description in the valence bond, coupled to Bloch waves in the conduction band~\cite{Osika2017}.

If the effective Fermi-liquid non-interacting quasi-electron description is a good approximation, then the response of the solid to a strong laser pulse is well-described by a theory similar to the Keldysh framework and to the SFA.  On the other hand, if we deal with strongly-correlated systems that cannot be described by the effective mean-field-like description, then one has to rely on much more rigid theoretical tools, including the effects of Coulomb interactions. This is a challenging area, with only a few \textit{ab initio} numerical studies~\cite{Silva2018, Takayoshi2019}, though some additional headway can be made using DFT~\cite{Bauer2018, Drueke2019}.

Within the effective single-active-electron SFA theory, however, there is substantially more that one can say analytically, at least on an approximate footing. Here, the current can be expressed~\cite{Chacon2018} in the form
\begin{align}
J^{(i)}_\mathrm{er}(t)
& =
-\mathrm{i}\sum_{j}{\frac{d}{dt}}
\int_{t_0}^{t} dt' \int_{\rm  \overline{BZ}} d^3 {\bf K}\,
|d^{(i)}_{cv}\left({\bf K} - e{\bf A}(t)/c\right)| \
|d^{(j)}_{cv}\left({\bf K} - e{\bf A}(t')/c\right)|\, E^{(j)}(t')
\label{eqn:inter2}
\\ &  \nonumber \qquad \qquad \times
e^{
  -\mathrm{i}S({\bf K},t,t')/\hbar
  -(t-t')/T_2
  + i\left(\phi_{cv}^{(j)}({\bf K},t)-\phi_{cv}^{(i)}({\bf K},t)\right)
  }
+\mathrm{c.c.},
\end{align}
where $S({\bf K},t,t')$ is the so called quasi-classical action for the electron-hole and is defined according to:
\begin{equation}
S({\bf K},t,t')
=
\int_{t'}^{t}\left[\varepsilon_g({\bf K} - e{\bf A}(t'')/c)
+ e{\bf E}(t'')\cdot{\bm \xi}_g({\bf K} - e{\bf A}(t'')/c)
- \hbar\frac{d}{dt''}\phi_{cv}^{(j)}({\bf K} - e{\bf A}(t'')/c) \right]dt'' . \label{eqn:Action1}
\end{equation}
This forms the heart of the SFA description of high-harmonic emission in a crystalline solid, and it contains all of the semiclassical dynamics for the electron trajectories in the conduction band as well as the hole's trajectory in the valence band~\cite{Vampa2014, Vampa2015, Vampa2015a, Vampa2017}. More recent theoretical developments concern the effects of multiple conduction bands~\cite{Hawkins2015}.

Traditional treatments of HHG in solids have worked in one-dimensional configurations where the geometrical-phase element of \eqref{eqn:Action1} can be set to zero using an appropriate choice of gauge for the Bloch-function basis, in which case the kinematics of the electron and hole wavepackets are fully determined by the band structure~\cite{Vampa2015, Hawkins2013, Higuchi2014}. However, there are materials where this gauge transformation is not possible, due to the existence of a Berry curvature on one or more of the bands, and the geometric-phase terms in \eqref{eqn:Action1} cannot be neglected. This Berry curvature is crucial for a wide array of solid-state effects~\cite{vonKlitzing1986, Haldane2017, Xiao2010}, and it is the driving ingredient of many nontrivial phenomena.
Its presence in the SFA harmonic-emission current~\eqref{eqn:inter2} means that it can in principle be measured via HHG observables, and indeed recent experimental~\cite{Liu2017, Luu2018} and theoretical~\cite{Chacon2018, Silva2018a} works show that this is the case; we showcase in Fig.~\ref{fig:alexis-paper} some recent results on that front. A review on high-harmonic generation from solids by Ghimire and Reis was published very recently in Ref.~\citealp{Ghimire2019}.

\newlength{\haldaneFigureHeight}
\setlength{\haldaneFigureHeight}{45mm}
\begin{figure}[t!]
  \begin{tabular}{ccc}
  \includegraphics[height=\haldaneFigureHeight]{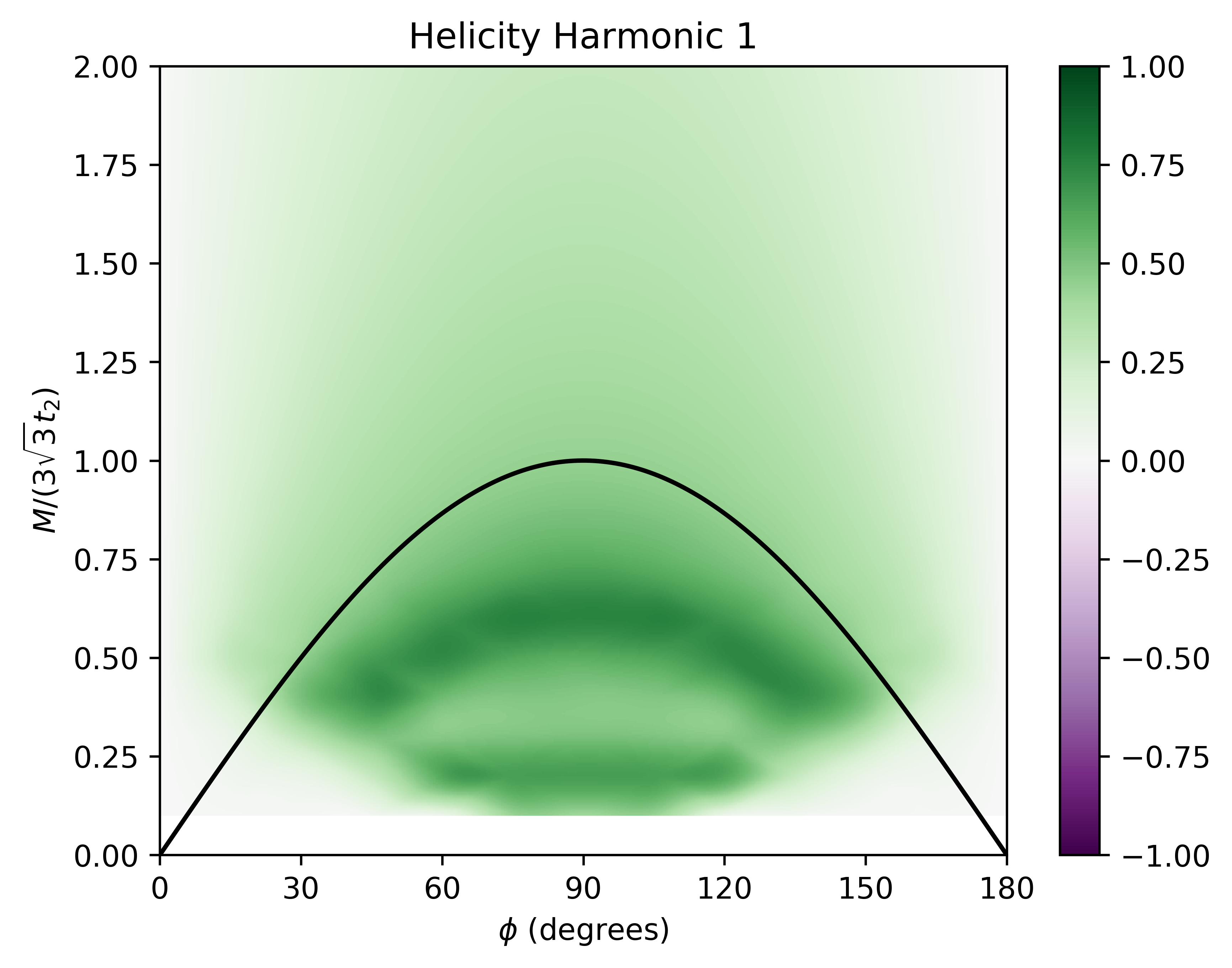} &
  \includegraphics[height=\haldaneFigureHeight]{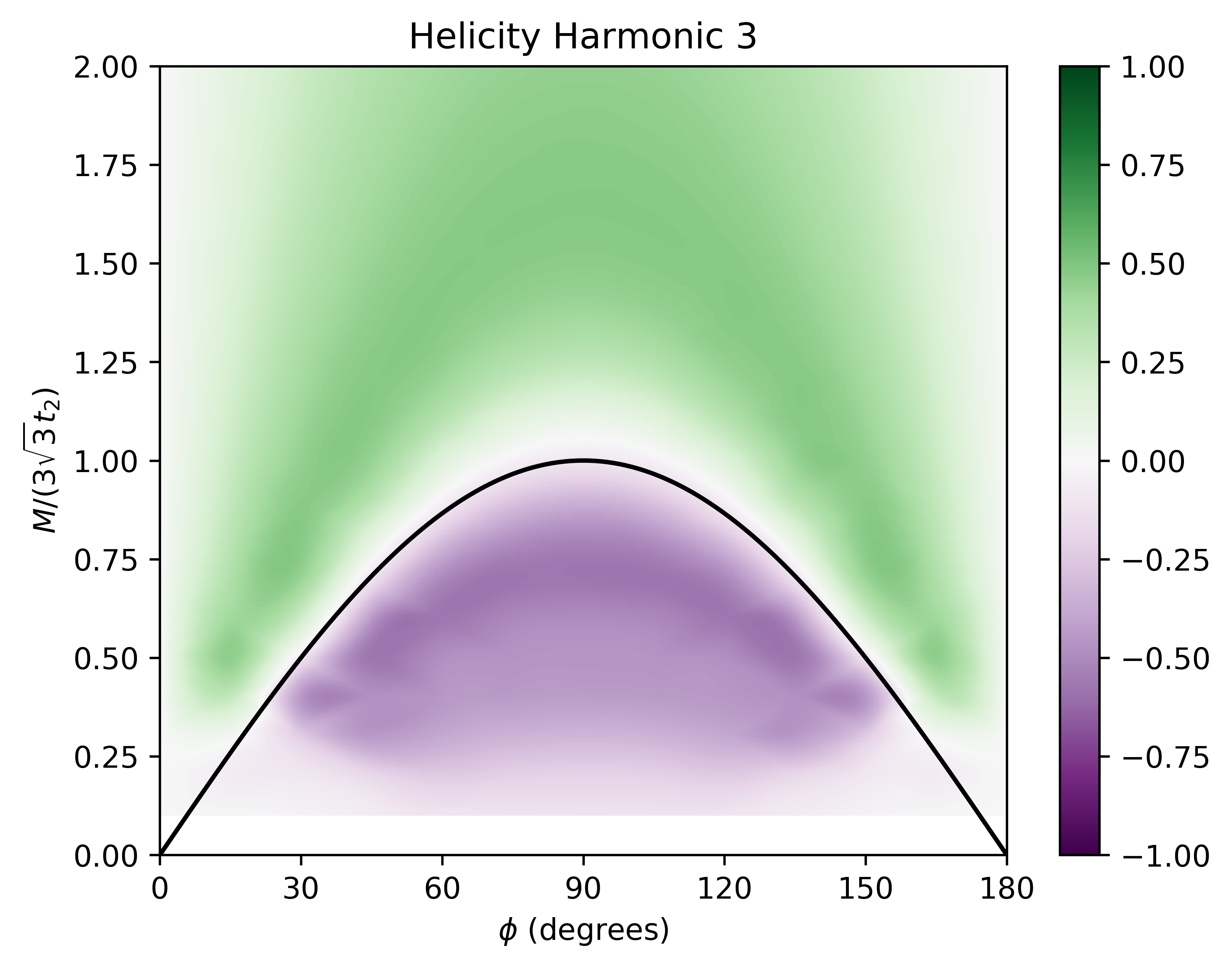} &
  \includegraphics[height=\haldaneFigureHeight]{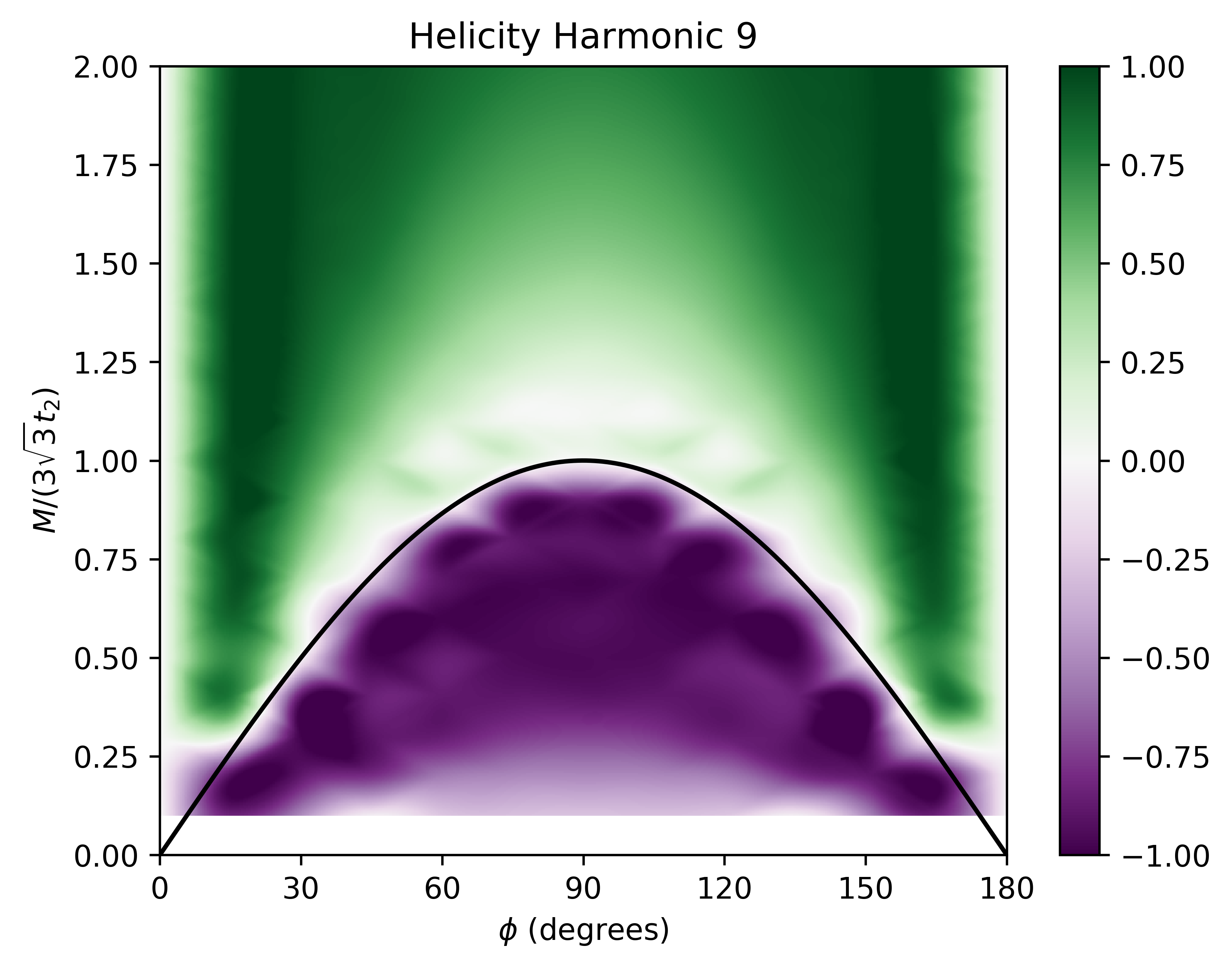} 
  \end{tabular}
  \caption{
  Traces of a topological phase transition in HHG from a two-dimensional material. When the Haldane model is driven with a linearly-polarized pulse, the optical response at the fundamental and its harmonics is elliptical, and the helicity (shown as the color scale) has a different relationship between the harmonics and the fundamental depending on whether the material is in the topologically trivial or non-trivial phase; this showcases the potential of high-harmonic spectroscopy for topological-phase discrimination.
  Figure adapted from \cite{Silva2018a}.
  }
\label{fig:alexis-paper}
\end{figure}

\acknowledgments{
We thank Sougato Bose,  Paul Corkum, Anne L'Huillier, Yann Mairesse and Pascal Sali\`eres for enlightening discussions. 
This work was supported by  National Science Centre Poland-Symfonia Grant No.\ 2016/20/W/ST4/00314.  
We acknowledge also the project Advanced Research Using High Intensity Laser Produced Photons and Particles (No.\ CZ.\ 02.1.01/0.0/0.0/16\texttt{\_}019/0000789) from the European Regional Development Fund (ADONIS),
Ministerio de Econom\'{\i}a y Competitividad through Plan Nacional 
(Grant No. FIS2016-79508-P FISICATEAMO, and 
Severo Ochoa Excellence Grant No.\ SEV-2015-0522), 
funding from the European Union Horizon 2020 Research and Innovation Programme under the Marie Sklodowska-Curie Grant Agreement No. 641272
Laserlab-Europe (Grant No.\ EU-H2020 654148),
Fundaci\'o Privada Cellex, and 
Generalitat de Catalunya (Grant No.\ SGR 1341 and CERCA Programme). 
N.S. was supported by the Erasmus Mundus Doctorate Program Europhotonics (Grant No.\ 159224-1-2009-1-FR-ERAMUNDUS-EMJD). 
N.S., A.C., A.D., E.P., and M.L.\ acknowledge ERC AdG OSYRIS and NOQIA, and EU FETPRO QUIC. 
K.A.\ and J.B.\ additionally acknowledge ERC AdG Transformer (788218), Ministerio de Econom\'{\i}a y Competitividad through Plan Nacional FIS2017-89536-P and Agencia de Gesti{\'o} d'Ajuts Universitaris i de Recerca (AGAUR) with 2017 SGR 1639.
E.P.\ acknowledges Cellex-ICFO-MPQ Fellowship funding.
A.D.\ is financed by a ``Juan de la Cierva'' fellowship (IJCI-2017-33180).
A.P.\ acknowledges funding from Comunidad de Madrid through TALENTO grant ref.\ 2017-T1/IND-5432.
J.A.P.-H.\ acknowledges Spanish Ministerio de Econom\'ia, Industria y Competitividad (MINECO) through the PALMA Grant No.\ FIS2016-81056-R, from LaserLab Europe IV Grant No.\ 654148, and from the Regional Government, Junta de Castilla y Le\'on, Grant No.\ CLP087U16 and the Consolidated Research Unit 167, recognized by the Junta de Castilla y Le\'on.
C.F.M.F.\ and A.S.M.\ acknowledge funding by the UK Engineering and Physical Sciences Research Council (EPSRC) under grants EP/J019143/1 and EP/P510270/1, the latter within the remit of the InQuBATE Skills Hub for Quantum Systems Engineering.
A.Z.\ acknowledges funding from UK EPSRC under the grant EP/J002348/1.
}

\appendix

\section{Time dependent ADK rates}
\label{app:A-TD-ADK}

This is the first section in which we remind the reader about a possible way to evaluate/estimate the amplitude of the ground state, $a(t)$, which is the essential ingredient of our formulation of the SFA. The ADK rates apply in principle for arbitrary atoms, but the theory can be extended to molecules~\cite{Tong2002}.
This approach can be used for pulses which are not too short: the rate has to have time to ``define itself''. The static expressions for the ADK rates are
 \begin{equation}
W_\mathrm{ADK}= |C_{n^*l^*}|^2\sqrt{\frac{6}{\pi}}f_{lm}I_p(2(2I_p)^{3/2}/F)^{2n^*-|m|-3/2}\exp(-2(2I_p)^3/2/3F),
\end{equation}
where $F$ is the peak value of the laser electric field, $n^*=Z/\sqrt{I_p}$, and $Z$ being the charge of the atomic/ionic core.
The other symbols are
\begin{equation}
|C_{n^*l^*}|^2=\frac{2^{2n^*}}{n^*\Gamma(n^*+l^* +1)\Gamma(n^*-l^*)},
\end{equation}
with $\Gamma(\cdot)$ denoting the Gamma function, $l^*=n^*-1$, and
\begin{equation}
f_{lm}=\frac{(2l+1)(l+|m|)!}{2^{|m|}|m|!(l-|m|)!},
\end{equation}
with $l,\, m$ denoting electrons initial orbital and magnetic quantum numbers.

For shorter pulses, we obtain $W_\mathrm{ADK}(t)$ replacing $F\to F(t)= \mathcal{E}_0\:f(t)$ (the rates change adiabatically with the pulse envelope). If the laser frequency is even smaller, and the pulse shortened we can even replace $F\to F(t)= |\mathbf{E}(t)|$, i.e.~the actual value of the electric field.

\section{Ground state amplitude according to SFA}
\label{app:B-ground-state}

 Inserting the expression \eqref{Eq:IntDirecTerm} in the equation for $a(t)$, we obtain
\begin{equation}
\frac{da(t)}{dt}=\int_0^t dt'\gamma(t,t') a(t'),
\label{gamma}
\end{equation}
 where
\begin{equation}
\begin{split}
\gamma(t,t') 
= & 
\int d^3 
{\bf p}
\textbf{E}(t)\cdot \textbf{d}\left( \textbf{p}-e\textbf{A}(t)/c\right) 
\: \textbf{E}(t^{\prime})\cdot \textbf{d}\left( \textbf{p}-e\textbf{A}(t^{\prime})/c\right)
\\
&\times 
\exp\left(
-\textit{i} 
\int_{t'}^{t}dt''[(\textbf{p}-e\textbf{A}(t'')/c)^2/2 +I_p]/\hbar
\right)
. 
\label{Eq:IntDirecTermBB}
 \end{split}
\end{equation}
In principle solving equation \eqref{gamma} in the static, quenched or even self-consistent case presents no basic difficulties. If the pulse is longer, and the rate of change of $a(t)$ slow, we may replace $a(t') \to a(t)$ and obtain explicitly
\begin{equation}
a(t) = \exp{(-W_\mathrm{SFA}(t))}a(0),
\end{equation}
 where the SFA rate is
\begin{equation}
W_\mathrm{SFA}= \int_0^t \gamma(t,t') \, dt'.
\end{equation}

\section{Model atom and molecule} %$$$$$$$$$$$$$$$$$$$$$$
\label{app:C-atom-molecule-model}

The paradigm examples of separable potentials used in atomic physics are zero-range potentials. 
In 1D, a Dirac delta potential can be used for this purpose. 
In 3D, the Dirac delta must be regularized---that is why the celebrated pseudo-potential must be used. 
It has found multiple applications in the many-body theory of ultracold atomic gases (cf.\ Refs.~\citealp{StringariPitaevski, LewensteinQuantumSimulators});  in strong-field physics it was elaborated by W. Becker and his collaborators \cite{Becker1994a}. 
In 2D the situation is more complex due to the logarithmic divergences (cf.\ Ref.~\citealp{Busch1998} and references therein).

One should stress, however, that the use of separable potentials has a long history in strong-field physics~\cite{Faisal1987, Faisal1987b, Faisal1989, Faisal1990, Faisal1991, Faisal1993, Faisal1993b, Radozycki1993, Kaminski1989, Klaiber2005, TetchouNganso2007, TetchouNganso2011, TetchouNganso2013, Galstyan2017, Galstyan2018}. 
Many of these papers deal with zero-range models, some with general separable potentials, but typically using laser fields of constant strength and circular polarization. 
For such a case, one can transform the problem to a rotating frame in which the Hamiltonian is time-independent. 
The recent papers by Galstyan et al.~\cite{Galstyan2017, Galstyan2018} are perhaps the closest to the approach developed by us in Refs.~\citealp{Noslen1, Noslen2, Noslen3, Noslen4, Noslen-diss}.

In this section  we discuss a useful  non-local separable  potential  with the purpose of applications for atoms, but most importantly for  large molecules. The idea will be to compute both the direct and the re-scattering  transition amplitudes~\cite{Lewenstein1995}. These terms involve the dipole and the continuum-continuum matrix elements  defined by Eqs.~(\ref{Eq:dv}) and (\ref{Eq:CCM1}). 
Then, our main task will be devoted to find analytically the wavefunctions for the ground  and scattering  states  of our model potential. The Hamiltonian, $\hat{H}(\textbf{p},\textbf{p}^{\prime}) $,  of the atomic system in the momentum representation can be written as:
 \begin{equation}
\hat{H}(\textbf{p},\textbf{p}^{\prime}) = \frac{{\bf p}^2}{2m}\delta(\textbf{p}-\textbf{p}^{\prime}) + \hat{V}(\textbf{p},\textbf{p}^{\prime}),
\end{equation}
where the first term  on the right-hand side is the kinetic energy operator,  and the second one is the  non-local potential $\hat{V}(\bf{p},{\bf p}')$. We use non-local {\it separable} potentials that can be understood as sums of projectors on certain states. They generally have the form
\begin{equation}
\hat{V}({\bf p},{\bf p}')= -\gamma\sum_{i=1}^M \phi_i(\bf{p})\phi_i^*({\bf p}'),
\end{equation}
where $\gamma$ is a coupling constant.

When we model molecules, each of the orbitals is typically centered in real space at the positions of the nuclei, ${\bf R}_i$. In the momentum representation it translates to $  \phi_i(\bf{p}) \exp(i\bf{p}\cdot {\bf R}_i\tilde \phi_i({\bf p})$, where $\tilde \phi_i({\bf p})$ is a ``smooth'' function, with the Fourier transform centered at ${\bf R}=0$. The above potential has generically $M$ bound states, so one can model with its help not only the ground state of the molecule in question, but even some of its excited states. Alternatively it may be used to model multielectron molecules.

\subsection*{Ground state}
In the present Appendix we will consider $M=1$ only, so the models contain a single bound state, but we will explore the full richness of the orbital $\phi({\bf p})$ to model  multicentered ground states of, in principle, arbitrary molecules. By using the non-local separable Hamiltonian, we write the stationary  Schr\"odinger equation as follows:
\begin{equation}
E\: \Psi(\textbf{p})=\frac{p^2}{2m}  \Psi(\textbf{p})  -\gamma \phi(\textbf{p})\: \int{\textit{d}^3 \textbf{p}^{\prime}\phi^*(\textbf{p}^{\prime})\Psi(\textbf{p}^{\prime})},
\end{equation}
where $E$ denotes the energy of the wavefunction $\Psi({\bf p})$.
Note that we have defined the non-local potential as
$\hat{V}(\textbf{p},\textbf{p}^{\prime})= -\gamma \phi(\textbf{p})\: \phi(\textbf{p}^{\prime})$, 
which describes the attraction between the electron  and  the nucleus~\cite{Lewenstein1995}. This  potential  has been chosen such that it assures analytical  solutions of the continuum  or scattering  states,  i.e.~for states  with energies $E>0$. 
Note that the  ground  state  can  also be  calculated  analytically.
The parameter $\gamma $ is a constant that, as we will see, determines  the energy of the ground state.  The shape of the ground state, however, can be controlled to a high degree by the choice of a suitable auxiliary function $\phi({\bf p})$, which may correspond to a multicenter molecular orbital of arbitrary shape.

For the ground  state, $\Psi_0({\bf p})$, we solve the stationary Schr\"odinger equation in the momentum representation:
\begin{equation}
\frac{p^2}{2m}\: \Psi_0(\textbf{p})  -\gamma\phi({\bf p}) \int \textit{d}^3 \textbf{p}^{\prime} \phi^*({\bf p}^{\prime}) \Psi_0(\textbf{p}^{\prime})  = -I_p\: \Psi_0(\textbf{p}),
\label{Eq:Sch2}
\end{equation}
From Eq.~(\ref{Eq:Sch2}) we easily determine the ground state:
  \begin{equation}
 \Psi_0(\textbf{p}) = \frac{\mathcal{N}\phi({\bf p})}{(\frac{p^2}{2m} +I_p)}
 \label{Eq:WF1}
\end{equation}
where, $\mathcal{N}$ denotes a normalization constant. Multiplying the last formula by $\phi({\bf p})^*$, and taking the volume integral on $\textbf{p}$, we obtain an equation that determines the ground state energy,
\begin{equation}
1 = \gamma   \int{\frac{\textit{d}^3 \textbf{p}|\phi({\bf p})|^2}{ (\frac{p^2}{2m} + I_p)} }. 
\label{Eq:Ncond}
\end{equation}
The solution of the last integral in Eq.~(\ref{Eq:Ncond}) gives us the relation between the parameters $I_p$, and $\gamma$ given $\phi({\bf p})$. The normalization constant fulfils
\begin{equation}
1 =  \mathcal{N}^2 \int{\frac{\textit{d}^3 \textbf{p}|\phi({\bf p})|^2}{ (\frac{p^2}{2m} + I_p)^2} }. 
\label{Eq:Ncond2}
\end{equation}

\subsection*{Scattering waves}
Let us consider the scattering wave, $\Psi_{{\bf p}_0}({\bf p})$, with asymptotic momentum ${\bf p}_0$, as a coherent superposition of a plane wave and an extra correction $\delta\Psi_{{\bf p}_0}({\bf p})$:
\begin{align}
\Psi_{\textbf{p}_0}(\textbf{p})  = \delta(\textbf{p}-\textbf{p}_0) +  \delta\Psi_{{\bf p}_0}(\textbf{p}).
\end{align}

This state has an energy $E={{\bf p}_0^2}/2m$. Then, the Schr\"odinger equation in momentum representation reads:
\begin{align}
\frac{p_0^2}{2m}\:\Psi_{\textbf{p}_0}(\textbf{p}) 
&=
\frac{{\ p}^2}{2m}\: \Psi_{\textbf{p}_0}(\textbf{p})  -\gamma \phi({\bf p}) \int \textit{d}^3 \textbf{p}^{\prime} \: \phi^*({\bf p}^{\prime})\Psi_{\textbf{p}0}(\textbf{p}) , 
\nonumber\\
 \bigg(\frac{{ p}^2}{2m} - \frac{p_0^2}{2m} \bigg)\ \delta\Psi_{{\bf p}_0}(\textbf{p}) 
  &= 
  \gamma \phi({\bf p})\phi^*({\bf p}_0)+ \gamma \phi({\bf p}) \int \textit{d}^3 \textbf{p}^{\prime} \: \phi^*({\bf p}^{\prime})\delta\Psi_{{\bf p}_0}(\textbf{p}^{\prime}).
\end{align}
To solve  analytically  the last equation, we apply elementary algebra and the following Dirac delta distribution properties:
$(\frac{p^2}{2m} -
\frac{p_0^2}{2m})\:\delta(\textbf{p}-\textbf{p}_0) = 0
$, and 
$(\frac{p^2}{2m} -
\frac{p_0^2}{2m})\:\delta(\frac{p^2}{2} -
\frac{p_0^2}{2m}\pm i\epsilon) = 0$.
  Finally, the
correction $\delta \Psi_{{\bf p}_0}$
reads:
\begin{align}
\delta\Psi_{{\bf p}_0}(\textbf{p})  
= 
\frac{
  2m\gamma \phi({\bf p})\phi^*({\bf p}_0)
  }{
  (1-A(\textit{i} p_0+\epsilon))
  (p^2-(p_0 -\textit{i}\epsilon)^2 ) 
  }
.
\label{Eq:SCorrection0}
\end{align}
Here, $\epsilon$, is the (positive) regularization  parameter to avoid the divergence at ${\bf p}={\bf p}_0$, and $A(ip_0+\epsilon)$ is the function
$
A(k) = 2m\gamma \int \frac{|\phi({\bf p})|^2}{p^2 + k^2} \textit{d}^3 \textbf{p}^{\prime}
$
evaluated at $k=ip_0+\epsilon$. The sign in front of $\epsilon$ defines the asymptotic behaviour of the scattering solutions. For the present choice, the scattered part of the wavefunction in the position representation behaves asymptotically as $\exp(-\textit{i}p_0 r)$, i.e.\ as a incoming spherical wave, which is the correct behaviour for the scattering wave functions, describing the states with asymptotic outgoing momentum ${\bf p}_0$. The final expression for the scattering wave functions is:
\begin{align}
\Psi_{\textbf{p}_0}(\textbf{p}) 
= 
\delta(\textbf{p}-\textbf{p}_0) 
+ \frac{
    2m\gamma \phi({\bf p})\phi^*({\bf p}_0)
    }{
    (1-A(\textit{i} p_0+\epsilon))
    (p^2-(p_0 -\textit{i}\epsilon)^2) 
    }
.
\label{Eq:RescM}
 \end{align}

\subsection*{Dipole matrix element}
The dipole matrix element is
\begin{equation}
{\bf d}({\bf p}_0)
=
e\langle \Psi_{{\bf p}_0}|i\hbar{\bf \nabla}_{\bf p}|\Psi_0\rangle
.
\end{equation}
Tedious, but elementary algebra leads to:
\begin{align}
{\bf d}({\bf p}_0)
&=
i\hbar e\mathcal{N}\left[  \frac{{\bf \nabla}_{\bf p}\phi({\bf p}_0)}{\frac{p_0^2}{2m} + I_p} -\frac{{\bf p}\:\phi({\bf p}_0)}{\left(\frac{p_0^2}{2m} + I_p\right)^2} \right]
\\ &\quad +
2i\hbar em\gamma \mathcal{N}\phi({\bf p}_0)
\int
\frac{
  \phi^*({\bf p})
  \left[ \left(\frac{p^2}{2m} + I_p\right){\bf \nabla}_{\bf p}\phi({\bf p}) -{\bf p}\:\phi({\bf p}) \right]
  }{
  (1-A(-\textit{i} p_0+\epsilon))
  (p^2-(p_0 +\textit{i}\epsilon)^2)
  \left(\frac{p^2}{2m} + I_p\right)^2
  }
\textit{d}^3 \textbf{p}
 %
%\\ &\quad +
%
%
%\left\{\int  \textit{d}^3 \textbf{p}\frac{\left[({\bf \nabla}_{\bf p}\phi({\bf p}))\phi^*({\bf p})+ 2{\bf p}|\phi({\bf  p})|^2\right]}{(1-
%A(-\textit{i} p_0+\epsilon))(p^2+2I_p)
%(p^2-(p_0-\textit{i}\epsilon)^2)}\right\}
.\nonumber
\end{align}

\subsection*{Continuum-continuum transition matrix element} %/////////////////////
Let us consider the scattering waves obtained in Eq.~(\ref{Eq:RescM}) and evaluate the continuum-continuum transition matrix element of Eq.~(\ref{Eq:CCM1}), i.e.
\begin{equation}
e\langle {\bf p}_1|{\bf x}| {\bf p}_2 \rangle=  ie \hbar {\nabla}_{{\bf p}_1}\delta({\bf p}_1-{\bf p}_2) + \hbar {\bf g}({\bf p}_1,{\bf p}_2).
\end{equation}
Again after tedious (though straightforward) calculations we obtain:
{%
\allowdisplaybreaks%
\begin{align}
\textbf{g}(\textbf{p}_1, \textbf{p}_2) 
&= 
2im\gamma{\bf \nabla}_{\bf p}\left[\frac{\phi({\bf p})\phi^*({\bf p}_2)}{(1-A(\textit{i} p_0+\epsilon))((p^2-(p_2-\textit{i}\epsilon)^2)}\right]\bigg{|}_{{\bf p}={\bf p}_1} 
\nonumber \\ & \qquad 
- 2im\gamma{\bf \nabla}_{\bf p}\left[\frac{\phi^*({\bf p})\phi({\bf p}_1)}{(1-A(-\textit{i} p_1+\epsilon))((p^2-(p_1+\textit{i}\epsilon)^2)}\right]\bigg{|}_{{\bf p}={\bf p}_2}
\nonumber \\ & \qquad +
4m^2\gamma^2 \int  \textit{d}^3 \textbf{p}\left[\frac{\phi({\bf p})\phi^*({\bf p}_2)}{(1-A(\textit{i} p_0+\epsilon))(p^2-(p_2-\textit{i}\epsilon)^2)}\right]
%\cdot
%\nonumber \\ & \qquad \qquad \qquad \qquad \qquad 
{\bf \nabla}_{\bf p}\left[\frac{\phi^*({\bf p})\phi({\bf p}_1)}{(1-A(-\textit{i} p_1+\epsilon))(p^2-(p_1+\textit{i}\epsilon)^2)}\right]
.
\end{align}
}%
All of the above formulae are directly and easily generalized to the case when the Hamiltonian is time-dependent.

\section{Dipole matrix elements}
\label{app:D-dipoles}
In the remaining appendices, we turn to the objects of the two-electron theory and their properties.

\subsection{Dipole Matrix Elements from the ground state}

The dipole matrix element given by the function $\mathbf{d}$ deals with dipole transitions from the two electron ground state into all other admissible two electron states. This is the analogue of the function by the same name for one electron case, given by Eq.~\eqref{Eq:dv}, but now there are three non-zero variants. The first, bearing the strongest resemblance to the one electron case, can be expressed as a sum of two terms
\begin{equation}
	\mathbf{d}(\mathbf{p})=e\braket{0|\hat{\mathbf{r}}_1|\mathbf{p},0}+e\braket{0|\hat{\mathbf{r}}_2|\mathbf{p},0}.
	\label{Eq:2e-d(p)}
\end{equation}
Each term will have two contributions, (1) direct laser induced ionization of the electron acted on directly by the operator, while the other electron remains bound in a ground state and (2) correlated ionization, where the the action of laser on one electron is transferred to the other via electron-electron interaction, this could be through elastic collision. This last contribution is expected to be small and will be zero in the non-interacting case, where Eq.~\eqref{Eq:2e-d(p)} simplifies to
\begin{align}
	\mathbf{d}(\mathbf{p})&=\frac{e}{\sqrt{2}}\left(\bra{0}\bra{0}\hat{\mathbf{r}}_1\otimes\mathbb{I}_2\ket{\mathbf{p}}\ket{0}+\bra{0}\bra{0}\mathbb{I}_1\otimes\hat{\mathbf{r}}_2\ket{0}\ket{\mathbf{p}}\right)\\
	%&=\frac{e}{\sqrt{2}}\left(\braket{0|(\hat{\mathbf{r}}_1+\hat{\mathbf{r}}_2)|\mathbf{p}}\right)
	&=\frac{2e}{\sqrt{2}}\braket{0_1|\hat{\mathbf{r}}|\mathbf{p}}
\end{align} The final expression is written in terms of 1-electron states and operators. This dipole matrix element is vital to most strong field processes as it will describe the initial tunnelling step, hence it is necessary to model NSDI for both the EI and RESI mechanism. The subscript denotes whether it is the first or second electron to be ionized, this is significant as the second will have a much larger ionization potential, which will make it less probable and for both mechanisms of NSDI we will neglect this contribution.

The other two variants of the $\mathbf{d}$ dipole matrix element require at least some electron interaction in all contributing processes and so will both be zero in the non-interacting case. The element $\mathbf{d}(\mathbf{p},\eta)$ %can be expanded to
%\begin{equation}
%	\mathbf{d}(\mathbf{p},\eta)=e\braket{0|\hat{\mathbf{r}}_1|\mathbf{p},\eta}+e\braket{0|\hat{\mathbf{r}}_2|\mathbf{p},\eta}
%\end{equation}
will again have two contributions, in the first the operator acts on an electron to ionize it and the electron-electron correlation causes the other electron to be excited in a `shake-up' process. This would be the main term involved in the previously proposed shake-off mechanism for NSDI \cite{Fittinghoff1992}, which has since fallen out of favour in preference of the re-collision mechanisms EII and RESI. This will be the dominant of the two contributions. 
%In the second \textcolor{red}{(presumably)} less likely transition, the laser excites an electron and the electron-electron correlation causes the second electron to be fully ionized. 
In the second (presumably less likely) transition, the laser excites an electron and the electron-electron correlation causes the second electron to be fully ionized. 
In both these possibilities the energy transfer between electron could be through direct collision or any other electron-electron interaction, however in the first only some of the electron's energy is transferred, while in the second scenario most of the electron's energy will be transferred to the other one.

For the dipole matrix element $\mathbf{d}(\mathbf{p},\mathbf{p}')$ the only transition is one where the laser-induced ionization of an electron through interactions between the electrons causes the other electron to ionize as well---in fact, this matrix element, when dressed in the laser field, leads to collective tunnelling, as discussed in Ref.~\citealp{Lewenstein1986a}.

\subsection{Dipole Matrix Elements from the scattering and ground state}
\begin{figure}
	\includegraphics[width=0.45\textwidth]{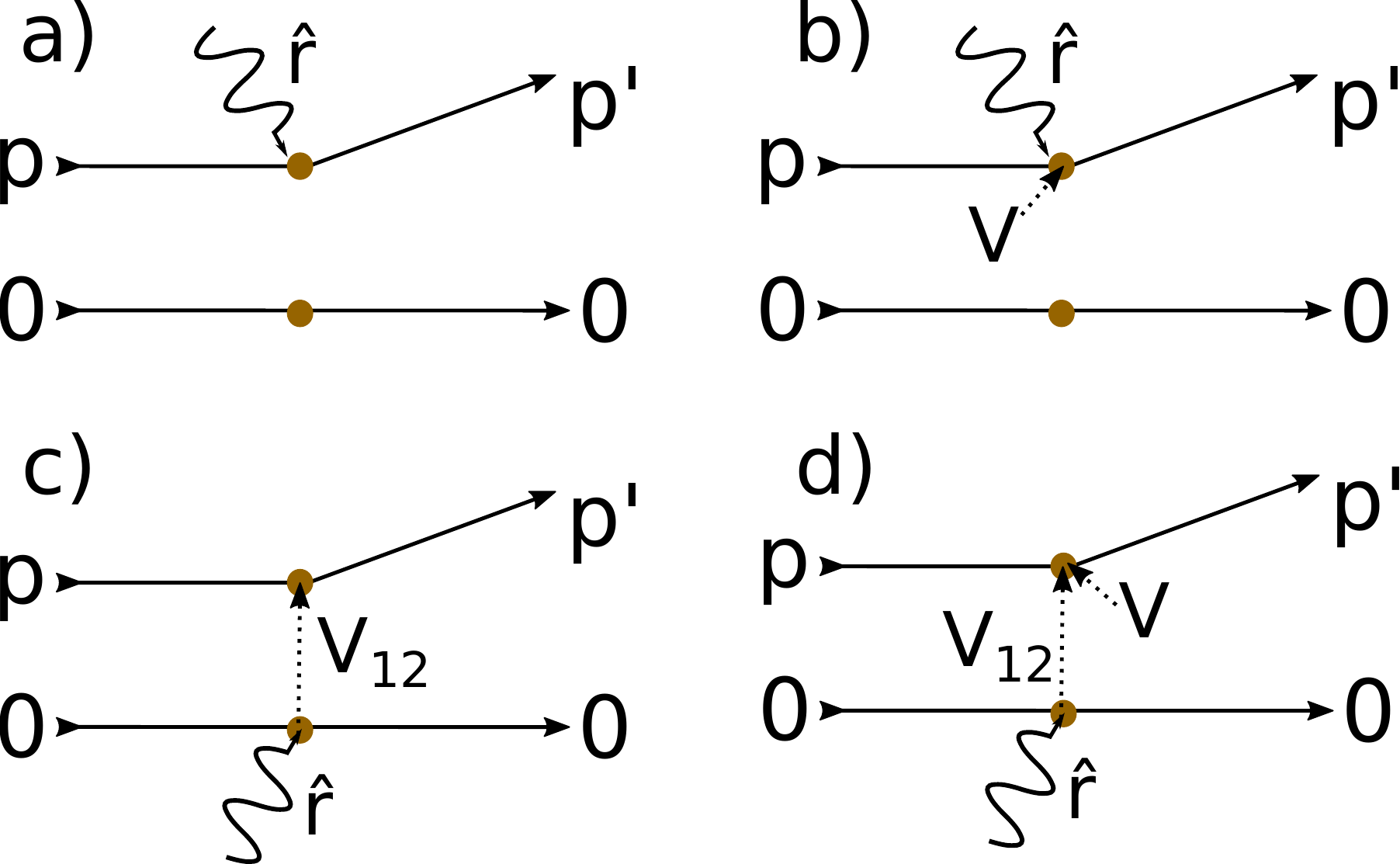}
	\caption{Four main process for the matrix element $g(\mathbf{p},\mathbf{p}')$, the top row [panels a) and b)] show processes where the laser dipole acts on the continuum state, the bottom row [panels c) and d)] show cases where the laser dipole acts on the ground state and electron interaction transfers the energy to the continuum state. The right hand column includes the continuum electrons interaction with the core potential}
	\label{Fig:2e-Feynman1}
\end{figure}

The dipole matrix elements from the two-electron scattering-ground state $\ket{\mathbf{p},0}$ are given by the function $\mathbf{g}$, there are three possible variants. The first, $\mathbf{g }(\mathbf{p},\mathbf{p}')$ relates to the transition from the two electron continuum-ground state $\ket{\mathbf{p},0}$ to an alternative continuum-ground state $\ket{\mathbf{p}',0}$. The leading contribution will typically come from the laser induced change of momentum of the continuum state, this term should play a strong role as we expect strong coupling between the laser and continuum electrons. Continuum-continuum transitions can, as in the one electron case, also involve contributions via interaction through the single electron potential. In Eq.~\eqref{Eq:CCM1} this was described by splitting of the continuum-continuum matrix element into two parts in the one-electron case. Alternatively, there is the strongly correlated and less likely process, where the laser acts on the bound electron which through electron-electron interaction changes the momentum of the scattering state for the other electron, without changing the state of the original bound electron. This would be quite an exotic case and generally it is a reasonable approximation to assume the two electrons in this state are somewhat physically separated. Of course, when an electron recollides with its parent atom or molecule then there will be much overlap and this term could contribute to elastic recollision processes such as high-order above-threshold ionization (HATI). In Fig.~\ref{Fig:2e-Feynman1} the four pathways discussed are shown in the form of Feynman diagrams; the non-interacting cases are given by panels a) and b).

This dipole moment can be considerably simplified if we consider non-interacting electrons
\begin{align}
	\mathbf{g}(\mathbf{p},\mathbf{p}')&=\frac{e}{2}\left(
	\bra{\mathbf{p}}\bra{0}\hat{\mathbf{r}}_1\ket{\mathbf{p}'}\ket{0}
	+\bra{0}\bra{\mathbf{p}}\hat{\mathbf{r}}_2\ket{0}\ket{\mathbf{p}'}
	\right),
\intertext{which can be written in terms of one particle state and operators as,}
\mathbf{g}_1(\mathbf{p},\mathbf{p}')&=e\braket{\mathbf{p}|\hat{\mathbf{r}}|\mathbf{p}'}.
\end{align}This can be treated as before by Eq.~\eqref{Eq:CCM1},

\begin{figure}
	\includegraphics[width=0.45\textwidth]{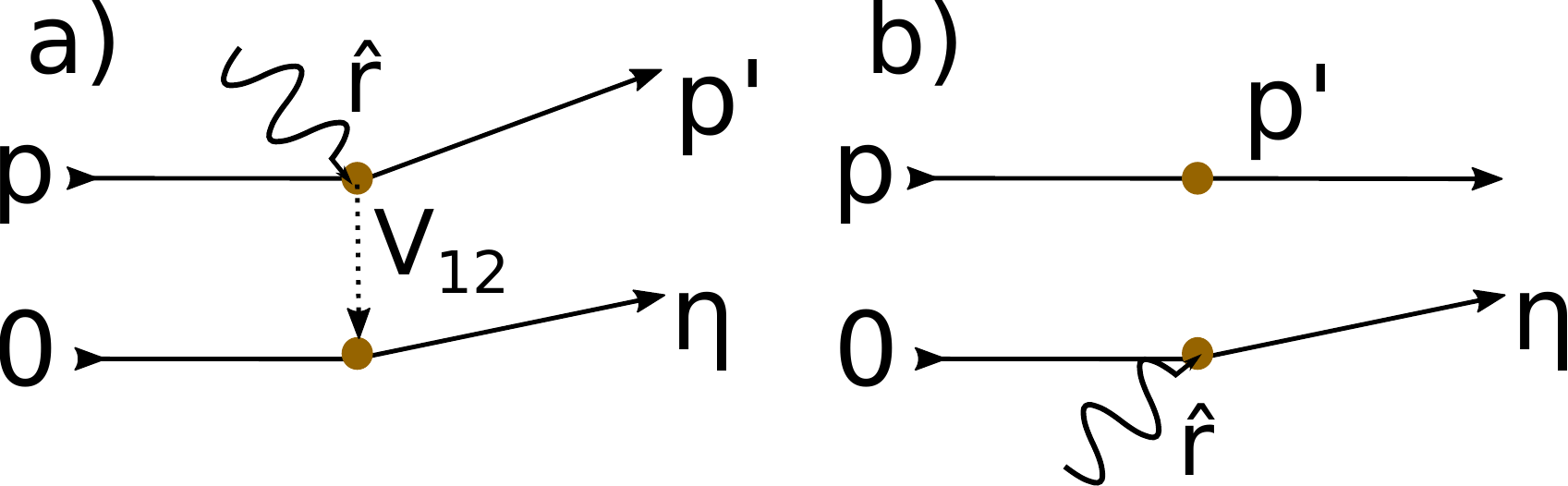}
	\caption{Two process from the matrix element $g(\mathbf{p},\mathbf{p}',\eta)$, panel a) shows the excitation process via electron interaction required in RESI, panel b) shows the laser induced excitation of the bound state while the continuum state remains unaffected.}
	\label{Fig:2e-FeynmanExcitation}
\end{figure}
The dipole matrix element $\mathbf{g}(\mathbf{p},\mathbf{p}',\eta)$ deals with transitions from $\ket{\mathbf{p},0}$ to $\ket{\mathbf{p}',\eta}$. This will have contributions from scattering states interacting with the laser to change the moment from $\mathbf{p}$ to $\mathbf{p}'$, which simultaneously through the electron interactions also leads to excitation of the other electron from its ground state into an excited state with principle quantum number $\eta$. It is crucial to include this to allow for the excitation step of the second electron in the RESI mechanism of NSDI. Alternatively, there is the contribution that the excitation of an electron via the laser also results in the momentum change of the scattering state from $\mathbf{p}$ to $\mathbf{p}'$.  In the non-interacting case the first pathway is has no contribution as the ground state electron cannot be excited without electron correlation. The second pathway is non-zero in the case $\mathbf{p}=\mathbf{p}'$ and the dipole matrix element can be written as
\begin{equation}
	\mathbf{g}(\mathbf{p},\mathbf{p}',\eta)=e\delta(\mathbf{p}-\mathbf{p}')\braket{0_1|\hat{\mathbf{r}}|\eta}
\end{equation}
The excitation via electron interaction and non-interacting pathways are depicted in panel a) and b) of Fig.~\ref{Fig:2e-FeynmanExcitation}, respectively.

\begin{figure}
	\includegraphics[width=0.45\textwidth]{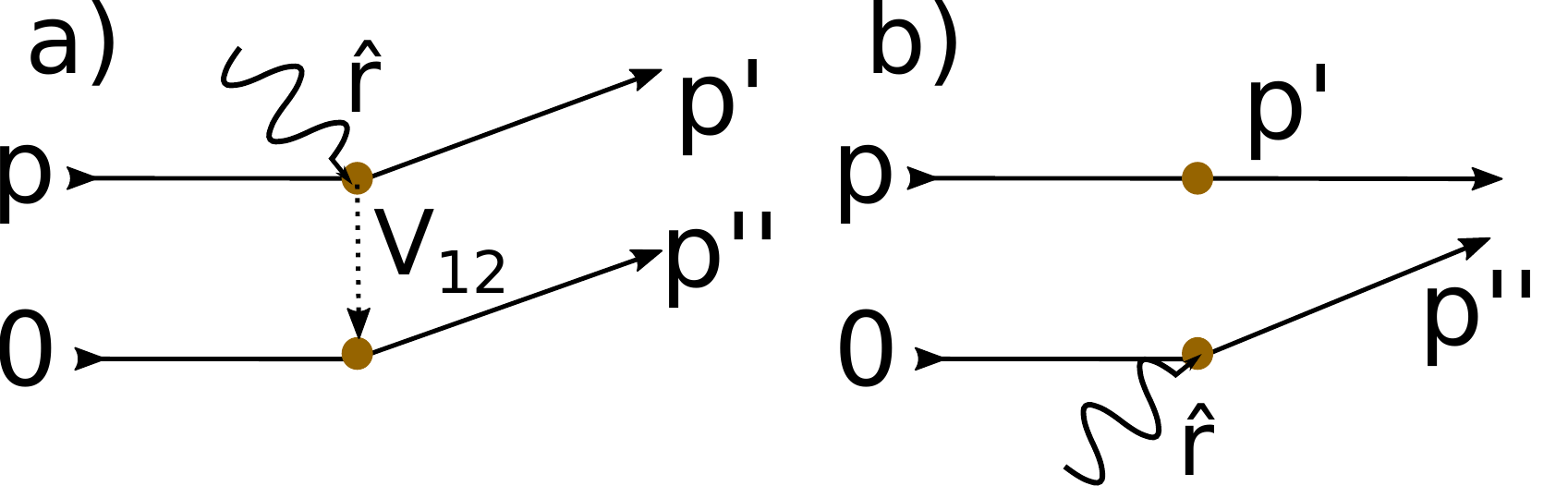}
	\caption{Two process from the matrix element $g(\mathbf{p},\mathbf{p}',\mathbf{p}'')$, panel a) shows the recollision ionization process from the EI of NSDI, panel b) shows the laser induced ionization of the bound state while the continuum state remains unaffected.}
	\label{Fig:2e-FeynmanRecollisionIonisation}
\end{figure}

It is a very similar situation for the dipole matrix element $\mathbf{g}(\mathbf{p},\mathbf{p}',\mathbf{p}'')$, except the excited state is replaced by a scattering state with momentum $\mathbf{p}''$. There are the same kind of contributions, one where the change of momentum of the scattering states leads through electron-electron interaction to ionization of the bound state, this is the crucial ionization step of the second electron in the EI mechanism of NSDI. Another possibility is that the laser induced ionization of the bound electron leads to change of momentum of the scattering state, which could happen through the collision of the two ionized electron. In the non-interacting case the first contribution is zero, while the second is non-zero again if $\mathbf{p}=\mathbf{p}'$ which means the matrix element can be simplified to
\begin{equation}
\mathbf{g}(\mathbf{p},\mathbf{p}',\mathbf{p}'')=e\delta(\mathbf{p}-\mathbf{p}')\braket{0_2|\hat{\mathbf{r}}|\mathbf{p}''}+e\delta(\mathbf{p}-\mathbf{p}'')\braket{0_2|\hat{\mathbf{r}}|\mathbf{p}'}.
\end{equation}
The subscript two denotes that this is the laser induced ionization of a second electron, i.e.\ this electron comes from a +1 ion, thus it is much less probable than the first ionization and for our purposes, where we consider NSDI, we will not consider such contributions.
In Fig.~\ref{Fig:2e-FeynmanRecollisionIonisation} we show the EI recollision excitation pathway and also the non-interacting pathway, in panels a) and b), respectively.

\subsection{Dipole Matrix Elements from the scattering and excited states}
The dipole matrix elements from the two electron continuum-excited state $\ket{\mathbf{p},\eta}$ are given by the function $\mathbf{h}$, there are two variants. The function $\mathbf{h}(\mathbf{p},\eta,\mathbf{p}',\eta')$ deals with transitions to alternate continuum-excited states. This will have contributions such as the laser induced recollision of the scattering state results in the excited electron moving to a another excited state. This will be non-zero in the non-interacting case if either $\mathbf{p}=\mathbf{p}'$ or $\eta=\eta'$ and the dipole matrix element can be written as
\begin{equation}
	\mathbf{h}(\mathbf{p},\eta,\mathbf{p}',\eta')=e\delta_{\eta \eta'}\braket{\mathbf{p}|\hat{\mathbf{r}}|\mathbf{p}'}+e\delta(\mathbf{p}-\mathbf{p}')\braket{\eta|\hat{\mathbf{r}}|\eta'}.
\end{equation}
The other function $\mathbf{h}(\mathbf{p},\eta,\mathbf{p}',\mathbf{p}'')$ deals with transition to continuum-continuum state $\ket{\mathbf{p}',\mathbf{p}''}$.  This will have a contribution where the laser induced recollision of the scattering state changes the momentum from $\mathbf{p}$ to $\mathbf{p}'$ and the excited electron is ionized through electron-electron interaction. Alternatively, the laser induced ionization of the excited electron can change the momentum of the continuum electron through their interaction.  The first contribution vanishes in the non-interacting case, while the second contribution is non-zero if $\mathbf{p}=\mathbf{p}'$. This what leads to the final ionization step in the RESI mechanism of NSDI, where the excited electron tunnel ionizes via the dipole interaction. This dipole matrix element can be written as
\begin{equation}
\mathbf{h}(\mathbf{p},\eta,\mathbf{p}',\mathbf{p}'')=e\delta(\mathbf{p}-\mathbf{p}'')\braket{\eta|\hat{\mathbf{r}}|\mathbf{p}'}+e\delta(\mathbf{p}-\mathbf{p}')\braket{\eta|\hat{\mathbf{r}}|\mathbf{p}''}.
\end{equation}

\subsection{Dipole Matrix Elements from the scattering states}
This dipole matrix element is between two-electron scattering states. This will mostly contribute to the final evolution of the electrons before detection. It will have strong terms, where the laser induces a change in momentum in one electron and the other electron remains unaffected. But it also includes the strongly correlated contribution, where the laser induces recollision and interaction between the two electrons. The dipole matrix element in the non-interacting case is given by
\begin{equation}
\mathbf{i}(\mathbf{p},\mathbf{p}',\mathbf{p}',\mathbf{p}'',\mathbf{p}''')=
 e\delta(\mathbf{p}'-\mathbf{p}''')\braket{\mathbf{p}|\hat{\mathbf{r}}|\mathbf{p}''}
+e\delta(\mathbf{p}'-\mathbf{p}'')\braket{\mathbf{p}|\hat{\mathbf{r}}|\mathbf{p}'''}
+e\delta(\mathbf{p}-\mathbf{p}''')\braket{\mathbf{p}'|\hat{\mathbf{r}}|\mathbf{p}''}
+e\delta(\mathbf{p}-\mathbf{p}'')\braket{\mathbf{p}'|\hat{\mathbf{r}}|\mathbf{p}'''}.
\end{equation}

\section{Two-Electron Integro-Differential Equations}
\label{app:E-two-electron}
The time variations of $a(t)$, $b(\mathbf{p},t)$, $c(\mathbf{p},\eta,t)$ and $d(\mathbf{p}, \mathbf{p}', t)$ follow from the TDSE for two electrons and read:
{
\allowdisplaybreaks
\begin{align}
	%a
	\dot{a}(t)&=-\frac{i}{\hbar}\left[
	\mathbf{E}(t)\cdot\int d^3\mathbf{p}\;b(\mathbf{p},t)\mathbf{d}(\mathbf{p})
	+\mathbf{E}(t)\cdot\sum_{\eta \ne 0}\int d^3\mathbf{p}\;c(\mathbf{p},\eta,t)\mathbf{d}(\mathbf{p},\eta)
	\right. \notag \\ & \qquad \qquad \qquad \qquad \qquad \qquad \qquad \qquad \qquad \qquad \qquad \qquad \left.
	+\mathbf{E}(t)\cdot\iint d^3\mathbf{p}  d^3\mathbf{p}'\; d(\mathbf{p},\mathbf{p}',t)\mathbf{d}(\mathbf{p},\mathbf{p}')
	\right]\notag\\
	%b
	\dot{b}(\mathbf{p},t)&=-\frac{i}{\hbar}\Bigg[
	\left(\frac{p^2}{2m}+E_0-E_{10}\right)b(\mathbf{p},t)
	+\mathbf{E}(t)\cdot a(t) \mathbf{d}^*(\mathbf{p})
	+\mathbf{E}(t)\cdot\int d^3\mathbf{p}'\;b(\mathbf{p}',t)\mathbf{g}(\mathbf{p},\mathbf{p}')
	\notag\\& \hspace{1cm}
	+\mathbf{E}(t)\cdot\sum_{\eta \ne 0}\int d^3\mathbf{p}'\;c(\mathbf{p}',\eta,t)\mathbf{g}(\mathbf{p},\mathbf{p}',\eta)
	+\mathbf{E}(t)\cdot\iint d^3\mathbf{p}'  d^3\mathbf{p}''\; d(\mathbf{p}',\mathbf{p}'',t)\mathbf{g}(\mathbf{p},\mathbf{p}',\mathbf{p}'')
	\Bigg]\notag\\
	%c
	\dot{c}(\mathbf{p},\eta,t)&=-\frac{i}{\hbar}\Bigg[
	\left(\frac{p^2}{2m}+E_0-E_{1\eta}\right)c(\mathbf{p},\eta,t)
	+\mathbf{E}(t)\cdot a(t) \mathbf{d}^*(\mathbf{p},\eta)
	+\mathbf{E}(t)\cdot\int d^3\mathbf{p}'\;b(\mathbf{p}',t)\mathbf{g}^*(\mathbf{p}',\mathbf{p},\eta)
	\notag\\& \hspace{1cm}
	+\mathbf{E}(t)\cdot\sum_{\eta' \ne 0}\int d^3\mathbf{p}'\;c(\mathbf{p}',\eta',t)\mathbf{h}(\mathbf{p},\eta,\mathbf{p}',\eta')
	+\mathbf{E}(t)\cdot\iint d^3\mathbf{p}'  d^3\mathbf{p}''\; d(\mathbf{p}',\mathbf{p}'',t)\mathbf{h}(\mathbf{p},\eta,\mathbf{p}',\mathbf{p}'')
	\Bigg]\notag\\
	%d
	\dot{d}(\mathbf{p},\mathbf{p}',t)&=-\frac{i}{\hbar}\Bigg[
	\left(\frac{p^2}{2m}+\frac{p'^2}{2m}+E_0\right)d(\mathbf{p},\mathbf{p}',t)
	+\mathbf{E}(t)\cdot a(t) \mathbf{d}^*(\mathbf{p}',\mathbf{p})
	+\mathbf{E}(t)\cdot\int d^3\mathbf{p}''\;b(\mathbf{p}'',t)\mathbf{g}^*(\mathbf{p}'',\mathbf{p},\mathbf{p}')
	\notag\\& \hspace{1cm}
	+\mathbf{E}(t)\cdot\sum_{\eta \ne 0}\int d^3\mathbf{p}''\;c(\mathbf{p}'',\eta,t)\mathbf{h}^*(\mathbf{p}'',\eta,\mathbf{p},\mathbf{p}')
	+\mathbf{E}(t)\cdot\iint d^3\mathbf{p}''  d^3\mathbf{p}'''\; d(\mathbf{p}'',\mathbf{p}''',t)\mathbf{i}(\mathbf{p},\mathbf{p}',\mathbf{p}'',\mathbf{p}''')
	\Bigg]
	\label{Eq:2e-New5}
\end{align}
}

\section{HHG from a quenched molecule: A toy model}
\label{app:F-toy-model}
In this appendix we analyze a toy 1D model of a quenched H$_2^+$ molecule, or any similar diatomic molecule, stripped of one or more electrons, and treated in the SAE approximation. The remaining (or SAE)  electron is after the rapid stripping process (say by an ultrashort XUV pulse) in  the ground state of the model separable Hamiltonian. 
In the position representation, the electron feels a potential that consists of two Dirac-delta peaks at positions $\pm R/2$; in the momentum (wave vector) representation, that Hamiltonian reads
\begin{equation}
H(k,k') 
= 
\frac{\hbar^2k^2}{2m } \delta(k-k')
-\tilde\lambda\cos((k-k')R/2),
\end{equation}
where $R$ is the instantaneous distance between the nuclei.
Setting $ \frac{\hbar^2\kappa^2}{2m }=E_0=-I_p$, the ground-state wave function has the form
\begin{equation}
\Phi_0(k)= {\cal N}\frac{\cos(kR/2)}{k^2+\kappa^2}.
\end{equation}
By further denoting
\begin{equation}
I(\kappa,R)
=
\int\, dk\, \frac{\cos(kR)}{k^2+\kappa^2}
=\frac{\pi}{\kappa}\exp(-\kappa R),
\end{equation}
we obtain the equation that determines the energy (through $\kappa$) as
\begin{equation}
1
= \lambda\left(I(\kappa,0)+ I(\kappa,R)\right) 
= \lambda\left(1+\exp(-\kappa R) \right),
\end{equation}
where $\lambda=\pi\tilde\lambda m/\hbar^2$. The normalization  of the ground-state wave function is therefore fixed by
\begin{equation}
\frac{{\cal N}^2}{2}\left(-\frac{\partial}{\partial \kappa^2}\right)\left[I(\kappa,0)+ I(\kappa, R)\right]
=1
.
\end{equation}

Similarly, for this simple model we can calculate the wave functions of the continuum states with the fixed outgoing momentum $p$, and the dipole moment $d(p)=\langle\Phi_p|x|\Phi_0\rangle$.
Here we are interested in approximate qualitative/semi-quantitative description of HHG, so we will calculate the dipole moment $d(p)$ using plane waves, but employing the careful approach of Refs.~\citealp{Noslen1,Noslen2,Noslen3,Noslen4,Noslen-diss}. 
This amounts, essentially, to neglecting ``unphysical'' terms $\propto R$ that arise because of the non-orthogonality of plane waves and the ground state, $|\Phi_0\rangle$;
from that approach, we get
\begin{equation}
d(p)= -2i{\cal N}  \frac{p\cos(pR/2\hbar)}{\left((p/\hbar)^2 + \kappa^2\right)^2}.
\end{equation}

The motion of nuclei in the simplest case can be treated classically with Coulombic repulsion. Denoting the nuclear positions as $x_1$, $x_2$, with $x_1-x_2=r$ and $|r|=R$, and neglecting the electronic energy $\hbar^2\kappa(R)^2/2m$, we obtain
\begin{equation}
\frac{d r}{dt}=\frac{2Zr}{(r^2+a^2)^3/2},
\end{equation}
where $Z$ is a ``charge'', and $a$ is the smoothing parameter in the ``Rochester'' 1D version of the Coulomb potential. 
All of these parameters can and, in fact, should be adjusted to mimic real molecules.

\begin{figure}[b!]
\subfigure[]{%
  \includegraphics [width=0.45\textwidth] {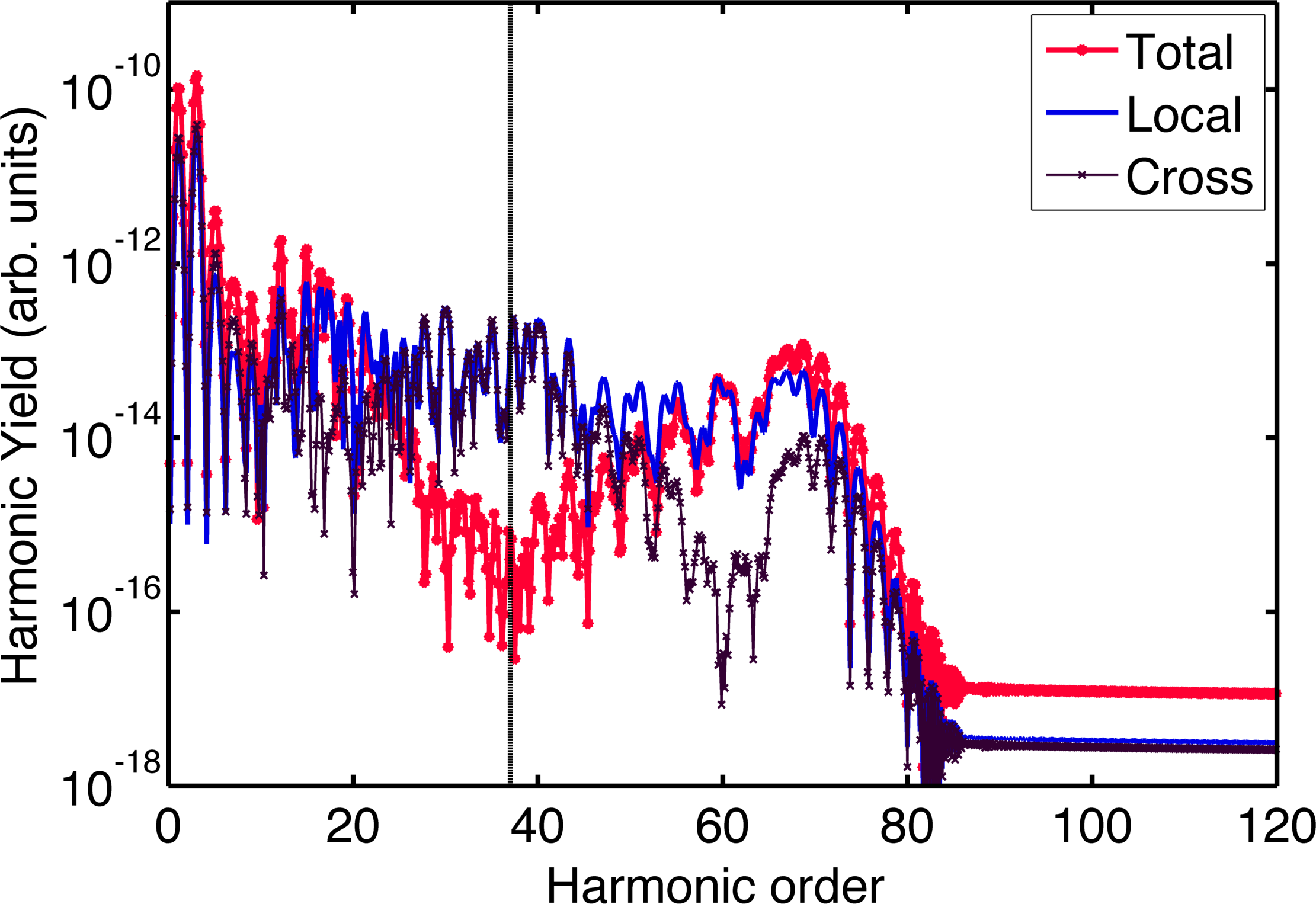}
  \label{fig:subfiga}
  }
\subfigure[]{
  \includegraphics[width=0.45\textwidth]{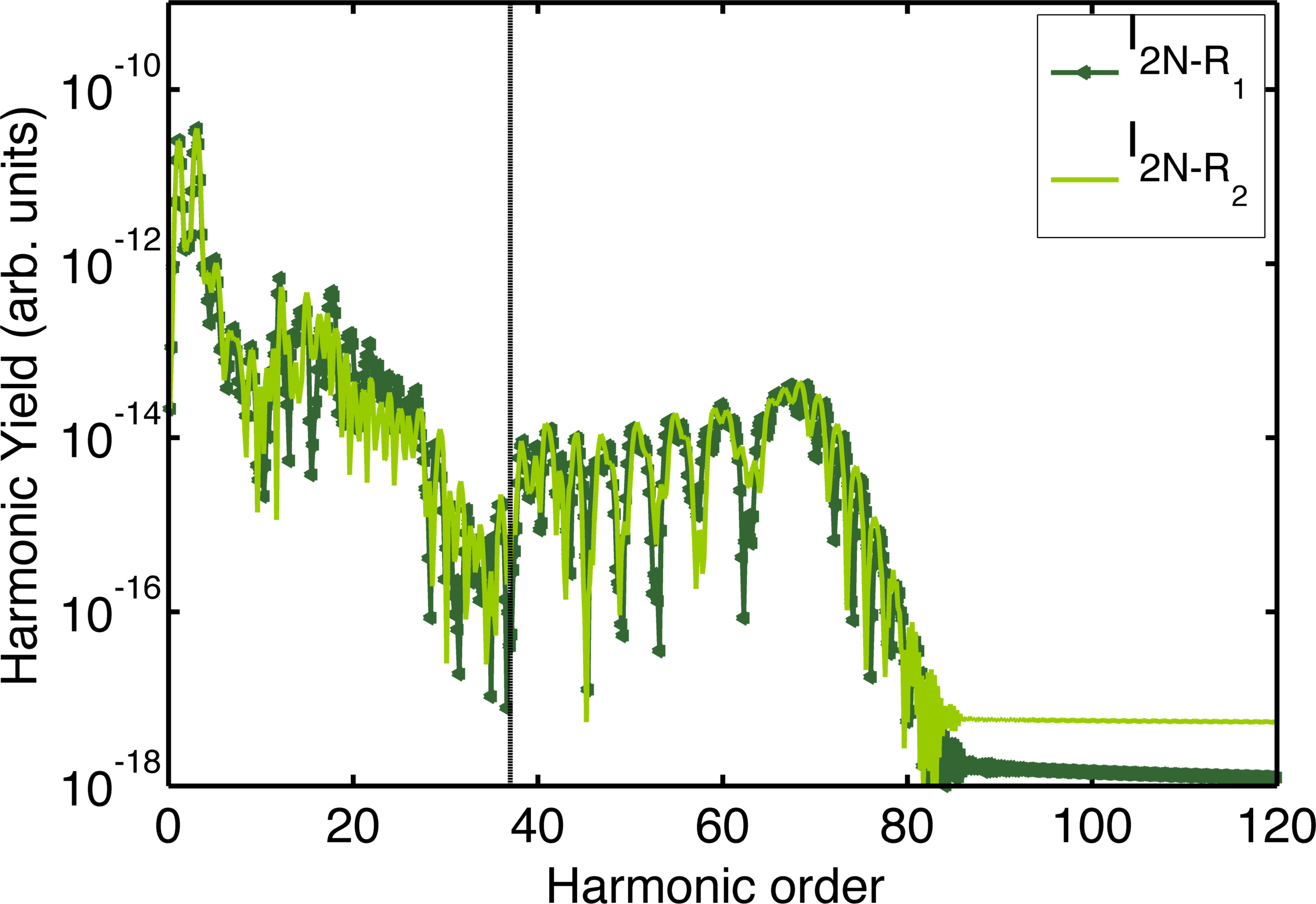}
  }
\caption{%
%(color online) %
Harmonic spectra $I_{2N}(\omega)$ (in logarithmic scale) of an H$_2^+$ molecule as a function of the harmonic order calculated using our quasi-classical SFA and for an orientation angle $\theta= 20^{\circ}$. 
(a) Local, cross and total contributions to the HHG spectrum; 
(b) contributions depending on the recombination atom. 
Green circle line: recombination at $\textbf{R}_1$ and light green line: recombination at $\textbf{R}_2$. The vertical lines indicate the position of the interference minima (see the text for details).
Adapted from Ref.~\citealp{Noslen3}.
}
\label{Fig:H2_ContM}
\end{figure}

In order to get some qualitative/semi-quantitative results, let us make some additional simplifying assumptions and observations:
\begin{itemize}
\item 
The femtosecond pulse is not too short, so its variation  within a laser period can be neglected -- we take it into account at the end integrating over the time-dependent $U_p(t)$.

\item 
During the dissociation of  our toy quenched molecule, $\kappa R$ changes from values $\ll 1$ to the ones $\gg 1$. Note that $R$ for the quenched molecule at the time $t=0$ is much smaller the the equilibrium size of the H$_2^+$ molecule (compare Fig.~\ref{Fig:H2_ContM} and discussion below). Correspondingly, the ionization potential $I_p$ changes from $I_p\simeq I_p(0)$ to $I_p\simeq I_p(\infty)=I_p(0)/4$. In other words, it undergoes a radical reduction by a factor of 4.

\item 
The HHG spectrum cutoff for ``local'' contributions, i.e.\ electronic trajectories starting at $x_i$ and ending at $x_i$,  changes thus from $3.2U_p + I_p(0)$ to $3.2U_p + I_p(0)/4$, which for large $I_p$ is quite significant. If dissociation is very fast, only the cutoff $3.2U_p + I_p(0)/4$ will be observed.

\item 
Similarly, the efficiency and cutoff position for the ``cross'' contribution, i.e.\ electronic trajectories starting at $x_1$ and ending at $x_2$ (and vice versa) also changes significantly with $R$.

\item 
At small $R$ (in the present terms $\kappa R<1$), the cutoffs for ``local'' and ``cross'' contributions are in distinguishable. Their efficiency is comparable, but they contribute with different phases, which leads to destructive interference (see Fig.~\ref{Fig:H2_ContM}, Refs.~\citealp{Noslen1,Noslen2,Noslen3,Noslen4,Noslen-diss} and references therein).

\item 
As $R$ grows, we expect the cutoff of ``cross'' terms to grow, but their efficiency to decrease (see discussion below).

\end{itemize}

As commented above, from this figure we can clearly see that there is a deep minimum for both terms, and that it is located at the same position as for the total HHG spectra. This means that, for the case of the recombination on $\textbf{R}_1$ (dark green circle line), the electron wave packet ionized at $\textbf{R}_1$ interferes with the one coming from $\textbf{R}_2$ and the other way around. These minima are then generated by the destructive interference of such electron wavepackets.  
On the other hand, we have seen that our short-range non-local potential is unable to accurately reproduce the positions of the interference minima for some of the molecular orientation angles. 
We note, however, that these minima are typically washed out when an average over the molecular orientation is considered, a configuration that is commonly used in molecular HHG experiments.

Let us finally discuss more quantitatively the position of the HHG cutoff for the `local' and `cross' terms. For the local one, the quasi-classical theory predicts $3.17U_P + I_p$ (or $3.17U_P + xI_p$, with $x\simeq 1.3$  for small $I_p$, and $x\simeq 1$ for large $I_p$~\cite{Lewenstein1994}.

To this aim we repeat the saddle-point calculations from Ref.~\citealp{Lewenstein1994} for the cross trajectories, setting in this last example $e=1$, $c=1$ and $\omega=1$, to make the expressions essentially equal to the ones used in the 1994 paper. The equation for the momentum gives
\begin{equation}
\int_{t-\tau}^t dt' (p_\mathrm{st}- A(t'))dt'= pm R, 
\end{equation}
which leads to
\begin{equation}
p_\mathrm{st}(t, \tau, R) = p_{st}(t, \tau, 0)+R/\tau,
\end{equation}
where we have fixed the trajectory to the one going from $-R/2$ to $R/2$. 
We can now calculate the gain of kinetic energy which should be equal to the emitted photon energy
\begin{equation}
E_\mathrm{kin}(t)-E_\mathrm{kin}(t-\tau) = (p_\mathrm{st}(t, \tau, 0)+R/\tau - A(t)/2 -A(t-\tau)/2)(A(t-\tau)-A(t)).
\end{equation}
Taking $E(t)=\cos(t)$, $A(t)=-\sin(t)$, we get 
\begin{equation}
\Delta E_\mathrm{kin}(t, \tau) 
= 2U_p\sin(2t-\tau)\left[\sin(\tau)-4\frac{\sin(\tau/2)^2}{\tau}\right] 
+ 2\sqrt{U_p}R\sin(\tau/2)\cos(t-\tau/2).
\end{equation}
Obviously, to estimate the cutoff, we need to maximize the above expression first with respect to $t$,  or better to say $t-\tau/2$, and then with respect to $\tau$.  The first term gives exactly $3.17 U_p$ as it does for the ``local'' trajectories. The second can give maxima which are much larger, depending on $R$, but they occur at completely ``dephased'' instants of $t-\tau/2$ -- note that when the second term in maximal, $\cos(t-\tau/2)=\pm 1$, the first term vanishes. This is why, although  the cutoff for the  the ``crossed'' trajectories might be very long, their efficiency for large $R$ is not.

\section{RESI and EII using perturbative expressions for matrix elements}
\label{app:G-RESI-EII}

\subsection{RESI}
We have left the dipole matrix element $\mathbf{g}(\mathbf{p},\mathbf{p}',\eta)$ unevaluated as we will deal with this next. We need to take into account interactions here and to do this we can use first order perturbation theory. 
In the one-electron case, to include a direct and `rescattering'-like contribution a perturbative series was used on the coefficients themselves. In this case in order to get a more explicit form of the electron-electron contribution we will use perturbation theory on the states themselves. We can split the states into a non-interacting part and a first order approximation to the interacting part as shown here
\begin{align}
\ket{\mathbf{p},0}&=\ket{\psi_0(\mathbf{p})}+\ket{\psi_1(\mathbf{p})}
,
\end{align}
where $\ket{\psi_0(\mathbf{p})}$ is given by the non-interacting states for $\ket{\mathbf{p},0}$ and 
\begin{align}
\ket{\psi_1(\mathbf{p})}
&=
\sum_{\eta' \ne 0}\int d^3\mathbf{p}''\; \frac{2m \braket{\psi_0(\mathbf{p}',\eta)|V_{12}|\psi_0(\mathbf{p})}}
{\hbar^2( p^2- p'^2)-2m(E_{10}-E_{1\eta})}\ket{\psi_0(\mathbf{p}'',\eta')},
\label{Eq:2e-psi_1}
\end{align}
where $\ket{\psi_0(\mathbf{p}',\eta)}$ is given by the non-interacting states for $\ket{\mathbf{p},\eta}$. In the basis used in the perturbative expansion we have used excited states but also including the grounds state, i.e. $\eta=0$ as a basis to expand in. This is key to including transition from the ground state to excited states via electron interaction. Substituting this into the matrix element gives,
\begin{align}
	\mathbf{g}(\mathbf{p},\mathbf{p}',\eta)&=\braket{\mathbf{p},0|e(\hat{\mathbf{r}}_1+\hat{\mathbf{r}}_2)|\mathbf{p}',\eta}\label{Eq:2e-g2}\\
	&=\braket{\psi_0(\mathbf{p})|e(\hat{\mathbf{r}}_1+\hat{\mathbf{r}}_2)|\mathbf{p}',\eta}+\braket{\psi_1(\mathbf{p})|e(\hat{\mathbf{r}}_1+\hat{\mathbf{r}}_2)|\mathbf{p}',\eta}
\end{align}
The first part is just given by the non-interacting form that we previously calculated for this matrix element, while for the second we can insert Eq.~\eqref{Eq:2e-psi_1}
{
\allowdisplaybreaks
\begin{align}
\mathbf{g}(\mathbf{p},\mathbf{p}',\eta)
&=
e\delta(\mathbf{p}-\mathbf{p}')\braket{0_1|\hat{\mathbf{r}}|\eta}
	+\sum_{\eta' \ne 0}\int d^3 \mathbf{p}''\; \left(\frac{2me \braket{\psi_0(\mathbf{p}',\eta)|V_{12}|\psi_0(\mathbf{p})}}
	{\hbar^2( p^2- p'^2)-2m(E_{10}-E_{1\eta})}\right)^*
\underbrace{\braket{\psi_0(\mathbf{p}'',\eta')|e(\hat{\mathbf{r}}_1+\hat{\mathbf{r}}_2)|\mathbf{p}',\eta}}
_{\mathbf{h}(\mathbf{p}'',\eta',\mathbf{p}',\eta)}
\\
&=
e\delta(\mathbf{p}-\mathbf{p}')\braket{0_1|\hat{\mathbf{r}}|\eta}
+\sum_{\eta' \ne 0}\int d^3 \mathbf{p}''\;\left(
\delta_{\eta' \eta}\left(
ie \hbar \nabla_{\mathbf{p}''}\delta(\mathbf{p}''-\mathbf{p}')+\hbar\tilde{\mathbf{g}}(\mathbf{p}'',\mathbf{p}')
\right)
+e \delta(\mathbf{p}''-\mathbf{p}')\braket{\eta'|\hat{\mathbf{r}}|\eta}
\right)
\notag\\ &\hspace{5cm}\times 
\left(\frac{2me \braket{\psi_0(\mathbf{p})|V_{12}|\psi_0(\mathbf{p}',\eta)}}
{\hbar^2( p^2- p'^2)-2m(E_{10}-E_{1\eta})}\right)
\\
&=
e\delta(\mathbf{p}-\mathbf{p}')\braket{0_1|\hat{\mathbf{r}}|\eta}
+\nabla_{\mathbf{p}'}\frac{2me \braket{\psi_0(\mathbf{p})|V_{12}|\psi_0(\mathbf{p}',\eta)}}
{\hbar^2( p^2- p'^2)-2m(E_{10}-E_{1\eta})}
\notag\\
&\hspace{3.12cm}
+\int d^3\mathbf{p}''\frac{2me \braket{\psi_0(\mathbf{p})|V_{12}|\psi_0(\mathbf{p}',\eta)}}
{\hbar^2( p^2- p'^2)-2m(E_{10}-E_{1\eta})}\tilde{g}(\mathbf{p}'',\mathbf{p}')
\notag\\
&\hspace{3.12cm}
+
\sum_{\eta' \ne 0}\frac{e}{\Delta E}\braket{\psi_0(\mathbf{p})|V_{12}|\psi_0(\mathbf{p}',\eta')}\braket{\eta'|\hat{\mathbf{r}}|\eta}
\end{align}
}
The last line shows the different pathways we have revealed by this expansion. Each term in the above equation relates to a different physical process: 
the first is where the ground state is excited by the laser while the scattering state is unaffected; 
the second term relates to the continuum electron being driven by the laser to interact with the ground-state electron, causing it to be excited -- this is the most relevant one for RESI. 
These first two are shown in Fig.~\ref{Fig:2e-FeynmanExcitation}. 
The third term is similar to the second except there is an additional interaction with the core for the continuum electron, this could be from re-scattering after the RESI process has taken place. 
The final term excites the second electron in two steps through electron interaction and the laser via an intermediate state. 
An additional pathway can be revealed if the same expansion is applied to the right ket state of Eq.~\eqref{Eq:2e-g2}. 
This pathway is the same as the latter, except the order of the interactions is reversed and the laser excites the electron into the intermediate state from the ground state. We did not consider this, given these pathways will not contribute to the core of the RESI process.

Now we make the assumption that there will be no laser induced transitions between bound states and the matrix element simplifies it to
\begin{equation}
	\mathbf{g}(\mathbf{p},\mathbf{p}',\eta)=\nabla_{\mathbf{p}'}\frac{2me \braket{\psi_0(\mathbf{p})|V_{12}|\psi_0(\mathbf{p}',\eta)}}
	{\hbar^2( p^2- p'^2)-2m(E_{10}-E_{1\eta})}.
\end{equation}
This assumption also simplifies the integro-differential equations and we can insert the calculation we performed for $\mathbf{g}(\mathbf{p},\mathbf{p}',\eta)$, this yields
{
\allowdisplaybreaks
\begin{align}
%a
\dot{a}(t)&=-\frac{2ie}{\hbar}\mathbf{E}(t)\cdot\int d^3\mathbf{p}\; \braket{0_1|\hat{\mathbf{r}}|\mathbf{p}}b(\mathbf{p},t)\notag\\
%b
\dot{b}(\mathbf{p},t)&=-\frac{i}{\hbar}\Bigg[
\left(\frac{\hbar^2p^2}{2m}+E_0-E_{10}\right)b(\mathbf{p},t)
+ \frac{2 e \mathbf{E}(t)}{\sqrt{2}}a(t)\braket{\mathbf{p}|\hat{\mathbf{r}}|0_1}+ie \hbar\mathbf{E}(t)\cdot\mathbf{\nabla}_{\mathbf{p}}b(\mathbf{p},t)
\notag\\& \hspace{1cm}
+\mathbf{E}(t)\cdot\sum_{\eta \ne 0}\int d^3\mathbf{p}'\; \nabla_{\mathbf{p}'}\frac{2me \braket{\psi_0(\mathbf{p})|V_{12}|\psi_0(\mathbf{p}',\eta)}}
{\hbar^2( p^2-p'^2)-2m(E_{10}-E_{1\eta})}c(\mathbf{p},\eta,t)
\Bigg]\notag\\
%c
\dot{c}(\mathbf{p},\eta,t)&=-\frac{i}{\hbar}\Bigg[
\left(\frac{\hbar^2p^2}{2m}+E_0-E_{1\eta}\right)c(\mathbf{p},\eta,t)
+\mathbf{E}(t)\cdot\int d^3\mathbf{p}'\; \nabla_{\mathbf{p}}\frac{2me \braket{\psi_0(\mathbf{p},\eta)|V_{12}|\psi_0(\mathbf{p}')}}
{\hbar^2( p'^2 - p^2)-2m(E_{10}-E_{1\eta})}b(\mathbf{p}',t)
\notag\\& \hspace{1cm}
+ie\hbar \mathbf{E}(t)\mathbf{\nabla}_{\mathbf{p}}c(\mathbf{p},\eta,t)
+2 e \mathbf{E}(t)\cdot\int d^3\mathbf{p}'\; \braket{\eta|\hat{\mathbf{r}}|\mathbf{p}'}d(\mathbf{p},\mathbf{p}',t)
\Bigg]\notag\\
%d
\dot{d}(\mathbf{p},\mathbf{p}',t)&=-\frac{i}{\hbar}\Bigg[
\left(\frac{\hbar^2p^2}{2m}+\frac{\hbar^2p'^2}{2m}+E_0\right)d(\mathbf{p},\mathbf{p}',t)
+e\mathbf{E}(t)\cdot\sum_{\eta \ne 0}(c(\mathbf{p}',\eta,t)\braket{\mathbf{p}|\hat{\mathbf{r}}|\eta}+c(\mathbf{p},\eta,t)\braket{\mathbf{p}'|\hat{\mathbf{r}}|\eta})
\notag\\& \hspace{1cm}
+\mathbf{E}(t)\cdot(\mathbf{\nabla}_{\mathbf{p}}+\mathbf{\nabla}_{\mathbf{p}'})(d(\mathbf{p},\mathbf{p}',t)+d(\mathbf{p}',\mathbf{p},t))
\Bigg]
\label{Eq:2e-New5-RESI2}
\end{align}
}
Now we select only terms that contribute to the final probability flow into double ionized states and ignore the flow in the opposite direction, as such we will assume $a(t)=1$ is unity throughout. This simplifies the integro-differential equations into a solvable form,
{
\allowdisplaybreaks
\begin{align}
%b
	\dot{b}(\mathbf{p},t)&=-\frac{i}{\hbar}\Bigg[
	\left(\frac{\hbar^2p^2}{2m}+E_0-E_{10}\right)b(\mathbf{p},t)
	+ \frac{2 e \mathbf{E}(t)}{\sqrt{2}}\braket{\mathbf{p}|\hat{\mathbf{r}}|0_1}
	+ie \hbar\mathbf{E}(t)\cdot\mathbf{\nabla}_{\mathbf{p}}b(\mathbf{p},t)
	\Bigg]\\
%c
	\dot{c}(\mathbf{p},\eta,t)&=-\frac{i}{\hbar}\Bigg[
	\left(\frac{\hbar^2p^2}{2m}+E_0-E_{1\eta}\right)c(\mathbf{p},\eta,t)
	+\mathbf{E}(t)\cdot\int d^3\mathbf{p}'\; \nabla_{\mathbf{p}}\frac{2me \braket{\psi_0(\mathbf{p},\eta)|V_{12}|\psi_0(\mathbf{p}')}}
	{\hbar^2( p'^2 - p^2)-2m(E_{10}-E_{1\eta})}b(\mathbf{p}',t)
	\notag\\& \hspace{1cm}
	+ie\hbar \mathbf{E}(t)\mathbf{\nabla}_{\mathbf{p}}c(\mathbf{p},\eta,t)
	\Bigg]\\
%d
	\dot{d}(\mathbf{p},\mathbf{p}',t)&=-\frac{i}{\hbar}\Bigg[
	\left(\frac{\hbar^2p^2}{2m}+\frac{\hbar^2p'^2}{2m}+E_0\right)d(\mathbf{p},\mathbf{p}',t)
	+e\mathbf{E}(t)\cdot\sum_{\eta \ne 0}(c(\mathbf{p}',\eta,t)\braket{\mathbf{p}|\hat{\mathbf{r}}|\eta}+c(\mathbf{p},\eta,t)\braket{\mathbf{p}'|\hat{\mathbf{r}}|\eta})
	\notag\\& \hspace{1cm}
	+\mathbf{E}(t)\cdot(\mathbf{\nabla}_{\mathbf{p}}+\mathbf{\nabla}_{\mathbf{p}'})(d(\mathbf{p},\mathbf{p}',t)+d(\mathbf{p}',\mathbf{p},t))
	\Bigg]
\end{align}
}

\subsection{EII}
Here, we neglect the dipole matrix element for the recollision-excitation (RESI) contribution given by $\mathbf{g}(\mathbf{p},\mathbf{p}',\eta)$, and instead include in a similar way the matrix element $\mathbf{g}(\mathbf{p},\mathbf{p}',\mathbf{p}'')$. We can proceed as before splitting the wavefunction into an non-interacting part and an interacting perturbation, where the perturbation is given by
\begin{equation}
	\ket{\psi_1(\mathbf{p})}=\iint d^3\mathbf{k} d^3 \mathbf{k}'\; \frac{2 m\braket{\psi_0(\mathbf{k},\mathbf{k}')|V_{12}|\psi_0(\mathbf{p})}}{\hbar^2 (\mathbf{p}-\mathbf{k}-\mathbf{k}')-E_{10}}\ket{\psi_0(\mathbf{k},\mathbf{k}')}.
\end{equation}
Then removing  dipole transitions for the second electron from the bound state to the continuum to keep only the EI mechanism for the $\mathbf{g}(\mathbf{p},\mathbf{p}',\mathbf{p}'')$ matrix elements yields
\begin{equation}
	\mathbf{g}(\mathbf{p},\mathbf{p}',\mathbf{p}'')=2(\nabla_{\mathbf{p}'}+\nabla_{\mathbf{p}''})\frac{2m \braket{\psi_0(\mathbf{p}',\mathbf{p}'')|V_{12}|\psi_0(\mathbf{p}))}}{\hbar^2(p^2-p'^2-p''^2-E_{10})}
	%+2\nabla_{\mathbf{p}''}\frac{2m\braket{\psi_0(\mathbf{p}'',\mathbf{p}'|V_{12}|\psi_0(\mathbf{p}))}}
	%{\hbar^2(p^2-p''^2-p'^2-E_{10})}
\end{equation}

Now the integro-differential equation for $d(\mathbf{p},\mathbf{p}',t)$ can be written out, this time it only depends on 	$b(\mathbf{p}'',t'')$ and is given by
\begin{align}
	\dot{d}(\mathbf{p},\mathbf{p}',t)&=-\frac{i}{\hbar}\Bigg[
	\left(\frac{\hbar^2p^2}{2m}+\frac{\hbar^2p'^2}{2m}+E_0\right)d(\mathbf{p},\mathbf{p}',t)
	+2\mathbf{E}(t)\cdot \int d^3 \mathbf{p}''\;
	b(\mathbf{p}'',t'')(\nabla_{\mathbf{p}}+\nabla_{\mathbf{p}'})
	\frac{2 m \braket{\psi_0(\mathbf{p}'')|V_{12}|\psi_0(\mathbf{p},\mathbf{p}')}}{\hbar(p''^2-p^2-p'^2)-E_{10}}
	\notag\\&\hspace{3cm}
	+ie\hbar \mathbf{E}(t)\cdot(\mathbf{\nabla}_{\mathbf{p}}+\mathbf{\nabla}_{\mathbf{p}'})(d(\mathbf{p},\mathbf{p}',t)+d(\mathbf{p}',\mathbf{p},t))
	\Bigg].
\end{align}
Then the solution can be written as
\begin{align}
 	d(\mathbf{p},\mathbf{p}',t)&=i\int_0^{t}dt''\;
 	2e\mathbf{E}(t'')\cdot\int d^3 \mathbf{p}''\; b(\mathbf{p}'',t'')(\mathbf{\nabla}_{\mathbf{p}}+\mathbf{\nabla}_{\mathbf{p}'})
 	\frac{2 m \braket{\psi_0(\mathbf{p}'')|V_{12}|\psi_0(\mathbf{p}-\frac{e}{c}\mathbf{A}(t''),\mathbf{p}'-\frac{e}{c}\mathbf{A}(t''))}}{\hbar(p''^2-(p-\frac{e}{c}\mathbf{A}(t''))^2-(p'-\frac{e}{c}\mathbf{A}(t''))^2)-E_{10}}
 	\notag\\&\hspace{4cm}
 	\times\exp\left[\frac{i}{\hbar} S_d\left(\mathbf{p},\mathbf{p}',t'',t\right)\right]
 	.
\end{align}
The same equations for $b(\mathbf{p}'',t'')$ and $S_d\left(\mathbf{p},\mathbf{p}',t'',t\right)$ can be used as before. Additionally, both the EII and RESI mechanisms can be included in $d(\mathbf{p},\mathbf{p}',t)$, and it will still be integrable and, as expected, will simply be equal to the sum of these two solutions.

\bibliographystyle{arthur} 
\bibliography{references}{}

\begin{thebibliography}{100}
\bibitem{footnote}
The workshop was famous also for the fact that there was a flood of the river
  and there was no fresh water, so that the participants had to fight their
  thirst by drinking Belgian beer in large quantities.
\bibAnnoteFile{footnote}
\end{thebibliography}

\end{document}